\documentclass{article}

\usepackage{arxiv}

\usepackage[utf8]{inputenc} 
\usepackage[T1]{fontenc}    
\usepackage{hyperref}       
\usepackage{url}            
\usepackage{booktabs}       
\usepackage{amsfonts}       
\usepackage{nicefrac}       
\usepackage{microtype}      
\usepackage{lipsum}		    
\usepackage{doi}
\usepackage{caption}
\usepackage{subcaption}
\usepackage{graphicx}
\usepackage{proba}
\usepackage{bm}
\usepackage{bbm}
\usepackage{enumitem}
\usepackage{xcolor}
\usepackage{afterpage}
\usepackage[authoryear]{natbib}
\usepackage[symbol]{footmisc}
\usepackage{amsmath}
\usepackage{cleveref}
\usepackage{lscape}
\usepackage{mwe}
\usepackage{adjustbox}
\usepackage{multirow}
\usepackage[flushleft]{threeparttable}
\usepackage{lineno}
\usepackage{soul}
\usepackage{array}
\usepackage{algpseudocode}
\usepackage{algorithm}
\usepackage[authoryear]{natbib}

\usepackage{xr-hyper}
\newcolumntype{A}{ >{${}} r <{$} @{} >{${}} l <{$} } 

\title{Assessing Driving Risk Through Unsupervised Detection of Anomalies in Telematics Time Series Data}

\author{
    Ian Weng Chan\thanks{Department of Statistical Sciences, University of Toronto. Ontario Power Building, 700 University Avenue, 9th Floor, Toronto, ON M5G 1Z5, Canada. Email addresses: \texttt{ianweng.chan@mail.utoronto.ca} (Ian Weng Chan), \texttt{andrei.badescu@utoronto.ca} (Andrei L.~Badescu), \texttt{sheldon.lin@utoronto.ca} (X. Sheldon Lin).}\\ 
	\and
	{Andrei L.~Badescu$^*$} \\
        \and
	{X. Sheldon Lin$^*$} \\
}

\date{March, 2025}

\renewcommand{\headeright}{}
\renewcommand{\shorttitle}{Assessing Driving Risk Through Unsupervised Detection of Anomalies in Telematics Time Series Data}

\hypersetup{
pdftitle={Trip Level Telematics},
pdfsubject={},
pdfauthor={},
pdfkeywords={},
}

\begin{document}
\maketitle

\begin{abstract}

Vehicle telematics provides granular data for dynamic driving risk assessment, but current methods often rely on aggregated metrics (e.g., harsh braking counts) and do not fully exploit the rich time-series structure of telematics data.  In this paper, we introduce a flexible framework using continuous-time hidden Markov model (CTHMM) to model and analyze trip-level telematics data.  Unlike existing methods, the CTHMM models raw time-series data without predefined thresholds on harsh driving events or assumptions about accident probabilities.  Moreover, our analysis is based solely on telematics data, requiring no traditional covariates such as driver or vehicle characteristics.  Through unsupervised anomaly detection based on pseudo-residuals, we identify deviations from normal driving patterns --- defined as the prevalent behaviour observed in a driver's history or across the population --- which are linked to accident risk.  Validated on both controlled and real-world datasets, the CTHMM effectively detects abnormal driving behaviour and trips with increased accident likelihood.  In real data analysis, higher anomaly levels in longitudinal and lateral accelerations consistently correlate with greater accident risk, with classification models using this information achieving ROC-AUC values as high as 0.86 for trip-level analysis and 0.78 for distinguishing drivers with claims.  Furthermore, the methodology reveals significant behavioural differences between drivers with and without claims, offering valuable insights for insurance applications, accident analysis, and prevention.

\end{abstract}


\newpage
\section{Introduction}
\label{Introduction}

Traditional auto-insurance ratemaking relies on driver and vehicle characteristics --- such as driver's age, driving experience, and vehicle specifications --- for risk classification, claim prediction, and premium determination.  These `traditional' covariates provide a static representation of risk but fail to capture real-time driving behaviour that significantly influence accident likelihood.  The rise of telematics, enabled by on-board diagnostics systems and smartphone applications, has introduced granular data on speed, acceleration, and GPS location.  Such data offers a dynamic perspective on risk and is increasingly adopted by insurers to supplement traditional models, with the potential to enhance pricing accuracy and fairness (\citet{eling_impact_2020}).

However, the use of telematics data faces two primary challenges.  First, identifying meaningful features from high-frequency telematics data is complex.  Traditional feature engineering aggregates metrics like harsh braking, cornering events, or time spent driving on specific road types (see \citet{paefgen_multivariate_2014}, \citet{verbelen_unravelling_2018}, \citet{bian_good_2018}, \citet{jin_latent_2018}, \cite{denuit_multivariate_2019}, \citet{ayuso_improving_2019}, \citet{huang_automobile_2019}, \citet{sun_assessing_2020}, \citet{longhi_car_2020}), but these approaches often neglect critical temporal patterns, trends, and interactions between telematics variables.  Emerging methods, such as speed-acceleration heatmaps (\citet{wuthrich_covariate_2017}, \citet{gao_claims_2019}, \citet{gao_boosting_2022}) and time-series modelling with neural networks (\citet{fang_mocha_2021}), have been proposed to analyze raw telematics data.  While promising, these techniques demand careful feature design and interpretation, particularly for insurance applications.

Second, integrating telematics into insurance models requires balancing the granular nature of telematics data with the sparsity of claims data.  Accidents are rare and claims are often aggregated over policy periods, whereas telematics data is recorded continuously and at high frequency.  Most existing studies address this mismatch by aggregating telematics metrics at the trip level and regressing annual claim probabilities and frequencies on these features.  However, these approaches often assume stable driving behaviour over time, failing to capture the variability and evolving patterns in risk.  These limitations necessitate new methods which preserve the richness of telematics data and perform risk evaluation on a more granular level.


We would like to highlight that this study is not a typical actuarial paper focused on modelling claim numbers or developing pricing frameworks.  Rather, it aims to understand driving behaviour using telematics data, detect anomalies, and evaluate risk at both the trip and driver levels.  While the insights gained can benefit actuarial applications such as pricing and risk classification, the primary objective is to advance methodological approaches for telematics-based risk analysis, rather than directly linking driving behaviour to claims frequency or premium determination.  Moreover, all the analysis presented in this paper is based solely on telematics data --- no classical covariates commonly used in Property and Casualty (P\&C) insurance are considered.  This is a significant contribution, as we demonstrate that meaningful insights and effective risk analysis can be achieved without relying on traditional covariates.  By shifting the focus to directly observable driving behaviour, our approach provides a data-driven alternative that may potentially avoid fairness-related concerns associated with demographic characteristics (e.g., gender, race) and socioeconomic status.  While fairness in insurance is often framed as a pricing problem, we emphasize that fairness begins at the risk assessment stage, which ultimately informs pricing decisions.  Studies have shown that telematics data can capture behavioural differences that correlate with risk, reducing the need for demographic proxies.  For example, \citet{ayuso_telematics_2016} demonstrated that average distance travelled per day can replace gender in risk assessment.  While we do not explicitly test such substitutions, our findings demonstrate that risk differentiation is possible using telematics data alone.  By focusing solely on driving behaviour, our approach provides a more individualized and transparent framework for risk evaluation, independent of classical covariates.  While integrating such methods into pricing models is beyond the scope of this study, future research could explore their potential role in fairer insurance pricing.


This work tackles the aforementioned challenges by addressing three specific problems.  First, we would like to define an anomaly index to quantify the level of anomalousness of each trip, capturing deviations from normal driving behaviour.  Driving behaviour such as aggressive driving, drunk driving, and fatigued driving are well-known contributors to accidents \citep{deng_review_2022}.  Existing studies on driving behaviour recognition and prediction often focus on distinguishing between specific behaviour, leveraging labelled datasets where driving conditions or behaviour are explicitly known.  However, in real-world scenarios, such labels are rarely available, limiting the applicability of these methods.  To overcome this, we do not target specific behaviour and instead, categorize diverse accident-prone behaviour broadly as `abnormal' behaviour.  This motivates the need for a generalized anomaly detection method that can operate without reliance on labelled data.



Second, by detecting anomalous driving behaviour, we aim to identify trips associated with accidents. When claims involving accidents are reported, they often include only the date of loss, leaving the specific trip unknown.  Pinpointing the exact trip enhances claims investigations, uncovers contributing factors, and supports preventative measures to reduce future occurrences.  Furthermore, accident detection improves insurance pricing by differentiating risk profiles, rewarding safer drivers, and pricing riskier behaviour appropriately.  From a fleet management perspective, identifying accident-prone trips enables proactive strategies such as targeted driver training, route optimization, and maintenance scheduling, ultimately leading to increased operational efficiency, reduced costs, and safer roads.  While our analysis in this work does not explicitly include location data, future research could extend our model to incorporate GPS coordinates, enabling route optimization based on trip duration, route length, and anomalous driving behaviour.  Similarly, for maintenance scheduling, our anomaly detection framework could flag trips that are outliers, signaling potential accidents or vehicle issues.  A rising trend in overall anomaly levels could also suggest the need for vehicle inspections and repairs.  While these applications would benefit from additional data, such as real-time traffic conditions and vehicle characteristics, our approach provides a foundation for enhancing fleet management through data-driven insights.



Third, we would like to evaluate risk at both the trip and driver levels.  Not all trips by a claimed driver are dangerous, nor are all trips by a no-claim driver safe.  A more granular, trip-based risk evaluation allows for dynamic updates to risk profiles, as well as improved risk classification that are not solely based on claims history.  Rather than solely identifying accident-related trips, our anomaly detection method assesses deviations from typical driving behaviour, providing a broader measure of driving risk.   Importantly, a trip without an accident does not necessarily mean it was completely safe --- risky manoeuvres and near-misses (e.g., \citet{arai_accidents_2001}, \citet{guillen_can_2020}, \citet{guillen_near-miss_2021}, \citet{sun_driving_2021}) can occur without resulting in a recorded crash.  These events are strong indicators of increased risk and should be accounted for when evaluating driving behaviour.  Moreover, risk assessment must extend beyond accident identification, as even claim-free drivers need to be ranked by their riskiness.  While a simpler classification model might suffice for distinguishing accident-related trips with precise accident labels, our approach applies to all trips, regardless of whether an accident occurs.  Accident-related trips are used to establish the relationship between the anomaly index and accident likelihood, allowing us to predict accident probability and assess trip riskiness for trips with no accidents.  This ensures that our method provides a more comprehensive evaluation of risk, capturing both actual accidents and high-risk driving patterns that could lead to future incidents.  Trip-level risk assessments can then be aggregated to the driver level by evaluating extreme behaviour and focusing on the right-tail of the risk distribution.  This approach enables better identification of claim-prone individuals and more effective differentiation of risk for insurance ratemaking purposes.

We tackle these problems by proposing a flexible framework that uses continuous-time hidden Markov models (CTHMMs) to model and analyze trip-level telematics time series.  By modelling multivariate telematics time series --- speed, longitudinal acceleration, and lateral acceleration --- the CTHMM exploits the time series structure of telematics and captures the dependencies between different variables.  This enables the detection of abnormal driving patterns and their connection to risk, providing detailed and novel insights into driver risk profiles.  Importantly, the CTHMM accommodates unevenly spaced time intervals in telematics data, fully utilizing data granularity without interpolation, which could otherwise distort driving behaviour patterns.

Hidden Markov models (HMMs), as dynamic mixture models supported by extensive statistical theories, offer significant flexibility and have been applied to various domains, including speech recognition, finance, and bioinformatics.  HMMs have also been widely used to study driving behaviour.  \citet{deng_review_2022} conducted a survey of HMM-based approaches in driving behaviour recognition and prediction, especially in the development of advanced driver assistance systems (ADAS), and we refer interested readers to an extensive lists of literature therein.  However, the complex structure of real-world telematics data presents unique challenges: irregular time intervals pose difficulties for data modelling, while boundary points complicate anomaly detection.  Most of the reviewed literature uses discrete-time HMMs, which struggle with these challenges, demanding modifications to the standard HMM framework.

In the context of actuarial science, the study of vehicle telematics at the trip level is relatively new.  To our best knowledge, \citet{jiang_auto_2024} is the only other work that also applied HMMs to model trip-level telematics data.  However, their approach and ours differ substantially.  \citet{jiang_auto_2024} employed discrete-time HMMs to model driving behaviour that have been categorized based on predefined thresholds (e.g., hard braking, sharp turns), and analyzed each trip's (latent) state durations with assumed probabilities for claim occurrences.  In contrast, our CTHMM models numerical telematics data directly in a continuous-time setting, accounting for the irregular time intervals while avoiding hard thresholds.  Additionally, our analysis is based on anomaly detection which does not require explicit state interpretations.  Instead, we link claims to the proportion of anomalies across telematics dimensions using logistic models, without assuming any direct relationship between claim probabilities and latent states.

We further developed a generalized anomaly detection method using a fitted CTHMM.  Traditional HMM-based anomaly detection methods rely on labelled normal sequences for training (\citet{khreich_combining_2009}, \citet{li_multivariate_2017}, \citet{leon_anomaly_2021}), and identify anomalies based on low likelihoods of observations (\citet{khreich_combining_2009}, \citet{florez_efficient_2005}, \citet{gornitz_hidden_2015}).  However, not only is labelled data often unavailable, defining thresholds for low likelihoods is also challenging, especially when telematics sequences vary in length.  Our method overcomes these limitations by unsupervisedly learning the most prevalent transitions and observations across trips.  We quantify anomaly level based on the proportion of outliers within each trip, introducing the concepts of the \textbf{`anomaly index'} and the \textbf{`normalized anomaly index'}.  Through a controlled study and a real data application, we show that the method is effective in identifying outliers and detecting deviations in driving behaviour, proving its practicality in telematics-based risk analysis.

Anomaly detection can be performed at either the driver level using an individual-specific CTHMM, or at the group level using a pooled CTHMM.  The individual-specific model captures each driver's unique, normal driving patterns and identifies anomalous trips that deviate from these norms.  For instance, we find that a higher anomaly index in longitudinal acceleration correlates with aggressive driving, while increased anomaly indices in both longitudinal and lateral accelerations are strongly linked to a higher probability of accident involvement.  Classification models based on these features achieve ROC-AUCs (areas under the receiver operating characteristic curve) of at least 0.8, demonstrating the effectiveness of our approach in identifying high-risk trips.  To the best of our knowledge, there are no existing classification models in the actuarial science literature specifically designed to identify accident-related trips.  Given this, our result is particularly notable, as values above 0.8 are generally considered to indicate excellent classification performance.

In the context of trip-level risk evaluation, \citet{meng_improving_2022} proposed an alternative scoring method using telematics data.  However, their approach imposes stricter preprocessing requirements, such as focusing only on speed ranges between 10 and 60 $km/h$, analyzing the first five minutes of each trip, and limiting the analysis to a fixed number of trips per driver.  In contrast, our methodology considers all non-zero speed intervals, evaluates entire trips, and accommodates varying trip lengths and counts, capturing richer temporal dynamics.  Furthermore, the training datasets used for trip-level evaluations differ fundamentally. \citet{meng_improving_2022} used Poisson GLM deviance residuals to classify drivers as risky or safe, labelling all trips from high-residual drivers as risky and those from high-exposure, claimless drivers as safe.  In comparison, our individual-specific CTHMM approach targets trips associated with accidents and at-fault claims, based on analysis on accelerations and speed.  While one claim occurred in each of the studied 3-day windows, we do not assume trips in these windows to be equally risky.


Conversely, the pooled model learns common, normal driving patterns across a population.  It detects anomalous trips among drivers, and facilitates the comparison and ranking of drivers based on their level of anomalousness.  Our real data analysis reveals notable differences in driving behaviour between drivers with at-fault claims and those without claims.  While trips by claimed drivers exhibit lower average anomaly indices, their maximum anomaly indices attained are consistently higher --- over time, drivers in the claimed group demonstrate more extreme behaviour than those in the no-claim group.  Using the right tail of each driver’s anomaly index empirical distribution to classify drivers, our models achieve ROC-AUCs ranging from 0.7 to 0.78, with improved results when incorporating exposure (number of trips).  These results are competitive with existing studies on policy-level claim classification incorporating telematics data, where most models report ROC-AUCs (if available) between 0.58 and 0.62 (\citet{paefgen_multivariate_2014}, \citet{baecke_value_2017}, \citet{bian_good_2018}, \citet{jin_latent_2018}, \citet{huang_automobile_2019}, \citet{duval_how_2022}, \citet{duval_enhancing_2023}).  The highest reported ROC-AUC in this domain is 0.80, achieved by \citet{li_driving_2023}, which also relies solely on telematics data, but incorporates a much broader set of features, including annual mileage, number of trips, exposure to different times of the day and days of the week, and harsh events, etc.  More broadly, models based only on traditional covariates generally perform the worst, followed by those relying only on telematics data, with the best results achieved when both sources are combined.  These comparisons further highlight the strength of our anomaly-based approach, as it achieves comparable or superior performance using only a minimal set of telematics features.  We emphasize that our risk evaluation relies exclusively on telematics data, as traditional covariates regarding the driver or vehicle are unavailable.  Nevertheless, our ROC-AUCs outperform the reported industry benchmark of 0.65, demonstrating the effectiveness of our methodology.


The remainder of this paper is organized as follows.  Section \ref{Data-Description-and-Challenges} gives an overview of the telematics dataset analyzed in this work and discusses the associated data challenges which motivate the proposed modelling.  Section \ref{CTHMM} introduces the continuous-time hidden Markov model (CTHMM) framework for modelling trip-level telematics data and outlines the proposed anomaly detection method.  Section \ref{Estimation} details the estimation algorithm.  Section \ref{Controlled-Study} validates the framework and methodology through a controlled study, while Section \ref{Real-Data-Analysis} applies the model to real-world data, focusing on accident-related trips and driver behaviour.  Finally, Section \ref{Conclusion} concludes the paper with a discussion of future research direction.

\section{Data Description and Challenges}
\label{Data-Description-and-Challenges}

This section describes the real-world telematics dataset we will study throughout this paper and the associated challenges which have motivated our proposed modelling.

\subsection{Overview of Data and Challenges}
\label{Overview-of-Data-and-Challenges}

We analyze a rental car dataset from Australia focused on medium- to long-term rentals, with a median duration of 1 month and a mean of 2 months.  For each trip, telematics information are recorded at \textbf{irregular} time intervals: timestamp, GPS locations (latitude and longitude) and speed.  This raw telematics data allows the calculation of time lapse (time since the last observation) and lateral and longitudinal accelerations.  The dataset also includes claims reported between January 1, 2019 and January 31, 2023.

The first challenge with this dataset is the lack of labels.  On one hand, we do not have labels regarding the driving behaviour (e.g., normal or aggressive) nor the riskiness of each trip.  On the other hand, in cases where the rental results in claims and the date of loss is known, we do not know exactly in which trip the accident or damage occurred.  Therefore, we will employ unsupervised learning for trip telematics modelling.

Moreover, this dataset poses another challenge, which is the lack of traditional covariates, such as driver-specific or vehicle-specific characteristics commonly used in risk modelling.  As a result, we focus solely on telematics data to assess driving risk.  First, we would like to use the proposed model to study and analyze various driving behaviour.  However, driving behaviour can be influenced by many factors, such as the driver's experience, habits and the current traffic environment, leading to considerable heterogeneity.  Furthermore, driving behaviour can be categorized into different driving styles, such as normal, cautious and aggressive \citep{ma_driving_2021}.  However, there is no unified standard to clearly distinguish these styles.  \cite{deng_review_2022} pointed out that in existing literature, the levels, terms, and concepts often depend on the authors' own definitions.  For this reason, instead of pre-specifying the different behaviour, we let the fitted model inform us what the various types of behaviour and the transitions between them are.  This is similar to the study of animal movements (\citet{michelot_movehmm_2016}, \citet{whoriskey_hidden_2017}, \citet{grecian_understanding_2018}).  Second, while claims records are available for the dataset under consideration, this is not typically the case for our industry collaborator's other datasets.  Yet, telematics datasets without claims records are exactly where we would like to apply our proposed methodology for risk evaluation.  The absence of claims or accidents does not imply an absence of risk, as driving risk is not constant even when no claims or accidents occur.  Hence, we require a model that can learn and analyze driving behaviour without supervision, and quantify the riskiness of each trip independently of claims data.  Using the available claims records, we further validate that the anomaly indices are positively associated with accident involvement, demonstrating that our method can effectively assess driving risk in both settings.


There are two more challenges regarding the structure of the telematics records.  Firstly, the time series are irregular, where the sequence of data points are recorded at unevenly spaced time intervals.  The reason is that our collaborator employs a patented curve logging technique in its telematics tracking system, which eliminates intermediate data points when changes are `insignificant', to reduce the burden of data transmission and storage.  In the dataset, this technique has been applied to both GPS locations logging and speed logging, and consequently the telematics time series become irregular.  Although this technique eases data transmission and storage burden, analysis and modelling of irregular time series are more challenging because traditional time series methods often assume a regular time interval between observations.  One of the simplest and most straight-forward solutions is to choose a regular time interval (e.g., per second) and fill in missing data via linear interpolation.  However, such method conflicts with the purpose of curve logging.  Moreover, while we can return to traditional time series analysis after linear interpolation, much of the modelling and computing effort will be placed on the interpolated points which provide no new information.  For these reasons, we opt against using linear interpolation.

\begin{figure}[ht]
    \centering
    \includegraphics[width=\textwidth]{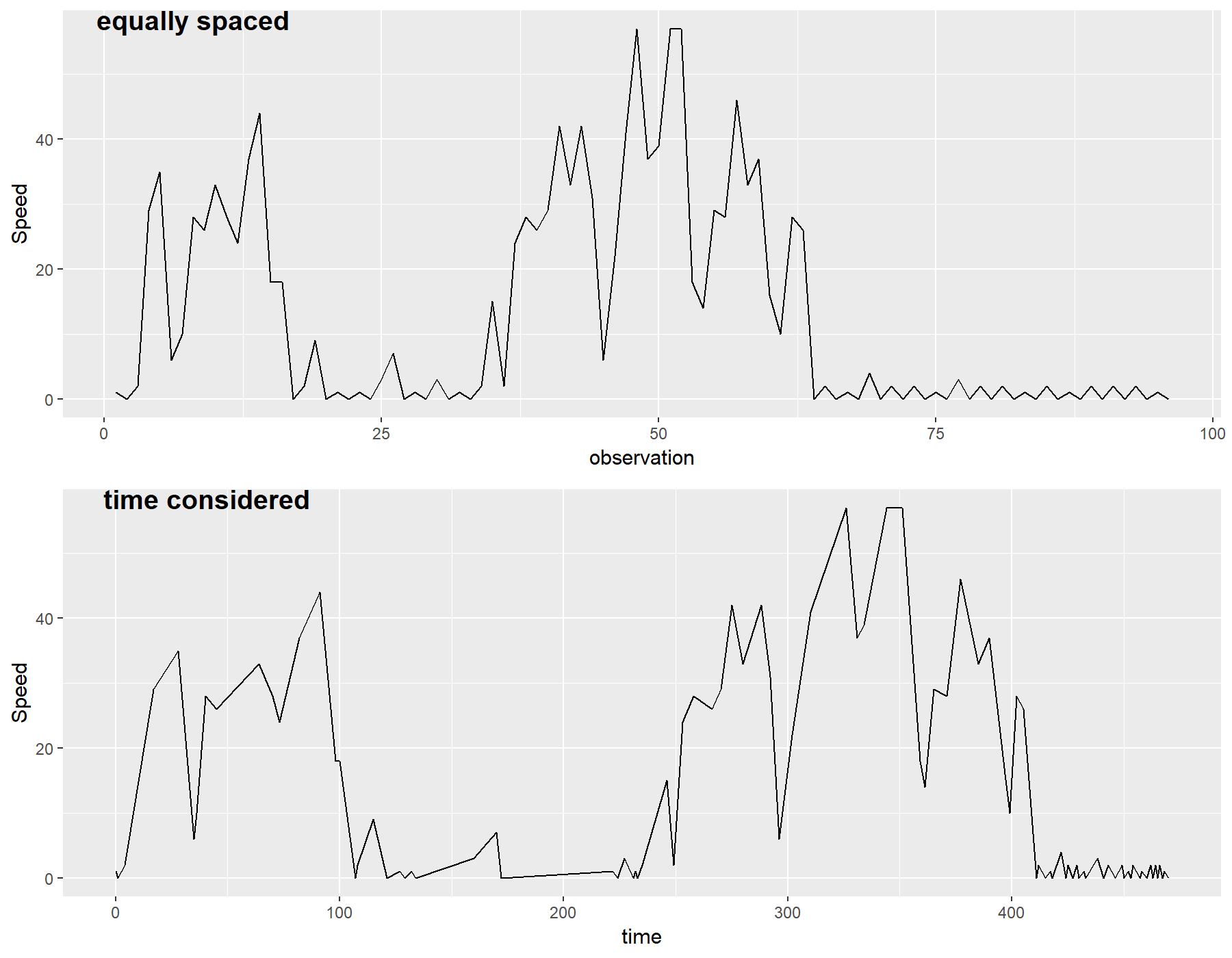}
    \caption{Speed time series (in $km/h$) of a sample trip. Top: assuming observations are recorded regularly; Bottom: considering the actual observation times (in seconds).}
    \label{fig: sample_trip}
\end{figure}

It is natural to question how important it is to consider the irregular time intervals, and how different the time series can be if these intervals are neglected.  For illustrative purpose, Figure \ref{fig: sample_trip} shows the speed time series of a sample trip.  The top sub-figure illustrates how the time series would appear if observations were assumed to be recorded at regular intervals.  While the $x$-axis is labelled `observation', it can also be interpreted as if the observations were collected at a fixed rate (e.g., one per second or at another constant interval).  The bottom sub-figure represents the time series considering the actual timestamps (in seconds) of each observation, hence accounting for the irregular time intervals between observations; this reflects the true speed time series as it occurred in reality.  For example, the small `V' shape at the start of the trip consists of observations recorded approximately one second apart, whereas the observations centered around 200 seconds have a time gap of 50 seconds between them.


We observe that the patterns are similar, since the `significant' changes should have been captured by curve logging.  However, the interpretation can be different, especially if we would like to study the different driving behaviour, such as the magnitudes of acceleration and braking.  In this example, we observe that first, changes can seem less or more abrupt than they actually are; and second, the small fluctuations at the end of the trip appear to have a much longer duration if observation times are not considered (around 40\% of the entire trip vs. less than 25\% of the trip).

Another challenge with the telematics records is that the trips (and their time series) are of different lengths.  A possible solution is Dynamic Time Warping (DTW), which is a technique to align, compare and measure the similarity of time series of different lengths.  However, DTW requires regular time series, hence it is not directly applicable to our data unless the time series have been linearly interpolated.

The three aforementioned challenges demand a flexible, unsupervised time series model which can incorporate time series irregularity and of varying lengths.  We propose the use of a time-homogeneous, continuous-time hidden Markov model (CTHMM), where different states of the latent Markov chain will represent various driving actions and can be inferred via the observed telematics data.  Together, all these states will describe the prevalent driving behaviour observed in the telematics data.


\subsection{Data Preparation}
\label{Data-Preparation}


In particular, we model the following telematics responses: time (implicitly incorporated in the hidden Markov chain transitions), speed, and acceleration, which we further decompose into longitudinal (acceleration and braking) and lateral (left and right turning) components.  Longitudinal acceleration reflects forward and backward movements, capturing braking and acceleration patterns, while lateral acceleration represents side-to-side movements, associated with cornering and swerving (\citet{rajamani_vehicle_2012}, \citet{gillespie_fundamentals_2020}).  This decomposition provides a more detailed characterization of driving behaviour, allowing us to distinguish between different driving styles and risk factors.  For example, high longitudinal acceleration may indicate aggressive driving or stop-and-go traffic, whereas high lateral acceleration could suggest sharp cornering or sudden lane changes, which may result from reckless maneuvering or defensive evasive actions.  Moreover, this separation enhances anomaly detection, enabling the model to identify distinct types of risky events rather than treating all acceleration events uniformly.

\begin{table}[h!]
    \centering
    \caption{Telematics response alignments.}
    \begin{tabular}{c|ccccc}
         \textbf{time}&  \textbf{GPS locations}&  \textbf{speed}&  \textbf{longitudinal acceleration}&  \textbf{lateral acceleration} \\\hline
         $t_0$&  $GPS_0$&  $v_0$&  missing&    missing\\
         $t_1$&  $GPS_1$&  $v_1$&  $a_{x,1}$&  $a_{y,1}$\\
         $t_2$&  $GPS_2$&  $v_2$&  $a_{x,2}$&  $a_{y,2}$\\
         $t_3$&  $GPS_3$&  $v_3$&  $a_{x,3}$&  $a_{y,3}$\\
         $t_4$&  $GPS_4$&  $v_4$&  $a_{x,4}$&  $a_{y,4}$\\
         $t_5$&  $GPS_5$&  $v_5$&  $a_{x,5}$&  $a_{y,5}$\\
         $t_6$&  $GPS_6$&  $v_6$&  missing&    missing\\
    \end{tabular}
    \label{tab: response-alignments}
\end{table}

For illustrative purposes, let us consider a trip starting at time $t_0$ and ending at time $t_6$.  We will employ the response alignments as shown in Table \ref{tab: response-alignments}.  To achieve the necessary structure, we first convert the GPS locations in latitude and longitude to Universal Transverse Mercator (UTM) coordinates, then we apply the following physics formula to obtain the longitudinal and lateral accelerations.

Let $\textbf{s} = (e, n)$ be the position of the vehicle at some time $t$, where $e$ denotes the Easting and $n$ denotes the Northing of the vehicle in UTM coordinates.  From two consecutive UTM coordinates and the time interval in between, we calculate the velocity of the vehicle as
$$\textbf{v} = \frac{d\textbf{s}}{dt} = \left(\frac{de}{dt}, \frac{dn}{dt}\right).$$  In the case of discrepancy between this velocity and the GPS speed, velocity is scaled such that the magnitude matches the latter speed.  Hence, velocity will be missing at the start of the trip.  Then from two consecutive velocities, the acceleration is given by
$$\textbf{a} = \frac{d\textbf{v}}{dt} = \left(\frac{d^2e}{dt^2}, \frac{d^2n}{dt^2}\right).$$

It is noteworthy that this acceleration is attributed to the initial timepoint $t_1$ as this is the (average) acceleration applied from time $t_1$ to $t_2$.  Consequently, the acceleration values at both the start and end of the trip will be missing.  Then the x-axis/longitudinal/tangential component of acceleration is
$$a_x = \frac{\textbf{v} \cdot \textbf{a}} {||\textbf{v}||}$$
and the y-axis/lateral/normal component of acceleration is
$$a_y = \sqrt{||\textbf{a}||^2 - a^2_x}.$$

\section{Modelling Framework: Continuous Time Hidden Markov Models}
\label{CTHMM}

In this section, we introduce the time-homogeneous, continuous-time hidden Markov model (CTHMM) used to model trip-level telematics data.  Then, we outline an anomaly detection method to quantify the degree of deviation in driving behaviour demonstrated by each trip.  Finally, we discuss some assumptions and adjustments made to ensure the CTHMM can effectively accommodate the complexities of telematics data and fulfill our analysis need.

We consider a population of $M$ drivers.  For driver $i$, $i = 1, 2, ..., M$, we have his/her telematics records over $N_i$ trips.  For trip $j$, $j = 1, ..., N_i$, the multi-dimensional telematics data $\mathbf{y}_{ijl}$ are recorded at $t_{ijl}$, $l = 0, 1, ..., T_{ij}$, time-points with irregular intervals.

\subsection{Model Description}
\label{Model-Description}

The CTHMM is driven by a hidden Markov chain.  A continuous-time hidden Markov chain with $S$ discrete latent states is defined by an initial state probability distribution $\pi$ and a state transition rate matrix $\mathbf{Q}$.  The elements $q_{uv}$ in $\mathbf{Q}$ describe the rate the process transitions from states $u$ to $v$ for $u \neq v$, $u, v = 1,...,S$ and $v = 1,...,S$, and $q_{uu} = - q_u = - \sum_{v\neq u} q_{uv}$.  The chain is time-homogeneous in the sense that $q_{uv}$ are independent of time $t$.  The probability that the process transitions from states $u$ to $v$ is $q_{uv}/q_u$, while the sojourn time in state $u$ follows an exponential distribution with rate $q_u$.

We assume drivers are independent, and each trip for a given driver is an independent time series generated from the CTHMM.  To simplify the notation, we first consider a single trip $j$.  At timepoint $t_l$, the available observed telematics data $\mathbf{y}_l$ depends on the current latent state $Z_l$ through the state-dependent distribution $f(\mathbf{y}_l | Z_l = z(t_l))$, where $z(t_l) \in \{1,2,...,S\}$ is the latent state at time $t_l$.  Denote $\boldsymbol{\Phi} = (\pi, \mathbf{Q}, \boldsymbol{\Theta})$ the set of model parameters, where $\pi$ is the initial state probability distribution, $\mathbf{Q}$ is the transition rate matrix and $\boldsymbol{\Theta}$ are the parameters of the state-dependent distributions.  If the continuous-time Markov model has been fully observed, meaning that one can observe every state transition time $\mathbf{T}' = (t'_0, t'_1, ..., t'_{T'})$, the corresponding states $\mathbf{Z}' = (z'_0 = z(t'_0), z'_1 = z(t'_1), ..., z'_{T'} = z(t'_{T'}))$, the observed data $\mathbf{Y} = (\mathbf{y}_0, \mathbf{y}_1, ..., \mathbf{y}_T)$ at observation timepoints $\mathbf{T} = (t_0, t_1, ..., t_T)$, then the complete data likelihood is
\begin{align*}
    \mathcal{L}(\boldsymbol{\Phi}; \mathbf{Z}', \mathbf{T}', \mathbf{Y}, \mathbf{T}) &= \pi_{z'_0} \prod_{k=0}^{T' - 1} \left(\frac{q_{z'_{k}, z'_{k+1}}}{q_{z'_{k}}}\right)\left(q_{z'_{k}} \exp(-q_{z'_{k}}(t'_{k+1} - t'_{k}))\right) \prod_{l=0}^T f(\mathbf{y}_l | Z_l = z(t_l)) \\
    &= \pi_{z'_0} \prod_{u=1}^{S} \prod_{v=1, v\neq u}^{S} q_{uv}^{n_{uv}} \exp(-q_u \tau_u) \prod_{l=0}^T f(\mathbf{y}_l | Z_l = z(t_l))
\end{align*}

\noindent
where $n_{uv}$ is the number of transitions from states $u$ to $v$ and $\tau_u$ is the total time the chain remains in state $u$.

However in a realization of a CTHMM, there are two levels of hidden information.  First, none of the states $\mathbf{Z}$ (at observation times) nor $\mathbf{Z}'$ (at transition times) are directly observed in the hidden Markov chain, and can only be inferred from the observed data $\mathbf{Y}$.  Second, the state transitions between two consecutive observations are also hidden --- note the difference between state transition times $\mathbf{T}'$ and observation times $\mathbf{T}$.  In other words, the chain may have multiple transitions before reaching a state which emits an observation.  As a consequence, both $n_{uv}$ and $\tau_u$ are also unobserved.  Here we assume the observed telematics process possesses the `snapshot' property, such that the observation made at time $t_l$ depends only on the state active at time $t_l$, instead of the entire state trajectory over the interval $(t_{l-1},t_l]$ \citep{patterson_statistical_2017}.  

We introduce two indicator functions: $\mathbbm{1}_{\{Z_l = u\}} = 1$ if the latent state $z_l$ at observation time $t_l$ is $u$, and 0 otherwise; and $\mathbbm{1}_{\{Z_l = u, Z_{l+1} = v\}} = 1$ if the latent states $z_l$, $z_{l+1}$ at observation times $t_l$ and $t_{l+1}$ are $u$ and $v$ respectively, and 0 otherwise.  The likelihood can be expressed instead as
\begin{equation*}
\mathcal{L}(\boldsymbol{\Phi}; \mathbf{Y}, \mathbf{T}) = \prod_{u=1}^S \pi_u ^{\mathbbm{1}_{\{Z_0 = u\}}} \prod_{l=0}^{T - 1} \prod_{u,v=1}^S \mathbf{P}_{uv}(\Delta_{l+1})^{\mathbbm{1}_{\{Z_l = u,Z_{l+1} = v\}}} \prod_{l=0}^T \prod_{u=1}^S f_u(\mathbf{y}_l)^{\mathbbm{1}_{\{Z_l = u\}}}
\end{equation*}

\noindent
where $\mathbf{P}(t) = \exp(\mathbf{Q}(t))$ with $\Delta_l = t_l-t_{l-1}$ and matrix exponential $\exp$, and $f_u(\mathbf{y}_l) = f(\mathbf{y}_l | Z_l = u)$.  $\mathbf{P}(t)$ is a state-transition matrix with the $(u,v)$-th entry $\mathbf{P}_{uv}(t)$, which is the probability that the (latent) Markov chain transitions from state $u$ to state $v$ after time $t$.  It accounts for all possible intermediate state transitions which emit no observations happening before time $t$.

In an individual-specific model, where a distinct CTHMM is fitted for each individual, the complete data likelihood for a driver $i$ with $N_i$ trips is given by
\begin{align*}
     \mathcal{L}_i(\boldsymbol{\Phi}; \mathbf{Z}'^*, \mathbf{T}'^*, \mathbf{Y}^*, \mathbf{T}^*) &= \prod_{u=1}^{S} \prod_{v=1, v\neq u}^{S} q_{uv}^{\sum_{j=1}^{N_i} n_{ij,uv}} \exp\left(-q_u \sum_{j=1}^{N_i} \tau_{ij,u}\right) \prod_{j=1}^{N_i} \pi_{z'_{ij0}} \prod_{l=0}^{T_{ij}} f(\mathbf{y}_{ijl} | Z_{ijl} = z(t_{ijl}))
\end{align*}
Alternatively, it can be represented as
\begin{equation*}
\mathcal{L}_i(\boldsymbol{\Phi}; \mathbf{Y}, \mathbf{T}) = \prod_{j=1}^{N_i} \left(\prod_{u=1}^S \pi_u ^{\mathbbm{1}_{\{Z_{ij0} = u\}}} \prod_{l=0}^{T_{ij} - 1} \prod_{u,v=1}^S \mathbf{P}_{uv}(\Delta_{ij,l+1})^{\mathbbm{1}_{\{Z_{ijl} = u,\;Z_{ij,l+1} = v\}}} \prod_{l=0}^{T_{ij}} \prod_{u=1}^S f_u(\mathbf{y}_{ijl})^{\mathbbm{1}_{\{Z_{ijl}= u\}}}\right)
\end{equation*}

In a pooled model, where a single CTHMM is trained to accommodate multiple individuals, each with $N_i$ trips for $i=1,...,M$, the complete data likelihood across all drivers is
\begin{align*}
     \mathcal{L}(\boldsymbol{\Phi}; \mathbf{Z}'^*, \mathbf{T}'^*, \mathbf{Y}^*, \mathbf{T}^*) = \prod_{i=1}^M \mathcal{L}_i(\boldsymbol{\Phi}; \mathbf{Z}'^*, \mathbf{T}'^*, \mathbf{Y}^*, \mathbf{T}^*)
\end{align*}
or equivalently
\begin{equation*}
\mathcal{L}_i(\boldsymbol{\Phi}; \mathbf{Y}, \mathbf{T}) = \prod_{i=1}^M \mathcal{L}_i(\boldsymbol{\Phi}; \mathbf{Y}, \mathbf{T})
\end{equation*}

The rationale for choosing between the individual-specific model and the pooled model will be discussed in Section \ref{Adapting-CTHMM-For-Telematics}.

While our modelling framework assumes independence across both drivers and trips by the same driver, the latter requires further consideration to assess its validity and implications.  Rather than modelling each trip separately, we fit a single individual-specific CTHMM per driver, allowing all trips to contribute to learning that driver's unique driving patterns.  This ensures that recurring behaviour, such as tendencies in acceleration, braking, and cornering, are reflected in the estimated state transitions.  As a result, while individual trips are treated as independent realizations from the same model, the model itself captures intra-driver dependencies by learning stable behavioural tendencies over time.

Although consecutive trips within a short time window (e.g., within the same day) may exhibit stronger correlations due to persistent road, weather, or driver conditions, variations still arise due to factors such as traffic conditions, driver fatigue, and trip purpose.  These sources of randomness support the assumption that trips are drawn from the same underlying distribution but are not necessarily dependent on one another.  Given that our focus is on long-term behavioural patterns and anomaly detection rather than short-term trip clustering, assuming independence across trips remains a reasonable and practical simplification for risk assessment.

If strong intra-driver dependence exists beyond what is captured by the driver-specific model, it may lead to an underestimation of variability in a driver's risk profile.  Future extensions could incorporate hierarchical models to explicitly account for temporal dependencies across trips, though at the cost of increased complexity.  For this study, our approach balances model complexity and interpretability while effectively capturing key aspects of driver-specific behaviour, as will be demonstrated in the applications in Sections \ref{Controlled-Study} and \ref{Real-Data-Analysis}.

\subsection{Anomaly Detection}
\label{Anomaly-Detection}



Interpreting the hidden states and their corresponding state-dependent distributions in a CTHMM can be challenging, especially when many states are needed to capture all significant combinations of the response dimensions.  To illustrate this, Table \ref{tab: sample-state-dependent-distributions} presents the state-dependent distributions from a 5-state individual-specific CTHMM for Driver 1 from the UAH-DriveSet (to be studied in detail in Section \ref{Controlled-Study}).  Each row represents a latent state, and for instance, the last row (state 5) can be roughly interpreted as a driving action of `left turn (negative lateral acceleration) and deceleration (negative longitudinal acceleration) at a lower speed range (with the lowest mean speed of 81 $km/h$, though with a large standard deviation)'.

As shown in this example, a CTHMM with just 5 latent states for 3 response dimensions already requires interpreting 15 distributions.  With fewer states, the distributions exhibit large standard deviations and significant overlap, making it unclear which driving action each state represents.  Furthermore, certain driving actions may not be captured, e.g., lower speed ranges appear underrepresented in this example.  Conversely, with more states, similar actions may be represented by multiple distributions, making it challenging to differentiate between them.  More importantly, interpreting these latent states does not directly help achieve our goal of risk assessment, as associating states with driving risk levels is even more difficult and often subjective, especially in the absence of collision data.  For example, \cite{jiang_auto_2024} assumed specific accident probabilities associated with hard braking and big angle changes and simulated accidents based on these assumptions.  As explained in Section \ref{Introduction}, instead of focusing on specific accident-prone driving behaviour, we classify them broadly as `abnormal' behaviour, and propose to perform anomaly detection under a (fitted) CTHMM.  Our proposed method is statistically supported and does not require explicit state interpretations, yet it can effectively identify outliers --- in our case, the deviating driving behaviour.

\begin{table}[h!]
\centering
\caption{UAH-DriveSet: Summary of state-dependent distributions from a 5-state individual-specific CTHMM for D1. SD: standard deviation.}
\begin{tabular}{c|rr|rr|rrrr|}
\cline{2-9}
\multicolumn{1}{r|}{} & \multicolumn{2}{c|}{lateral acceleration} & \multicolumn{2}{c|}{longitudinal   acceleration} & \multicolumn{4}{c|}{speed} \\ \hline
\multicolumn{1}{|r|}{distribution} & \multicolumn{2}{c|}{Normal} & \multicolumn{2}{c|}{Normal} & \multicolumn{4}{c|}{Gamma} \\ \hline
\multicolumn{1}{|c|}{state} & \multicolumn{1}{c|}{mean} & \multicolumn{1}{c|}{SD} & \multicolumn{1}{c|}{mean} & \multicolumn{1}{c|}{SD} & \multicolumn{1}{c|}{shape} & \multicolumn{1}{c|}{scale} & \multicolumn{1}{c|}{mean} & \multicolumn{1}{c|}{SD} \\ \hline
\multicolumn{1}{|c|}{1} & \multicolumn{1}{r|}{0.0002} & 0.0216 & \multicolumn{1}{r|}{-0.0013} & 0.0147 & \multicolumn{1}{r|}{1822.3880} & \multicolumn{1}{r|}{0.0535} & \multicolumn{1}{r|}{97.5073} & 2.2841 \\ 
\multicolumn{1}{|c|}{2} & \multicolumn{1}{r|}{0.0012} & 0.0191 & \multicolumn{1}{r|}{-0.0014} & 0.0134 & \multicolumn{1}{r|}{1103.6513} & \multicolumn{1}{r|}{0.0972} & \multicolumn{1}{r|}{107.2352} & 3.2279 \\ 
\multicolumn{1}{|c|}{3} & \multicolumn{1}{r|}{-0.0002} & 0.0249 & \multicolumn{1}{r|}{0.0001} & 0.0158 & \multicolumn{1}{r|}{820.2069} & \multicolumn{1}{r|}{0.1097} & \multicolumn{1}{r|}{89.9889} & 3.1422 \\
\multicolumn{1}{|c|}{4} & \multicolumn{1}{r|}{-0.0004} & 0.0232 & \multicolumn{1}{r|}{-0.0021} & 0.0149 & \multicolumn{1}{r|}{217.7109} & \multicolumn{1}{r|}{0.5785} & \multicolumn{1}{r|}{125.9496} & 8.5360 \\ 
\multicolumn{1}{|c|}{5} & \multicolumn{1}{r|}{-0.0112} & 0.0752 & \multicolumn{1}{r|}{-0.0209} & 0.0573 & \multicolumn{1}{r|}{23.0331} & \multicolumn{1}{r|}{3.5364} & \multicolumn{1}{r|}{81.4553} & 16.9724 \\ \hline
\end{tabular}
\label{tab: sample-state-dependent-distributions}
\end{table}

Given a fitted CTHMM, we can perform anomaly detection on both entire trips and individual data points within a trip.  Specifically, we utilize the `forecast pseudo-residuals' introduced in \cite{zucchini_hidden_2016}.  For a trip $j$ from driver $i$ at timepoint $t$, the uniform forecast pseudo-residual $u_{ijt,d}$ for the $d$-th dimension of the response $\textbf{Y}_{ijt}$, denoted as $Y_{ijt,d}$, is computed under the fitted model as
$$u_{ijt,d} = P(Y_{ijt,d} \leq y_{ijt,d} \;|\; \textbf{Y}_{ij0} = \textbf{y}_{ij0}, \textbf{Y}_{ij1} = \textbf{y}_{ij1}, ..., \textbf{Y}_{ij,t-1} = \textbf{y}_{ij,t-1}).$$

In the context of HMMs, the analysis of pseudo-residuals serves two purposes: model checking and outlier detection, and we focus on the latter for this work.  The forecast pseudo-residuals help identify observations that are extreme relative to the model and all preceding observations, highlighting the abrupt changes in driving behaviour.  If a forecast pseudo-residual is extreme, it indicates that the corresponding observation is an outlier, or that the model no longer provides an acceptable description of the time series.

If the model fits the time series well, $u_{ijt,d}$ should follow a Uniform$(0, 1)$ distribution, with extreme observations indicated by residuals close to 0 or 1.  However, it can be hard to determine the extremity of a value under the uniform scale.  For example, while a value of 0.97 might seem extreme on its own, it is less so when compared to 0.999.  Consequently, \cite{zucchini_hidden_2016} suggest using the \textbf{normal forecast pseudo-residual}
\begin{equation}\label{eqn: normal-forecast-pseudo-residual}
    z_{ijt,d} = \Phi^{-1}(u_{ijt,d})
\end{equation}

\noindent
where $\Phi$ is the standard normal cumulative density function.  If the model is an accurate representation, $z_{ijt,d}$ should follow a standard normal distribution.  In our analysis, we propose to classify observations with normal residuals exceeding 3 standard deviations as outliers.  The 3-standard deviation threshold is widely used in statistical outlier detection, particularly for normally distributed data (\citet{barnett_outliers_1994}, \citet{aggarwal_outlier_2015}).  This threshold is based on the empirical rule (also known as the 68-95-99.7 rule) for normal distributions.  Flagging observations with residuals beyond 3 standard deviations --- representing a 0.3\% probability --- effectively identifies points that are truly extreme relative to the model and preceding observations.  We further examine the impact of threshold selection on anomaly detection performance in Section \ref{Controlled-Study-Individual-Specific-Model}.  Empirical results support the use of 3 standard deviations as the outlier threshold, as it effectively identifies anomalous trips and driving behaviour while maintaining a clear and consistent interpretation of what constitutes an extreme deviation from normal driving patterns.

Although we compute forecast pseudo-residuals $u_{ijt,d}$ and $z_{ijt,d}$ for each dimension $d$ separately, all response dimensions influence these values, as $u_{ijt,d}$ is conditioned on the history across all dimensions.  Consequently, the other dimensions also impact the probability of being in a particular latent state at each timepoint $t$.



In summary, our anomaly detection approach identifies deviations from learnt normal driving behaviour without predefining specific risky states (e.g., `aggressive' or `drowsy' driving).  Instead, the model estimates the states in which a driver is most likely to be at a given time based on the (multivariate) telematics response, such as speed, longitudinal acceleration, and lateral acceleration in our analysis.  Normal forecast pseudo-residuals $z_{ijt,d}$ are computed for each dimension $d$ of telematics observations, and residuals beyond 3 standard deviations are flagged as outliers.  At the trip level, we quantify the degree of anomaly by calculating the proportion of outliers in each dimension, defining this as the \textbf{`anomaly index'}.  Rather than classifying certain states as inherently risky, our approach evaluates whether an individual is in an unexpected state given their usual driving patterns.  This allows us to detect a broad range of abnormal behaviour without requiring predefined labels or assumptions about specific risk types, making the method both flexible and widely applicable.

\subsection{Adapting and Applying CTHMMs for Anomaly Detection in Telematics}
\label{Adapting-CTHMM-For-Telematics}

We perform anomaly detection at two levels to address two distinct objectives.  At the \textbf{driver level}, we focus on identifying anomalous behaviour and trips that deviate from a driver’s typical patterns, particularly those associated with accidents.  For this purpose, we use individual-specific models to account for heterogeneity in individual driving behaviour.  \cite{deng_review_2022} listed five factors influencing driving behaviour, which are driving styles, fatigue driving, drunk driving, driving skills, and traffic environment.  Additional factors, such as vehicle characteristics, locations, and weather conditions, also play a role but are not available for consideration in our modelling.  \cite{lefevre_driver_2015} compared the results of predicted acceleration and highlighted that a personalized/individualized model always outperforms an average/general model.  While prediction is not our primary focus, the forecast pseudo-residuals in our approach calculate the predictive probability that a response dimension at the next time point is less extreme than the observed value given all previous observations.  For reliable anomaly detection, the CTHMM must accurately capture a driver's typical behaviour.  A pooled model trained on data from multiple drivers may obscure individual anomalies, particularly for drivers whose deviations are minor within the group.  Therefore, we fit individual-specific models to each driver's trips to detect anomalies effectively.

At the \textbf{group level}, we aim to identify trips that deviate from the population norm and pinpoint drivers associated with at-fault claims.  For this, we use a pooled model.  While individual-specific models generate forecast pseudo-residuals to detect anomalies for each driver, these residuals may not be directly comparable across individuals.  A pooled model, trained on trips from multiple drivers, enables us to compare trips across drivers, identify population-level anomalies, and rank drivers by their relative anomalousness.

For driver $i$ with $N_i$ trips, the \textbf{anomaly index} for response dimension $d$ in trip $j$ with $T_{ij}+1$ observations is defined as:
\begin{equation}\label{eqn: anomaly-index}
    A_{ij,d} = \frac{\sum_{t} \mathbbm{1}(|z_{ijt,d}| \geq 3)}{T_{ij}}
\end{equation}
\noindent
Note that the denominator is $T_{ij}$ only as forecast pseudo-residuals are not calculated at the first observation.  To rank each driver's trips by their level of anomalousness, we further define the \textbf{normalized anomaly index}, which is the ratio of each trip's index to the maximum index for that driver:
\begin{equation}\label{eqn: normalized-anomaly-index}
    \tilde{A}_{ij,d} = \frac{A_{ij,d}}{\max_j (A_{ij,d})}
\end{equation}
\noindent

The (raw) anomaly indices for response dimension $d$ across $N_i$ trips form an empirical distribution $\{A_{ij,d}, j = 1,...,N_i\}$ with empirical cumulative distribution function $F_{i,d}(p)$.  For group-level analysis, we will focus on the right tail of this distribution, reflecting extreme anomaly levels in dimension $d$.  In particular, the maximum and $\alpha$th-percentiles will be of interest:
$$\max_{j}(A_{ij,d}) \quad \text{and} \quad P_{\alpha,d} = \inf\{p : F_{i,d}(p) \geq \alpha\}$$

Although HMM is a flexible time-series model and the anomaly detection method is statistically grounded, there are several assumptions and adjustments that we need to make to adapt the proposed framework for the complex telematics data and ensure effective analysis.
\begin{enumerate}
    \item We assume that driving behaviour is predominately normal.
    \begin{itemize}
        \item While we aim to identify all abnormal driving behaviour, instead of focusing on just one or a few, we still assume that driving behaviour is mostly normal and safe.  This means that for each driver, anomalous and risky trips are rare, and even within those risky trips (e.g., ones that result in accidents), abnormal driving instances occur infrequently.  Consequently, the CTHMM, no matter individualized or generalized, is fitted to the normal and safe driving behaviour, with anomaly detection used to identify deviating ones.  This assumption should be fair as accidents would no longer be rare events otherwise.  As will be shown in Section \ref{Real-Data-Analysis-Pooled-Model}, we tested pooled models with various portfolio compositions (having different claim rates), and the method proves robust regardless of the proportion of drivers with accidents the model is fitted on.
    \end{itemize}
    \item The CTHMM should have a reasonable number of states.
    \begin{itemize}
        \item The CTHMM should have sufficient (latent) states to capture the major combinations of response dimensions, which ensures accurate modelling of a driver's (in individual-specific models) and a group's (in a pooled model) normal driving behaviour.  However, it is equally important to avoid an excessive number of states, which can lead to model overfitting and mistakenly fitting well to abnormal behaviour as well.  As a general guideline, we suggest starting with a number of latent states greater than the number of response dimensions (e.g., at least 5 states for 3 response dimensions) and increasing the number of states incrementally.  The optimal number of states can then be selected based on Akaike information criterion (AIC) or Bayesian information criterion (BIC).
    \end{itemize}
    \item We divide each trip into intervals of non-zero speed and assume these sub-intervals of each driver are independent time series generated from the same CTHMM.
    \begin{itemize}
        \item The forecast pseudo-residuals we consider for anomaly detection are based on probability integral transform, which applies to continuous distributions.  While continuity adjustments can be made for discrete distributions, boundary points still pose problems.  \cite{harte_hiddenmarkov_2021} pointed out that pseudo-residuals are poor indicators of the model goodness of fit when observations are near or at the boundary of the domain.  Similarly, we find that under these conditions, pseudo-residuals are also ineffective at detecting outliers.  Although we can model all values of speed with a semi-continuous distribution (e.g., a mixture of a continuous distribution on the positive real line and a point mass at 0), the boundary point 0 is problematic for anomaly detection.  For this reason, we exclude 0-speed observations by dividing each trip into intervals of non-zero speed and non-missing accelerations, capturing moments when the driver is actually driving and the vehicle is moving.  Consequently, we assume these sub-intervals (instead of trips) of each driver are independent time series generated by the same model and the CTHMM is fitted to sub-intervals.  For analysis and anomaly detection, we still evaluate on a trip basis: we first compute the pseudo-residuals for each sub-interval, then aggregate the total number of outliers across all sub-intervals to compute the anomaly index (outlier proportion) for the trip; see Figure \ref{fig: sample_trip_shaded} and the description that follows for an illustration.  By cutting trips into shorter, non-zero speed intervals, we aim to standardize the time series and reduce trip heterogeneity due to varying trip duration, road and weather condition, etc.
        \item Periods of low speed (near zero) may correspond to stops at traffic lights or other circumstances beyond the driver's control, and therefore may not be relevant for identifying anomalous driving behaviour.  While excluding 0-speed intervals was necessary for effective outlier detection, this adjustment also mitigates the concern about such low-speed periods.  By focusing on non-zero speed intervals, representing moments when the driver is actively driving and the vehicle is moving, this approach ensures that stopping events, such as waiting at traffic lights, do not influence the anomaly detection process.
    \end{itemize}
\end{enumerate}

\section{Estimation Algorithm}
\label{Estimation}

The proposed CTHMM is fitted using the Expectation-Maximization (EM) algorithm.  As discussed in Section \ref{Adapting-CTHMM-For-Telematics}, we fit the model to trip sub-intervals --- either for individual driver $i$ in an individual-specific model or for a group of drivers $i=1,...,M$ in a pooled model.  In the following, we  only present the EM algorithm for the pooled model.  With the assumption that drivers are independent, we note that the individual-specific model is simply a special case where $M=1$.

\subsection{E-step}

At the $m$-th iteration, we first calculate the expected complete data loglikelihood 
\begin{equation}\label{eqn: ll-no-covariate}
\begin{split}
    Q(\boldsymbol{\Phi}; \mathbf{Y}^*, \mathbf{T}^*, \boldsymbol{\Phi}^{(m-1)}) &= 
    \mathbb{E}\left(\ell(\boldsymbol{\Phi}; \mathbf{Z}', \mathbf{T}', \mathbf{Y}, \mathbf{T}) | \mathbf{Y}^*, \mathbf{T}^*, \boldsymbol{\Phi}^{(m-1)}\right) \\
    &= \sum_{i=1}^M \sum_{j=1}^{N_i} \sum_{u=1}^S \mathbb{E}(\mathbbm{1}_{\{Z_{ij0} = u\}} | \mathbf{Y}^*, \mathbf{T}^*, \boldsymbol{\Phi}^{(m-1)}) \log(\pi_u) \\
    &\quad + \sum_{i=1}^M \sum_{j=1}^{N_i} \sum_{u=1}^S \sum_{v=1, v\neq u}^S \left\{ \log(q_{uv}) \mathbb{E}(n_{ij,uv}|\mathbf{Y}^*, \mathbf{T}^*, \boldsymbol{\Phi}^{(m-1)}) \right. \\
    &\left. \qquad - q_{u} \mathbb{E}(\tau_{ij,u}|\mathbf{Y}^*, \mathbf{T}^*, \boldsymbol{\Phi}^{(m-1)})\right\} \\
    &\quad + \sum_{i=1}^M \sum_{j=1}^{N_i} \sum_{l=0}^T \sum_{u=1}^S \mathbb{E}(\mathbbm{1}_{\{Z_{ijl} = u\}} | \mathbf{Y}^*, \mathbf{T}^*, \boldsymbol{\Phi}^{(m-1)}) \log(f_u(\mathbf{y}_{ijl}))
\end{split}
\end{equation}

\noindent
where $\mathbb{E}(\cdot|\mathbf{Y}^*, \mathbf{T}^*, \boldsymbol{\Phi}^{(m-1)})$ is the expected value given observed data $\mathbf{Y}^*$, observation times $ \mathbf{T}^*$ and the current estimate of model parameters $\boldsymbol{\Phi}^{(m-1)}$ resulted from the $(m-1)$-th iteration.

To simplify the notation, we first state the computation algorithms for a single trip sub-interval $j$ from a driver $i$.

\subsubsection{Conditional Expectation of $\mathbbm{1}_{\{Z_{l} = u\}}$}

To compute the terms $\mathbb{E}(\mathbbm{1}_{\{Z_{l} = u\}} | \mathbf{Y}^*, \mathbf{T}^*, \boldsymbol{\Phi}^{(m-1)})$ for $l = 0,1,...,T_j$, $u = 1,...,S$, one can apply the standard decoding methods --- forward-backward algorithm (soft decoding) or Viterbi algorithm (hard decoding).  Given the current model parameter estimates $\boldsymbol{\Phi}^{(m-1)}$, which include the initial state probabilities $\pi$, transition rate matrix $\mathbf{Q}$, state dependent distributions $f_u(\mathbf{y}_l)$, we will utilize the state-transition matrix $\mathbf{P}(\Delta_l)$, and let $\mathbf{f}(\mathbf{y}_l)$ be a diagonal matrix with the $u$-th diagonal element $f_u(\mathbf{y}_l)$.

\subsubsection*{Soft Decoding - Forward-Backward Algorithm}

We define the forward probabilities $\boldsymbol{\alpha}_l$, backward probabilities $\boldsymbol{\beta}_l$ and likelihood $L_{T_j}$ (at time $T_j$) recursively as follows:
\begin{align*}
    \boldsymbol{\alpha}_0 &= \pi \mathbf{f}(\mathbf{y}_0) \\
    \boldsymbol{\alpha}_l &= \boldsymbol{\alpha}_{l-1} \mathbf{P}(\Delta_l) \mathbf{f}(\mathbf{y}_l) \quad \text{for } l=1,2,...,T_j;\\
    \boldsymbol{\beta}_{T_j} &= \mathbf{1} \\
    \boldsymbol{\beta}'_l &= \mathbf{P}(\Delta_{l+1}) \mathbf{f}(\mathbf{y}_{l+1})\boldsymbol{\beta}'_{l+1} \quad \text{for } l=0,1,...,T_j-1;\\
    L_{T_j} &= \boldsymbol{\alpha}_l \boldsymbol{\beta}'_l \quad \text{for } l=0,1,...,T_j
\end{align*}

Then we can define the `smoother' for $l=0,1,...,T_j$
\begin{equation}\label{eqn: smoother-soft-decoding}
    \mathbb{E}(\mathbbm{1}_{\{Z_l = u\}} | \mathbf{Y}^*, \mathbf{T}^*, \boldsymbol{\Phi}^{(m-1)}) = \frac{\boldsymbol{\alpha}_l[u] \boldsymbol{\beta}_l[u]}{L_{T_j}}
\end{equation}
where $\boldsymbol{\alpha}_l[u]$ and $\boldsymbol{\beta}_l[u]$ are the $u$-th elements in the vectors $\boldsymbol{\alpha}_l$ and $\boldsymbol{\beta}_l$ respectively.

To this end, we also define the `two-slice marginal' for $l=0,1,...,T_j-1$, which will be used in the computation in Section \ref{section: learn-n-tau},
\begin{equation}\label{eqn: two-slice-marginal-soft-decoding}
    \mathbb{E}(\mathbbm{1}_{\{Z_l = u, Z_{l+1} = v\}} | \mathbf{Y}^*, \mathbf{T}^*, \boldsymbol{\Phi}^{(m-1)}) = \frac{\boldsymbol{\alpha}_l[u] \mathbf{P}(\Delta_{l+1})[u,v] f_v(\mathbf{y}_{l+1})\boldsymbol{\beta}_{l+1}[v]}{L_{T_j}}
\end{equation}
where $\mathbf{P}(\Delta_{l+1})[u,v]$ is the $(u,v)$-th element of $\mathbf{P}(\Delta_{l+1})$.

\subsubsection*{Hard Decoding - Viterbi Algorithm}

We first construct two $S$-by-$(T_j+1)$ matrices $\mathbf{V}_1$ and $\mathbf{V}_2$ with their $(u,l)$-th elements as follows:
\begin{align*}
    \mathbf{V}_1[u,0] &= \max_k (\pi_k f_u(\mathbf{y}_0)) \\
    \mathbf{V}_2[u,0] &= 0 \\
    \mathbf{V}_1[u,l] &= \max_k (\mathbf{V}_1[k,l-1] \mathbf{P}(\Delta_{l})[k,u] f_u(\mathbf{y}_l)) \quad \text{for } l=1,...,T_j\\
    \mathbf{V}_2[u,l] &= {\arg\max}_k (\mathbf{V}_1[k,l-1] \mathbf{P}(\Delta_{l})[k,u] f_u(\mathbf{y}_l)) \quad \text{for } l=1,...,T_j\\
\end{align*}

Then we backtrack to find the most likely sequence of latent states:
\begin{align*}
    \hat{z}_{T_j} &= {\arg\max}_k (\mathbf{V}_1[k,T_j]) \\
    \hat{z}_{l-1} &= \mathbf{V}_2(\hat{z}_l,l) \quad \text{for } l=1,...,T_j
\end{align*}

We can define the `smoother' for $l=0,1,...,T_j$
\begin{equation}\label{eqn: smoother-hard-decoding}
    \mathbb{E}(\mathbbm{1}_{\{Z_l = u\}} | \mathbf{Y}^*, \mathbf{T}^*, \boldsymbol{\Phi}^{(m-1)}) = \mathbbm{1}_{\{\hat{z}_l=u\}}
\end{equation}
and the `two-slice marginal' for $l=0,1,...,T_j-1$
\begin{equation}\label{eqn: two-slice-marginal-hard-decoding}
    \mathbb{E}(\mathbbm{1}_{\{Z_l = u, Z_{l+1} = v\}} | \mathbf{Y}^*, \mathbf{T}^*, \boldsymbol{\Phi}^{(m-1)}) = \mathbbm{1}_{\{\hat{z}_l=u, \hat{z}_{l+1}=v\}}.
\end{equation}

\subsubsection{Conditional Expectations of $n_{uv}$ and $\tau_{u}$}
\label{section: learn-n-tau}

The complication in the E-step lies in the calculation of $\mathbb{E}(n_{uv}|\mathbf{Y}^*, \mathbf{T}^*, \boldsymbol{\Phi}^{(m-1)})$ and $\mathbb{E}(\tau_{u}|\mathbf{Y}^*, \mathbf{T}^*, \boldsymbol{\Phi}^{(m-1)})$.  We make use of the \textit{Expm} algorithm proposed by \cite{liu_efficient_2015} which provides an efficient way for computing integrals of matrix exponentials; it is restated in Algorithm \ref{alg: Expm}.

The algorithm exploits the fact that the hidden Markov chain is time-homogeneous and the state-transition matrix at any timepoint $t_{l-1}$, $\mathbf{P}(\Delta_l)$, only depends on the time interval between now and the next observation timepoint $\Delta_l = t_l - t_{l-1}$.  As a consequence, the conditional expectations of $n_{ij,uv}$ and $\tau_{ij,u}$ only have to be evaluated for each distinct time interval rather than at each observation time.  We denote $\Delta'_k, k=1,...,r,$ the $r$ distinct values of $\Delta_l, l=1,...,T_j$, and create count tables $\mathbf{C}(\Delta'_k)$ with the two-slice marginals from Equation \ref{eqn: two-slice-marginal-soft-decoding} or Equation \ref{eqn: two-slice-marginal-hard-decoding} as
$$\mathbf{C}(\Delta'_k)[u,v] = \sum_{l: \Delta_l = \Delta'_k} \mathbb{E}(\mathbbm{1}_{\{Z_l = u, Z_{l+1} = v\}} | \mathbf{Y}^*, \mathbf{T}^*, \boldsymbol{\Phi}^{(m-1)}).$$

\begin{algorithm}
\caption{\textit{Expm} Algorithm for Computing $\mathbb{E}(\tau_{u}|\mathbf{Y}^*, \mathbf{T}^*, \boldsymbol{\Phi}^{(m-1)})$ and $\mathbb{E}(n_{uv}|\mathbf{Y}^*, \mathbf{T}^*, \boldsymbol{\Phi}^{(m-1)})$}
\label{alg: Expm}
\begin{algorithmic}
\For{states $u = 1,...,S$}
    \State $\mathbb{E}(\tau_{u}|\mathbf{Y}^*, \mathbf{T}^*, \boldsymbol{\Phi}^{(m-1)}) = 0$
    \For{distinct $\Delta'_k, k = 1,...,r$}
        \State $\mathbf{D}_u = \exp(\Delta'_k\mathbf{A})[(1:S),(S+1):(2S)], \text{where } \mathbf{A} = \begin{bmatrix}
        \mathbf{Q} & I(u,u)\\
        0 & \mathbf{Q}
        \end{bmatrix}$
        \State $\mathbb{E}(\tau_{u}|\mathbf{Y}^*, \mathbf{T}^*, \boldsymbol{\Phi}^{(m-1)}) += \sum_{u'=1}^S \sum_{v'=1}^S \mathbf{C}(\Delta'_k)[u',v']\frac{\mathbf{D}_u[u',v']}{\mathbf{P}(\Delta'_k)[u',v']}$
    \EndFor
\EndFor
\\
\For{states $u = 1,...,S$}
    \For{states $v = 1,...,S$}
        \State $\mathbb{E}(n_{uv}|\mathbf{Y}^*, \mathbf{T}^*, \boldsymbol{\Phi}^{(m-1)}) = 0$
        \For{distinct $\Delta'_k, k = 1,...,r$}
            \State $\mathbf{D}_{uv} = q_{uv} \exp(\Delta'_k\mathbf{A})[(1:S),(S+1):(2S)], \text{where } \mathbf{A} = \begin{bmatrix}
            \mathbf{Q} & I(u,v)\\
            0 & \mathbf{Q}
            \end{bmatrix}$
            \State $\mathbb{E}(n_{uv}|\mathbf{Y}^*, \mathbf{T}^*, \boldsymbol{\Phi}^{(m-1)}) += \sum_{u'=1}^S \sum_{v'=1}^S \mathbf{C}(\Delta'_k)[u',v']\frac{\mathbf{D}_{uv}[u',v']}{\mathbf{P}(\Delta'_k)[u',v']}$
        \EndFor
    \EndFor
\EndFor \\\\
where $I(u,v)$ is the matrix with a 1 in the $(u,v)$-th entry and 0 elsewhere.
\end{algorithmic}
\end{algorithm}

\subsection{M-step}

The goal of the M-step is to maximize $Q(\boldsymbol{\Phi}; \mathbf{Y}^*, \mathbf{T}^*, \boldsymbol{\Phi}^{(m-1)})$ with respect to $\boldsymbol{\Phi}$.  However, it is too computational costly to directly find the global maximum.  Instead, we aim to use a computational effective algorithm to find a near-maximum to update the parameters $\boldsymbol{\Phi}^{(m)}$ such that $Q(\boldsymbol{\Phi}^{(m)}; \mathbf{Y}^*, \mathbf{T}^*, \boldsymbol{\Phi}^{(m-1)}) \geq Q(\boldsymbol{\Phi}^{(m-1)}; \mathbf{Y}^*, \mathbf{T}^*, \boldsymbol{\Phi}^{(m-1)})$.  It follows from Equation \ref{eqn: ll-no-covariate} that $Q(\boldsymbol{\Phi}; \mathbf{Y}^*, \mathbf{T}^*, \boldsymbol{\Phi}^{(m-1)})$ can be decomposed into three parts
$$Q(\boldsymbol{\Phi}; \mathbf{Y}^*, \mathbf{T}^*, \boldsymbol{\Phi}^{(m-1)}) = Q^{(m)}_{\pi} + Q^{(m)}_{\mathbf{Q}} + Q^{(m)}_{\boldsymbol{\Theta}}$$
\noindent
where
\begin{align}
    Q^{(m)}_{\pi} &= \sum_{i=1}^M \sum_{j=1}^{N_i} \sum_{u=1}^S \mathbb{E}(\mathbbm{1}_{\{Z_{ij0} = u\}} | \mathbf{Y}^*, \mathbf{T}^*, \boldsymbol{\Phi}^{(m-1)}) \log(\pi_u); \label{eqn: Q-pi}\\
    Q^{(m)}_{\mathbf{Q}} &= \sum_{i=1}^M \sum_{j=1}^{N_i} \sum_{u=1}^S \sum_{v=1, v\neq u}^S \left\{ \log(q_{uv}) \mathbb{E}(n_{ij,uv}|\mathbf{X}^*, \mathbf{Y}^*, \mathbf{T}^*, \boldsymbol{\Phi}^{(m-1)}) - q_{u} \mathbb{E}(\tau_{ij,u}| \mathbf{Y}^*, \mathbf{T}^*, \boldsymbol{\Phi}^{(m-1)})\right\}; \text{and} \label{eqn: Q-Q}\\
    Q^{(m)}_{\boldsymbol{\Theta}} &= \sum_{i=1}^M \sum_{j=1}^{N_i} \sum_{l=0}^{T_j} \sum_{u=1}^S \mathbb{E}(\mathbbm{1}_{\{Z_{ijl} = u\}} | \mathbf{Y}^*, \mathbf{T}^*, \boldsymbol{\Phi}^{(m-1)}) \log(f_u(\mathbf{y}_{ijl})) \label{eqn: Q-emission}.
\end{align}

Clearly, $Q^{(m)}_{\pi}$ depends only on $\pi$, $Q^{(m)}_{\mathbf{Q}}$ depends only on $\mathbf{Q}$ and $Q^{(m)}_{\boldsymbol{\Theta}}$ depends only on $\boldsymbol{\Theta}$.  As a consequence, the problem reduces to maximize the three functions separately with respect to the corresponding subset of parameters.  While we can compute the E-step for each trip sub-interval separately, we have to consider all of them collectively in the M-step calculation.

\subsubsection{Initial State Probability Distribution $\pi$}

We constrain $\pi_u$ summing to unity by use of a Lagrange multiplier.  Then by taking the derivative of $Q^{(m)}_{\pi}$ with respect to $\pi_u$, we have the maximum likelihood estimator (MLE) in the $m$-th iteration as
$$\pi^{(m)}_u = \frac{\sum_{i=1}^M \sum_{j=1}^{N_i} \mathbb{E}(\mathbbm{1}_{\{Z_{ij0} = u\}} | \mathbf{X}^*, \mathbf{Y}^*, \mathbf{T}^*, \boldsymbol{\Phi}^{(m-1)})}{\sum_{u'=1}^S \sum_{i=1}^M \sum_{j=1}^{N_i} \mathbb{E}(\mathbbm{1}_{\{Z_{ij0} = u'\}} | \mathbf{X}^*, \mathbf{Y}^*, \mathbf{T}^*, \boldsymbol{\Phi}^{(m-1)})}.$$

\subsubsection{Transition Rate Matrix $\mathbf{Q}$}

By taking the derivative of $Q^{(m)}_{\mathbf{Q}}$ with respect to $q_{uv}$ for $u \neq v$, we have the MLE in the $m$-th iteration as
$$q^{(m)}_{uv} = \frac{\sum_{i=1}^M \sum_{j=1}^{N_i} \mathbb{E}(n_{ij,uv}|\mathbf{Y}^*, \mathbf{T}^*, \boldsymbol{\Phi}^{(m-1)})}{\sum_{i=1}^M \sum_{j=1}^{N_i} \mathbb{E}(\tau_{ij,u}|\mathbf{Y}^*, \mathbf{T}^*, \boldsymbol{\Phi}^{(m-1)})} \quad \text{and} \quad q^{(m)}_{uu} = - q^{(m)}_{u} = - \sum_{v\neq u} q^{(m)}_{uv}.$$

\subsubsection{State-Dependent Distributions $\boldsymbol{\Theta}$}

The maximization of $Q^{(m)}_{\boldsymbol{\Theta}}$ depends on the specific probability distributions that are employed.  Generally speaking, one will have to compute the weighted MLE of the distribution parameters, with weights for the $u$-th latent state $\mathbb{E}(\mathbbm{1}_{\{Z_{ijl} = u\}} | \mathbf{Y}^*, \mathbf{T}^*, \boldsymbol{\Phi}^{(m-1)})$.  In our proposed CTHMM, we assume the different dimensions $Y_{ijl,d}$ of the (multi-dimensional) response $\mathbf{Y}_{ijl}$ to be independent conditional on the latent state.  Then conditional on the latent state, one can compute the (weighted) MLE of each dimension separately.  In our Julia implementation, we incorporated the following distributions: Gamma, Laplace, Log-Normal, Normal, von Mises, zero-inflated Gamma, and zero-inflated Log-Normal.  The MLE of the parameters for each selected distribution are provided in Appendix A of the Supplementary Material.  The implementation is publicly available at https://github.com/ianwengchan/HMMToolkit.

\subsection{Joint Modelling of Telematics and Computational Cost}

In this work, we analyze multivariate time series of speed, longitudinal acceleration and lateral acceleration.  Joint modelling of telematics responses provides a more complete picture of driving behaviour while also enhancing anomaly detection.  Although it increases computational complexity, it provides a more comprehensive and robust risk assessment compared to analyzing each variable separately.

One key advantage of joint modelling is that it captures the interactions between different telematics responses, leading to a more complete representation of driving behaviour.  Speed and acceleration complement each other, offering richer context than when analyzed individually.  For instance, the same acceleration intensity at different speed ranges may indicate very different driving conditions: a harsh acceleration at low speeds could be a normal response in stop-and-go traffic, whereas the same acceleration at high speeds might signal aggressive driving.

Beyond improving behavioural modelling, joint modelling significantly enhances anomaly detection.  As will be demonstrated in Section \ref{Controlled-Study-Individual-Specific-Model} (Tables \ref{tab: D1-M-Aggressive-Snapshot} and \ref{tab: D2-M-Aggressive-Snapshot}), outlier identification depends not only on a single variable but on the combination of speed and acceleration components.  Since pseudo-residual calculations and anomaly detection are influenced by which latent state is the most probable at each time point, and it depends on the other response dimensions as well, excluding key response dimensions weakens the model's ability to identify true behavioural anomalies.

The trade-off between predictive accuracy and computational complexity primarily comes from an increase in the number of latent states.  For example, if speed and longitudinal acceleration are each categorized into three levels (low, mid, high), a univariate CTHMM requires only 3 states, while a multivariate model may require 9 states.  However, if both univariate and multivariate models are constrained to the same number of states, the multivariate model does not introduce significant additional computational costs.  Under the assumption of conditional independence of response dimensions given a state, likelihood calculations remain computationally efficient as they only involve the multiplication of density functions across dimensions.

\section{Controlled Study}
\label{Controlled-Study}

This section details the application of the proposed CTHMM to a labelled telematics dataset to validate its effectiveness in identifying deviating driving behaviour.

\subsection{Data Description}
\label{Controlled-Study-Data-Description}

Before applying our CTHMM-based anomaly detection to the real-world dataset which lacks labels for driving behaviour, we first evaluate its effectiveness using a labelled telematics dataset.  Specifically, we analyze the UAH-DriveSet \citep{romera_need_2016}, a publicly available telematics dataset.  This dataset consists of six drivers, each driving a different vehicle, performing three distinct driving behaviour (normal, aggressive, and drowsy) on two different roads (a motorway and a secondary road).  While the dataset contains extensive telematics information and even video recordings, our controlled study focuses on validating the proposed method’s ability to detect abnormal driving behaviour, ensuring that it can be applied to the real-world data for risk evaluation.  For this purpose, we limit our analysis to the same set of telematics features, which are speed, longitudinal and lateral accelerations only.

In the UAH-DriveSet, telematics recording starts when the vehicle enters the specific road and ends when it exits, hence speed is always non-zero, posing no issues for anomaly detection as discussed in Section \ref{Adapting-CTHMM-For-Telematics}.  Speed is collected from GPS at 1Hz, while accelerations are recorded from accelerometer at 10Hz.  According to \cite{romera_need_2016}, `Z' and `Y' denote the longitudinal and lateral accelerations respectively.  Since accelerometers are sensitive and measurements can be noisy, we consider `Z\_KF' and `Y\_KF' which have been filtered by a Kalman Filter and are readily available in the dataset.  We downsample the 3-dimensional responses such that they are all observed per second.  Although one can apply a discrete-time HMM to this data due to its regular time intervals, we apply the continuous-time version to also test that the model estimation algorithm runs correctly.

On the motorway, each driver performed three round-trips, simulating each of the three driving behaviour.  On the secondary road, each of them performed four one-way trips, consisting of both a departure and a return as normal, a departure as aggressive and a return as drowsy.  Hence each driver should have telematics records of seven trips.  In the following discussion, we omit driver D6 who drove an electric vehicle, since he did not perform the aggressive trip on the secondary road due to lack of autonomy.

While the road type is available in the UAH-DriveSet (in fact, drivers are not only driving on two specific road types but two designated roads), road types are not always available in real data.  In reality, trips tend to be longer and can involve various road types.  For example, a driver who commutes to work from home can have travelled on all of residential streets, primary to tertiary roads, and motorways, and it can be difficult to assign a single road type to such a trip.  For this reason, we do not consider road types in the controlled study; yet, we will show that results demonstrate strong model performance.

\subsection{Individual-Specific Model}
\label{Controlled-Study-Individual-Specific-Model}

We first test the ability of individual-specific models in identifying anomalous trips for each driver, with a particular focus on detecting the two aggressive trips as a representative example.  For each of the drivers D1 to D5, we fit a 20-state CTHMM to his/her seven trips. We tested a limited selection of models with 5, 10, 15, 20, and 25 states, with 20 states being optimal according to both AIC and BIC.  This choice ensures that the model captures all major combinations of speed and accelerations without overfitting.  More discussions on the robustness of the results with respect to the number of states are presented later in this subsection.

The state-dependent distributions are Gamma for speed, and Normal for both longitudinal and lateral accelerations.  These choices are made based on the domain and characteristics of each response dimension.  Gamma distribution is right-skewed and strictly positive, making it well-suited for modelling speed, which is often asymmetric with a long tail at higher values.  Additionally, it has two parameters (shape and scale), allowing it to adapt to different degrees of skewness and dispersion in speed distributions.  Normal distribution is appropriate for acceleration because acceleration values are often symmetrically distributed around zero, with both positive (acceleration) and negative (deceleration) values.  It is also widely used due to its tractable properties and well-established estimation methods.  In addition to optimizing the number of latent states, one can also explore alternative choices for state-dependent distributions.  For example, a 20-state CTHMM could be fitted using either Lognormal or Gamma distributions for speed, with model selection guided by AIC or BIC.  This flexibility allows for further refinement of the model based on empirical fit.

Initial guesses for parameter estimation are important in EM algorithm as they can influence the convergence of the algorithm and the final model fit, especially in the case when the algorithm converges to a local optimum instead of a global one.  Parameter initializations of $\pi$ and $\mathbf{Q}$ are random, while state-dependent distribution parameters $\boldsymbol{\Theta}$ are initiated with k-means clustering.  Users are encouraged to try multiple different initializations to reduce the chances of the model converging to a local optimum.  To robustify our analysis, we re-fit the model 100 times with different random initializations and analyze the average result.

After model fitting, we calculate the anomaly indices of each trip under the corresponding driver's model.  The results are reported in Table \ref{tab: UAH-individual-20}.  First, we observe that trips performed on motorway usually have some outliers in speed.  This is expected as the driving speed is generally higher on motorway than on secondary road.  However, speed alone does not reveal the driving behaviour and it is essential to consider the changes of it, which are the accelerations.  Second, we observe that the two aggressive trips are characterized by the higher anomaly index in the longitudinal acceleration (acceleration and braking).  This is in line with the experiment set-up: the drivers are told to push to the limit their aggressiveness without putting the vehicle at risk.  Furthermore, this is also in line with \cite{zylius_driving_2014}, which showed that aggressive and safe driving styles can be more effectively classified using longitudinal acceleration signals.  While longitudinal acceleration is clearly associated with aggressiveness, the relationship between lateral acceleration and driving behaviour is less straightforward.  However, as we will demonstrate in Section \ref{Real-Data-Analysis}, anomaly indices for both longitudinal and lateral accelerations show a positive correlation with accident probability.

\begin{table}[h!]
\caption{UAH-DriveSet: Anomaly index in each dimension of telematics observations, computed from individual-specific CTHMMs with 20 states for each driver and averaged over 100 random initializations.  Note: S stands for Secondary Road and M stands for Motorway; Z and Y are longitudinal and lateral accelerations, respectively.}
\centering
    \begin{tabular}{c|l|rrr}
    \textbf{Driver} & \textbf{Trip} & \textbf{Y\_KF} & \textbf{Z\_KF} & \textbf{Speed} \\\hline
    D1 & S-Normal 1 & 0.0115 & 0.0030 & 0.0000 \\
     & S-Normal 2 & 0.0058 & 0.0011 & 0.0000 \\
     & S-Aggressive & 0.0090 & \textbf{0.0080} & 0.0000 \\
     & S-Drowsy & 0.0099 & 0.0071 & 0.0000 \\
     & M-Normal & 0.0039 & 0.0013 & 0.0000 \\
     & M-Aggressive & 0.0046 & \textbf{0.0099} & 0.0049 \\
     & M-Drowsy & 0.0013 & 0.0035 & 0.0000 \\\hline
    D2 & S-Normal 1 & 0.0021 & 0.0003 & 0.0000 \\
     & S-Normal 2 & 0.0038 & 0.0003 & 0.0000 \\
     & S-Aggressive & 0.0018 & \textbf{0.0166} & 0.0009 \\
     & S-Drowsy & 0.0209 & 0.0050 & 0.0000 \\
     & M-Normal & 0.0033 & 0.0016 & 0.0011 \\
     & M-Aggressive & 0.0071 & \textbf{0.0133} & 0.0007 \\
     & M-Drowsy & 0.0139 & 0.0034 & 0.0000 \\\hline
    D3 & S-Normal 1 & 0.0045 & 0.0016 & 0.0000 \\
     & S-Normal 2 & 0.0074 & 0.0042 & 0.0000 \\
     & S-Aggressive & 0.0095 & \textbf{0.0133} & 0.0000 \\
     & S-Drowsy & 0.0051 & 0.0035 & 0.0000 \\
     & M-Normal & 0.0051 & 0.0017 & 0.0020 \\
     & M-Aggressive & 0.0037 & \textbf{0.0179} & 0.0001 \\
     & M-Drowsy & 0.0077 & 0.0003 & 0.0002 \\\hline
    D4 & S-Normal 1 & 0.0034 & 0.0027 & 0.0001 \\
     & S-Normal 2 & 0.0080 & 0.0016 & 0.0000 \\
     & S-Aggressive & 0.0083 & \textbf{0.0132} & 0.0003 \\
     & S-Drowsy & 0.0153 & 0.0036 & 0.0004 \\
     & M-Normal & 0.0016 & 0.0023 & 0.0017 \\
     & M-Aggressive & 0.0134 & \textbf{0.0168} & 0.0001 \\
     & M-Drowsy & 0.0045 & 0.0008 & 0.0000 \\\hline
    D5 & S-Normal 1 & 0.0010 & 0.0006 & 0.0000 \\
     & S-Normal 2 & 0.0015 & 0.0013 & 0.0037 \\
     & S-Aggressive & 0.0075 & \textbf{0.0099} & 0.0000 \\
     & S-Drowsy & 0.0060 & 0.0024 & 0.0000 \\
     & M-Normal & 0.0019 & 0.0029 & 0.0004 \\
     & M-Aggressive & 0.0105 & \textbf{0.0076} & 0.0000 \\
     & M-Drowsy & 0.0059 & 0.0011 & 0.0001 \\\hline
    \end{tabular}
\label{tab: UAH-individual-20}
\end{table}

While the anomaly index with the proposed outlier threshold of 3 standard deviations (SDs) provides strong anomaly detection results, we further examine the impact of threshold selection on performance.  Table \ref{tab: anoamly-index-threshold} presents the anomaly indices computed using alternative outlier thresholds of 2 and 4 SDs.  Using a lower (higher) threshold increases (decreases) the anomaly index.  We observe that with a 2-SD threshold, the two aggressive trips for each driver still exhibit higher anomaly indices in longitudinal acceleration and can be identified.  However, with a 4-SD threshold, many indices drop close to zero, as it becomes highly unlikely for a normal forecast pseudo-residual to exceed 4 in absolute magnitude.  As a result, the two aggressive trips can no longer be distinguished from other trips.  Furthermore, while the 2-SD threshold still identifies the aggressive trips, it results in anomaly indices that are approximately 10 times higher than those computed with a 3-SD threshold (Table \ref{tab: UAH-individual-20}).  This raises concerns about whether the trips are truly that anomalous and whether the anomaly index retains its intended meaning.  For instance, under the 2-SD threshold, normal trips on the secondary road have an average of 3.84\% outliers, whereas under the 3-SD threshold, this drops to 0.82\%, which more accurately reflects them as following normal driving behaviour.  These observations support the use of 3 standard deviations as the outlier threshold, as it effectively identifies anomalous trips and driving behaviour while maintaining interpretability.

\begin{table}[h!]
\centering
\caption{UAH-DriveSet: Alternative thresholds (2 and 4 standard deviations (SDs)) for calculating anomaly index in each dimension of telematics observations, computed from individual-specific CTHMMs with 20 states for each driver and averaged over 100 random initializations.  Note: S stands for Secondary Road and M stands for Motorway; Z and Y are longitudinal and lateral accelerations, respectively.}
\begin{tabular}{c|l|ccc|ccc}
 &  & \multicolumn{3}{c}{\textbf{2 SDs}}  & \multicolumn{3}{c}{\textbf{4 SDs}} \\
\textbf{Driver} & \textbf{Trip} & \textbf{Y\_KF} & \textbf{Z\_KF} & \textbf{Speed} & \textbf{Y\_KF} & \textbf{Z\_KF} & \textbf{Speed} \\\hline
D1 & S-Normal1 & 0.0740 & 0.0614 & 0.0126 & 0.0004 & 0.0000 & 0.0000 \\
D1 & S-Normal2 & 0.0690 & 0.0586 & 0.0038 & 0.0001 & 0.0000 & 0.0000 \\
D1 & S-Aggressive & 0.0667 & \textbf{0.0675} & 0.0010 & 0.0008 & \textbf{0.0004} & 0.0000 \\
D1 & S-Drowsy & 0.0657 & 0.0348 & 0.0100 & 0.0016 & 0.0021 & 0.0000 \\
D1 & M-Normal & 0.0399 & 0.0389 & 0.0047 & 0.0001 & 0.0000 & 0.0000 \\
D1 & M-Aggressive & 0.0565 & \textbf{0.0744} & 0.0327 & 0.0007 & \textbf{0.0011} & 0.0000 \\
D1 & M-Drowsy & 0.0341 & 0.0294 & 0.0035 & 0.0000 & 0.0000 & 0.0000 \\\hline
D2 & S-Normal1 & 0.0384 & 0.0185 & 0.0085 & 0.0000 & 0.0000 & 0.0000 \\
D2 & S-Normal2 & 0.0264 & 0.0299 & 0.0001 & 0.0008 & 0.0000 & 0.0000 \\
D2 & S-Aggressive & 0.0413 &\textbf{ 0.1120} & 0.0318 & 0.0000 & \textbf{0.0000} & 0.0000 \\
D2 & S-Drowsy & 0.1192 & 0.0431 & 0.0075 & 0.0025 & 0.0009 & 0.0000 \\
D2 & M-Normal & 0.0253 & 0.0280 & 0.0086 & 0.0009 & 0.0000 & 0.0010 \\
D2 & M-Aggressive & 0.0597 & \textbf{0.0999} & 0.0382 & 0.0012 & \textbf{0.0004} & 0.0000 \\
D2 & M-Drowsy & 0.0894 & 0.0378 & 0.0060 & 0.0009 & 0.0004 & 0.0000 \\\hline
D3 & S-Normal1 & 0.0349 & 0.0313 & 0.0028 & 0.0020 & 0.0000 & 0.0000 \\
D3 & S-Normal2 & 0.0341 & 0.0499 & 0.0042 & 0.0009 & 0.0000 & 0.0000 \\
D3 & S-Aggressive & 0.0500 & \textbf{0.1163} & 0.0080 & 0.0009 & \textbf{0.0001} & 0.0000 \\
D3 & S-Drowsy & 0.0728 & 0.0354 & 0.0024 & 0.0001 & 0.0000 & 0.0000 \\
D3 & M-Normal & 0.0417 & 0.0384 & 0.0112 & 0.0002 & 0.0001 & 0.0000 \\
D3 & M-Aggressive & 0.0528 & \textbf{0.0996} & 0.0528 & 0.0008 & \textbf{0.0007} & 0.0000 \\
D3 & M-Drowsy & 0.0709 & 0.0144 & 0.0053 & 0.0010 & 0.0000 & 0.0000 \\\hline
D4 & S-Normal1 & 0.0133 & 0.0433 & 0.0152 & 0.0000 & 0.0002 & 0.0000 \\
D4 & S-Normal2 & 0.0295 & 0.0344 & 0.0212 & 0.0000 & 0.0000 & 0.0000 \\
D4 & S-Aggressive & 0.0612 & \textbf{0.1135} & 0.0126 & 0.0000 & \textbf{0.0017} & 0.0000 \\
D4 & S-Drowsy & 0.0832 & 0.0436 & 0.0091 & 0.0016 & 0.0000 & 0.0000 \\
D4 & M-Normal & 0.0310 & 0.0247 & 0.0145 & 0.0000 & 0.0000 & 0.0000 \\
D4 & M-Aggressive & 0.0726 & \textbf{0.0943} & 0.0279 & 0.0011 & \textbf{0.0013} & 0.0000 \\
D4 & M-Drowsy & 0.0657 & 0.0222 & 0.0055 & 0.0000 & 0.0000 & 0.0000 \\\hline
D5 & S-Normal1 & 0.0220 & 0.0321 & 0.0032 & 0.0000 & 0.0000 & 0.0000 \\
D5 & S-Normal2 & 0.0504 & 0.0251 & 0.0105 & 0.0000 & 0.0000 & 0.0023 \\
D5 & S-Aggressive & 0.0480 & \textbf{0.0884} & 0.0143 & 0.0004 & \textbf{0.0000} & 0.0000 \\
D5 & S-Drowsy & 0.0655 & 0.0389 & 0.0053 & 0.0000 & 0.0000 & 0.0000 \\
D5 & M-Normal & 0.0415 & 0.0551 & 0.0030 & 0.0000 & 0.0000 & 0.0000 \\
D5 & M-Aggressive & 0.0823 & \textbf{0.0948} & 0.0055 & 0.0006 & \textbf{0.0001} & 0.0000 \\
D5 & M-Drowsy & 0.0425 & 0.0399 & 0.0127 & 0.0006 & 0.0001 & 0.0000 \\\hline
\end{tabular}
\label{tab: anoamly-index-threshold}
\end{table}

It is worth mentioning that pseudo-residuals consider more than just the magnitudes of the responses --- they also take into account information derived from the other dimensions of the response and how likely the observed transitions are.  Table \ref{tab: D1-M-Aggressive-Snapshot} includes a snapshot of D1's aggressive trip performed on the motorway, with pseudo-residuals of longitudinal acceleration Z\_KF.  First, we see that the longitudinal acceleration (-0.070) at timepoint 300 is an outlier, despite having a magnitude smaller than that (-0.102) at timepoint 334 which is not an outlier.  This is because the first transition is from a less extreme value (0.003) to a more extreme one, while the second transition is the other way around.  Second, we also observe that the longitudinal acceleration (-0.070) at timepoint 300 is an outlier while that (-0.086) at timepoint 383 is not despite having transitions of similar magnitude.  The reason is that the speed during the latter transition is in a different range, and such transitions are more common in the lower speed range for this driver. Furthermore, since model training and analysis are performed on a per-driver basis rather than collectively, how likely a transition is, and consequently the identification of outliers, also varies across drivers.  Outliers cannot be identified through direct comparison between drivers.  For example, Table \ref{tab: D2-M-Aggressive-Snapshot} provides a snapshot of D2's aggressive trip performed on the motorway.  If we compare this transition with D1's first (selected) transition, we see that despite having similar speed, comparable resulting longitudinal acceleration, and an arguably more aggressive transition (from 0.041 to -0.075 as compared to from 0.003 to -0.070), it is not an outlier while D1's transition is.  This is due to the fact that such transitions are more typical in D2's driving history but not as much for D1.

\begin{table}[h]
    \caption{Snapshot of D1's aggressive trip on motorway with pseudo-residuals of Z\_KF.  Note: Z and Y are longitudinal and lateral accelerations, respectively.}
    \centering
    \begin{tabular}{c|rrr|r}
         \textbf{Timestamp}&  \textbf{Y\_KF}&  \textbf{Z\_KF}&  \textbf{Speed}&  \textbf{Residuals}\\\hline
         &  &  \multicolumn{1}{c}{$\vdots$} &  &  \\
         299&  -0.010&  0.003&  96.4&  0.3056\\
         300&  0.006&  -0.070&  95.7&  -3.4650\\
         & & \multicolumn{1}{c}{$\vdots$} & &\\
         333&  0.030&  -0.179&  95.0&  -3.1200\\
         334&  0.049&  -0.102&  95.0&  -1.0931\\
         &  &  \multicolumn{1}{c}{$\vdots$} &  &  \\
         382&  -0.011&  0.006&  47.9&  0.4493\\
         383&  -0.033&  -0.086&  51.3&  -2.2369\\
         &  &  \multicolumn{1}{c}{$\vdots$} &  &  \\
    \end{tabular}
    \label{tab: D1-M-Aggressive-Snapshot}
\end{table}

\begin{table}[h]
    \caption{Snapshot of D2's aggressive trip on motorway with pseudo-residuals of Z\_KF.  Note: Z and Y are longitudinal and lateral accelerations, respectively.}
    \centering
    \begin{tabular}{c|rrr|r}
         \textbf{Timestamp}&  \textbf{Y\_KF}&  \textbf{Z\_KF}&  \textbf{Speed}&  \textbf{Residuals}\\\hline
         &  &  \multicolumn{1}{c}{$\vdots$} &  &  \\
         283&  -0.018&  0.041&  94.7&  0.9829\\
         284&  -0.003&  -0.075&  94.8&  -0.6250\\
         &  &  \multicolumn{1}{c}{$\vdots$} &  &  \\
    \end{tabular}
    \label{tab: D2-M-Aggressive-Snapshot}
\end{table}

While the above analysis uses CTHMMs with 20 states, we also explore the effect of different number of states on the results.  For simplicity, we consider a limited selection of 5, 10, 15, 20, and 25 states.  For each number of states, we re-fit the model 100 times with different random initializations and compute the average results.  Since the models with 25 states include degenerate state-dependent distributions for accelerations, we do not report their results and terminate the search at 20 states, which gives the lowest AIC and BIC.  With increasing number of states, we observe that the anomaly indices generally decrease, as more telematics combinations are described and fewer data points are then considered outliers.  The two aggressive trips are consistently characterized by more outliers in longitudinal acceleration, except for Driver 1, where there is a mix between the two aggressive trips and the drowsy trip completed on the secondary road when the number of states is lower.  Detailed results can be found in Appendix B.

Using the UAH-DriveSet, we have validated that the proposed individual-specific CTHMM, along with its anomaly detection method, effectively learns and distinguishes between different driving behaviour.  With minimal supervision --- specifying only the number of states and the state-dependent distribution families --- the fitted model is able to capture key combinations of speed and accelerations and identify deviating behaviour.  The model achieves this by considering not only speed, but also its changes, and more importantly their combinations.  We expect that speed inherently captures information regarding road type, and driving behaviour (normal, aggressive and drowsy) is then described by its combination with lateral and longitudinal accelerations.

Compared to the state-of-the-art methods applied to the UAH-DriveSet, this work introduces several methodological advancements and distinctions.  Existing studies primarily rely on supervised machine learning techniques, leveraging labelled data to classify aggressive driving behaviour using methods such as Convolutional Neural Networks (CNNs), Recurrent Neural Networks (RNNs), and ensemble models.  Moreover, these approaches typically incorporate diverse telematics inputs, including GPS data, gyroscope readings, and even video recordings, to enhance feature richness and classification accuracy (\citet{romera_need_2016}, \citet{moukafih_aggressive_2019}, \citet{liu_fmdnet_2024}).  In contrast, we introduce an unsupervised anomaly detection method under the CTHMM, addressing the practical challenge of label unavailability in real-world scenarios.  Furthermore, this work demonstrates the robustness of using only three fundamental telematics responses: speed, longitudinal acceleration, and lateral acceleration.  This much simpler yet effective approach highlights the adaptability of the proposed methodology to datasets with limited sensor information while maintaining strong capabilities in anomaly detection and aggressive trip identification.

\subsection{Pooled Model}
\label{Controlled-Study-Pooled-Model}

As shown in the previous section, individual-specific models can produce forecast pseudo-residuals that effectively highlight anomalous trips for each driver.  However, there is no guarantee that these residuals can be compared across different individuals as they are derived from distinct models.  Therefore, we employ a pooled model which enables trip comparison across individuals and driver rankings.

We train a 20-state CTHMM using all 35 trips from the five drivers D1 to D5.  The state-dependent distributions and parameter initializations match those used in the individual-specific models.  After model fitting, we assess the anomaly indices of each trip under the pooled model.  To enhance the robustness of our analysis, we also re-fit the model 100 times with different random initializations and present the average result in Table \ref{tab: UAH-pooled-20}.

\begin{table}[h!]
\caption{UAH-DriveSet: Anomaly index in each dimension of telematics observations, computed from pooled CTHMM with 20 states from all drivers and averaged over 100 random initializations.  Note: S stands for Secondary Road and M stands for Motorway; Z and Y are longitudinal and lateral accelerations, respectively.}
\centering
    \begin{tabular}{c|l|rrr|r}
    \textbf{Driver} & \textbf{Trip} & \textbf{Y\_KF} & \textbf{Z\_KF} & \textbf{Speed} & \textbf{Rank} \\\hline\
    D1 & S-Normal 1 & 0.0066 & 0.0009 & 0.0002 & 25 \\
     & S-Normal 2 & 0.0070 & 0.0014 & 0.0000 & 21 \\
     & S-Aggressive & 0.0058 & \textbf{0.0077} & 0.0000 & 7 \\
     & S-Drowsy & 0.0056 & \textbf{0.0063} & 0.0000 & 9 \\
     & M-Normal & 0.0024 & 0.0004 & 0.0000 & 29 \\
     & M-Aggressive & 0.0044 & \textit{0.0046} & 0.0001 & 11 \\
     & M-Drowsy & 0.0019 & 0.0038 & 0.0000 & 13 \\\hline
    D2 & S-Normal 1 & 0.0019 & 0.0010 & 0.0000 & 24 \\
     & S-Normal 2 & 0.0039 & 0.0004 & 0.0000 & 28 \\
     & S-Aggressive & 0.0001 & \textbf{0.0164} & 0.0000 & 1 \\
     & S-Drowsy & 0.0125 & 0.0068 & 0.0000 & 8 \\
     & M-Normal & 0.0026 & 0.0008 & 0.0007 & 26 \\
     & M-Aggressive & 0.0028 & \textbf{0.0124} & 0.0009 & 4 \\
     & M-Drowsy & 0.0070 & 0.0048 & 0.0000 & 10 \\\hline
    D3 & S-Normal 1 & 0.0023 & 0.0004 & 0.0000 & 30 \\
     & S-Normal 2 & 0.0052 & 0.0007 & 0.0000 & 27 \\
     & S-Aggressive & 0.0048 & \textbf{0.0028} & 0.0000 & 16 \\
     & S-Drowsy & 0.0043 & 0.0014 & 0.0000 & 22 \\
     & M-Normal & 0.0004 & 0.0000 & 0.0047 & 32 \\
     & M-Aggressive & 0.0025 & \textbf{0.0040} & 0.0001 & 12 \\
     & M-Drowsy & 0.0053 & 0.0000 & 0.0001 & 34 \\\hline
    D4 & S-Normal 1 & 0.0022 & 0.0017 & 0.0000 & 20 \\
     & S-Normal 2 & 0.0024 & 0.0024 & 0.0000 & 17 \\
     & S-Aggressive & 0.0027 & \textbf{0.0117} & 0.0000 & 5 \\
     & S-Drowsy & 0.0090 & 0.0029 & 0.0000 & 15 \\
     & M-Normal & 0.0000 & 0.0020 & 0.0047 & 18 \\
     & M-Aggressive & 0.0045 & \textbf{0.0132} & 0.0000 & 3 \\
     & M-Drowsy & 0.0025 & 0.0010 & 0.0000 & 23 \\\hline
    D5 & S-Normal 1 & 0.0000 & 0.0000 & 0.0000 & 33 \\
     & S-Normal 2 & 0.0018 & 0.0000 & 0.0058 & 34 \\
     & S-Aggressive & 0.0089 & \textbf{0.0113} & 0.0000 & 6 \\
     & S-Drowsy & 0.0070 & 0.0002 & 0.0000 & 31 \\
     & M-Normal & 0.0033 & 0.0037 & 0.0011 & 14 \\
     & M-Aggressive & 0.0204 & \textbf{0.0144} & 0.0010 & 2 \\
     & M-Drowsy & 0.0042 & 0.0018 & 0.0000 & 19 \\\hline
    \end{tabular}
\label{tab: UAH-pooled-20}
\end{table}

First, when comparing the 7 trips of each driver, we observe that the two aggressive trips again exhibit a higher anomaly index in longitudinal acceleration, except for D1, where there is once more a mix between the two aggressive trips and the drowsy trip driven on the secondary road.  Second, since the pseudo-residuals are derived from the same pooled models, we can now fairly compare them and rank the trips.  Table \ref{tab: UAH-pooled-20} also presents the ranking of each trip among all 35, in descending anomaly index in longitudinal acceleration.  D2's aggressive trip on the secondary road ranks highest in terms of anomalousness in longitudinal acceleration (or aggressiveness), followed by D5's aggressive trip on the motorway.  Lastly, due to heterogeneity in individual driving behaviour, it is possible for a driver's normal trip to be deemed more anomalous than another driver's abnormal trip.  For instance, D5's normal trip on the motorway exhibits a higher anomaly index in longitudinal acceleration compared to D3's aggressive trip on the secondary road.

While the intermediate rankings of trips may not be completely clear, it is obvious that D2 is the most aggressive driver, while D3 is the least.  Although both drivers are male and in the 20-30 age range, they drive different vehicles.  D2 drives a Mercedes B180 (2013), a compact yet relatively powerful vehicle, which could encourage more aggressive driving, particularly on secondary roads.  In contrast, D3 drives a Citroën C4 (2015), a smaller and potentially less powerful car, which could lead to a more cautious driving style.  These differences in vehicle characteristics likely influence their respective driving behaviors.

However, we lack ground-truth labels regarding the anomalousness or riskiness of each driver, as none has been involved in accidents.  While we have outlined a method in this section and demonstrated its use, we admit that we cannot fully validate the rankings presented above.  We keep this issue in mind, but the observed patterns --- such as D2's higher anomaly index indicating more aggressive driving among the group of drivers --- are suggestive.  We aim to validate our findings further in Section \ref{Real-Data-Analysis-Pooled-Model} using the rental car dataset.

\section{Real Data Analysis}
\label{Real-Data-Analysis}

This section details the application of the proposed CTHMM to the rental car dataset.  We analyze telematics data at two distinct levels.  At the driver/rental level using individual-specific modelling, we aim to identify accident-associated trips for rentals with at-fault claims.  At the group-level using pooled modelling, we analyze driving behaviour patterns across drivers with and without at-fault claims, focusing particularly on the anomaly index, to reveal population-level trends and behavioural differences.

\subsection{Data Description}
\label{Real-Data-Analysis-Data-Description}

We now apply our proposed CTHMM to the rental car dataset, which contains claims reported from January 1, 2019 to January 31, 2023.  As we would like to investigate the connection between driving behaviour and accidents, we focus on at-fault claims.  These claims are typically more directly related to driver actions and behaviour, and are expected to provide greater insight into how these factors contribute to accidents.  The portfolio includes 1,678 rentals with at least one at-fault claim and 16,495 rentals without claims.

Recall that one challenge with this dataset is the lack of precise labels for accident-associated trips.  Claims are reported with a date of loss but not the specific trip involved.  To approximate these labels, we rely on `potential accident triggers' provided by our collaborator, who specializes in both telematics recording hardware and software development.  In big lines, these triggers are derived based on predefined thresholds on accelerometer readings and further inspections on speed changes (e.g., whether the vehicle comes to a stop afterwards).

However, the method used by our collaborator is not perfect.  While these triggers are accurate in distinguishing between real crashes and harsh braking events, they suffer from a high false-negative rate due to curve logging (explained in Section \ref{Data-Description-and-Challenges}), which also applies to accelerometer readings.  Curve logging reduces noise in the accelerometer readings, but may unintentionally filter out some acceleration values that exceed the thresholds.  As a result, only about 10\% of the at-fault claims are matched with a corresponding trigger, leaving the remaining 90\% unassociated with specific trips --- even though we know these claims exist.

Given this limitation, we treat the detected triggers as ground truth.  Since the date of loss is customer-reported and may not be fully reliable, we extend the analysis to consider all trips completed within a 3-day window (the date of loss, plus the day before and after).  While we are certain that 1,678 trips are associated with accidents (corresponding to the 1,678 at-fault claims), only 94 of them are identified by our collaborator's method.  These 94 trips are referred to as `target trips' and are treated as the ground truth for our analysis.  The challenge we address with the individual-specific models is whether our method can correctly identify these 94 target trips as being accident-related.  The number of trips per 3-day window (around date of loss) ranges from 2 to 63, with an average of 18 trips, totalling 1,765 trips.  In summary, the goal of the individual-specific model is to identify the 94 target trips that correspond to actual claims and accidents from the pool of 1,765 trips.


\textit{Remark: While potential accident triggers are produced using accelerometer readings, we only consider telematics responses based on GPS recordings.  This is because we can compute acceleration values from GPS recordings, but not the other way around.  Additionally, accelerometer and GPS logging have their own curve logging logic and their readings do not always align.  Furthermore, identifying accident-associated trips is an extremely challenging problem, especially since our analysis is based solely on three telematics responses (speed, longitudinal acceleration and lateral acceleration).}

\begin{figure}[ht]
    \centering
    \includegraphics[width=\textwidth]{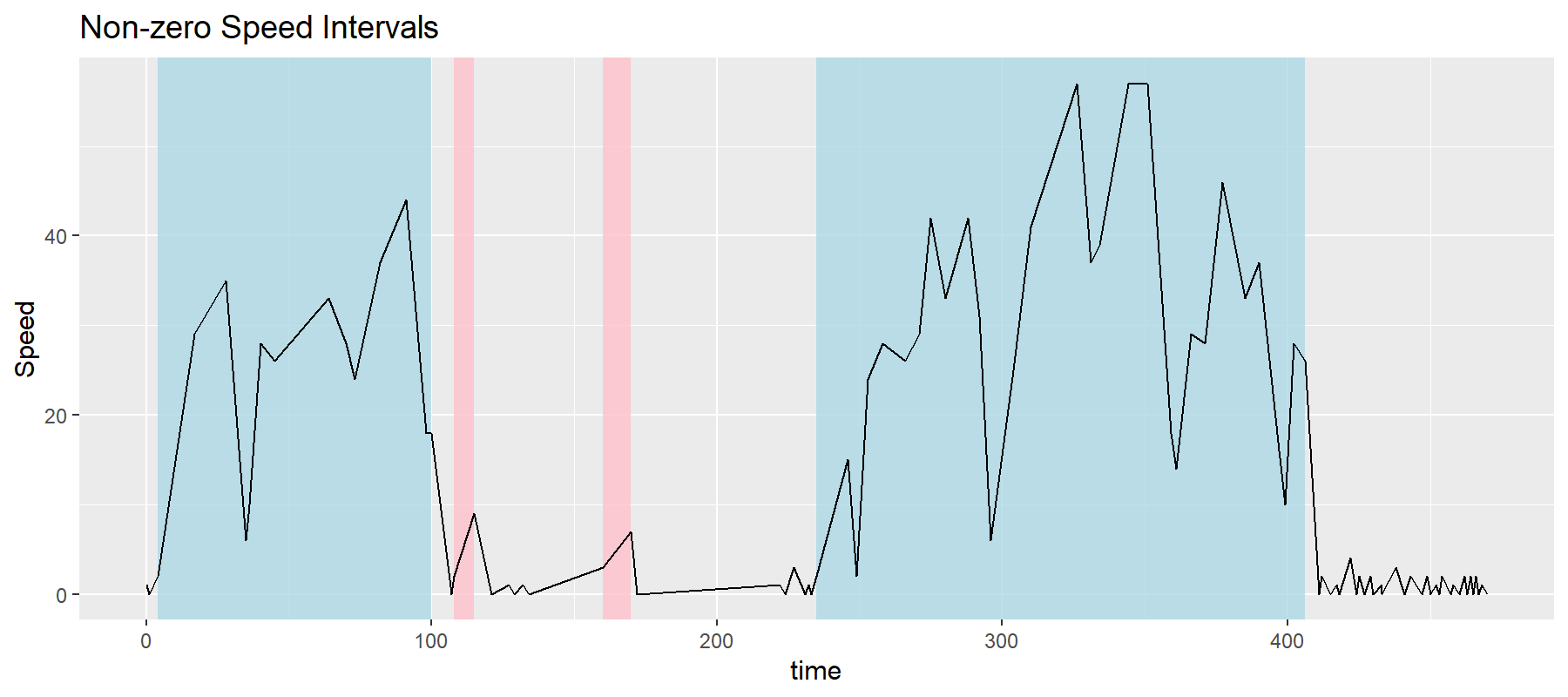}
    \caption{Speed time series (in $km/h$) of a sample trip. Shaded regions indicate intervals of non-zero speed, where all intervals are used for anomaly detection, but only the blue ones are used for model fitting.}
    \label{fig: sample_trip_shaded}
\end{figure}


For both individual-specific and pooled models, telematics data undergo the same preprocessing steps as follows.  First, we compute the longitudinal and lateral accelerations and align them with speed, as detailed in Section \ref{Data-Preparation}.  Next, we divide each trip into intervals of non-zero speed and valid (non-missing) accelerations, as shown by the shaded regions in Figure \ref{fig: sample_trip_shaded}.  For model training, trips must be at least 3 minutes long, and each trip sub-interval must contain at least 10 observations; these correspond to the blue shaded intervals in Figure \ref{fig: sample_trip_shaded}.  This ensures that the CTHMM is trained on longer, informative sequences.  For evaluation, the requirements are relaxed: trips must be at least 30 seconds long and contain at least 10 observations.  These trips are similarly divided into intervals of non-zero speed and valid accelerations, with sub-intervals needing only at least 2 observations so that forecast pseudo-residuals can be computed.  Hence, pseudo-residuals are computed for all shaded intervals (both the blue and pink ones).  Finally, all sub-intervals of a trip will be considered together to compute the anomaly index.

\textit{Remark: While we assume that all trips within each rental are performed by a single driver, it is possible for vehicles to have multiple drivers.  In such cases, the model captures the dominant driving behaviour exhibited collectively by all drivers of the vehicle rather than individual-specific patterns.  This can be viewed as a pooled model, which still accounts for vehicle-specific effects (e.g., horsepower) while incorporating driving patterns across multiple users.  As demonstrated in Section \ref{Controlled-Study-Pooled-Model}, a pooled model can still effectively identify anomalous trips, even when multiple drivers contribute to the data.}

\textit{A potential limitation arises when the dataset is imbalanced.  For example, if one driver completes most trips and another completes only a single trip, that one-off trip --- even if it is objectively safer --- may still be classified as anomalous if it significantly deviates from the dominant driving pattern.  However, this is an inherent characteristic of anomaly detection, and controlling for such cases would require additional contextual information.  If information on driver changes is available, the model could be adapted simply by clustering trips by driver and training a separate model for each driver, thereby reducing the influence of mixed driving patterns.}

\subsection{Individual-Specific Model}
\label{Real-Data-Analysis-Individual-Specific-Model}

We first apply individual-specific models to identify the 94 target trips.  For each of the 94 3-day window consisting of a potential accident triggers around the date of loss, we assume the processed trip sub-intervals are generated from the same model and a CTHMM is fitted on them.  The model structure is the same as that employed in the controlled study: the CTHMM has 20 states, with state-dependent distributions Gamma for speed and Normal for both acceleration components.  Although a CTHMM must be fitted for each rental, the computation time remains manageable (maximum of 10 minutes each using Julia) as we have restricted to 3-day windows.  Moreover, the process is highly scalable as CTHMMs can be fitted in parallel across rentals.

\begin{figure}[hp]
    \centering
    \begin{subfigure}[b]{0.45\textwidth}
        \centering
        \includegraphics[width=\textwidth]{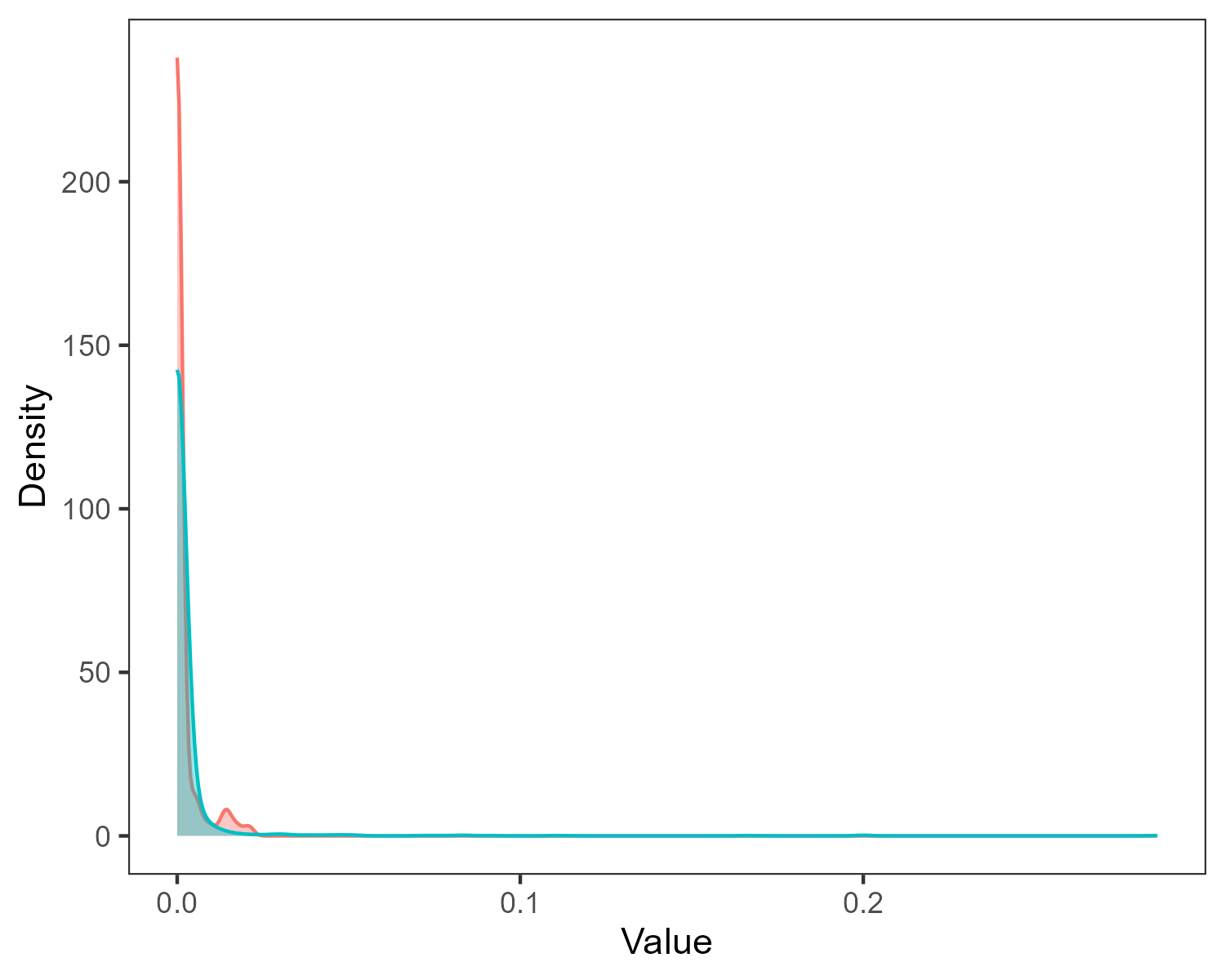}
        \caption{Raw index for speed}
        \label{fig:plot1}
    \end{subfigure}
    \begin{subfigure}[b]{0.45\textwidth}
        \centering
        \includegraphics[width=\textwidth]{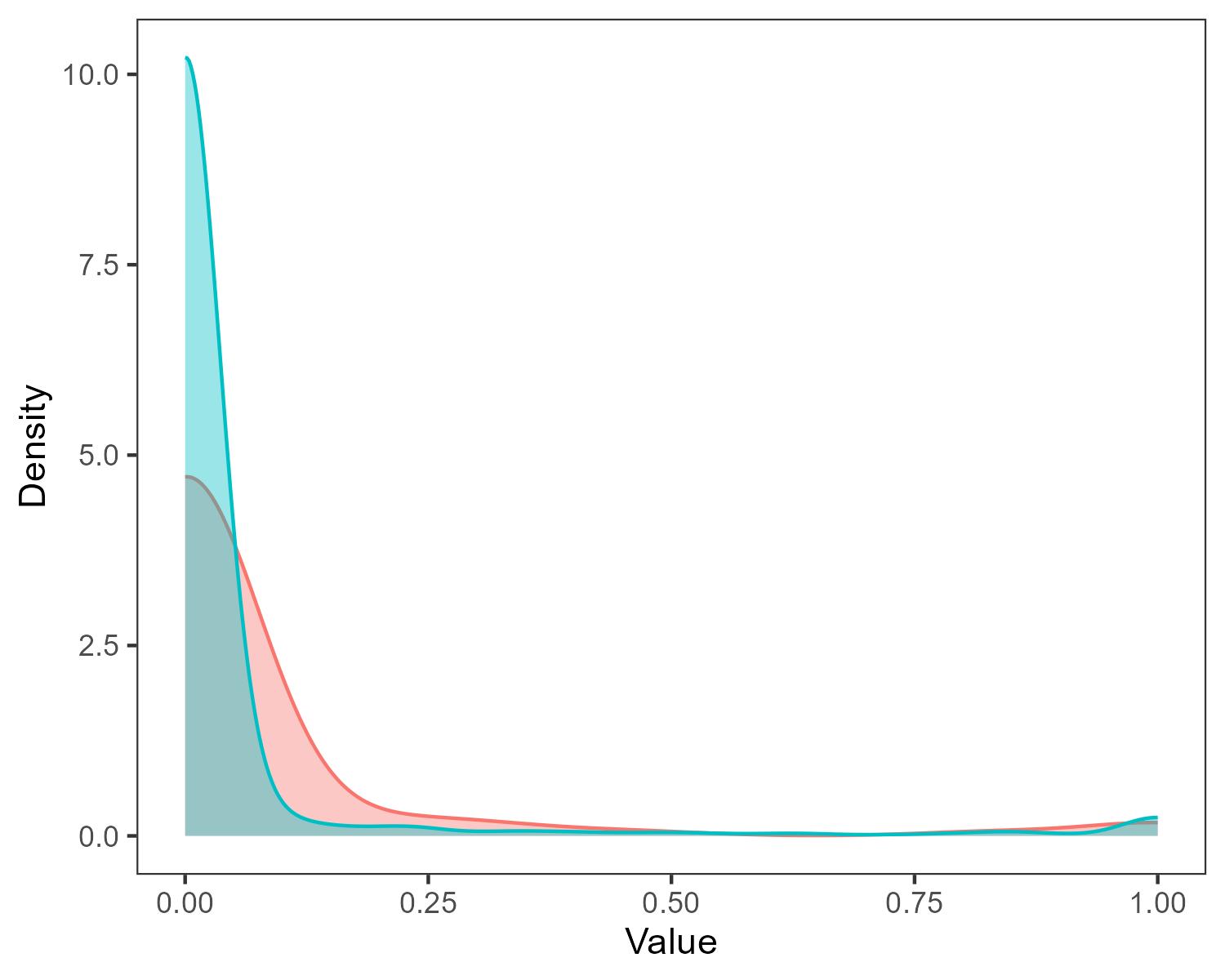}
        \caption{Normalized index for speed}
        \label{fig:plot2}
    \end{subfigure}
    \begin{subfigure}[b]{0.45\textwidth}
        \centering
        \includegraphics[width=\textwidth]{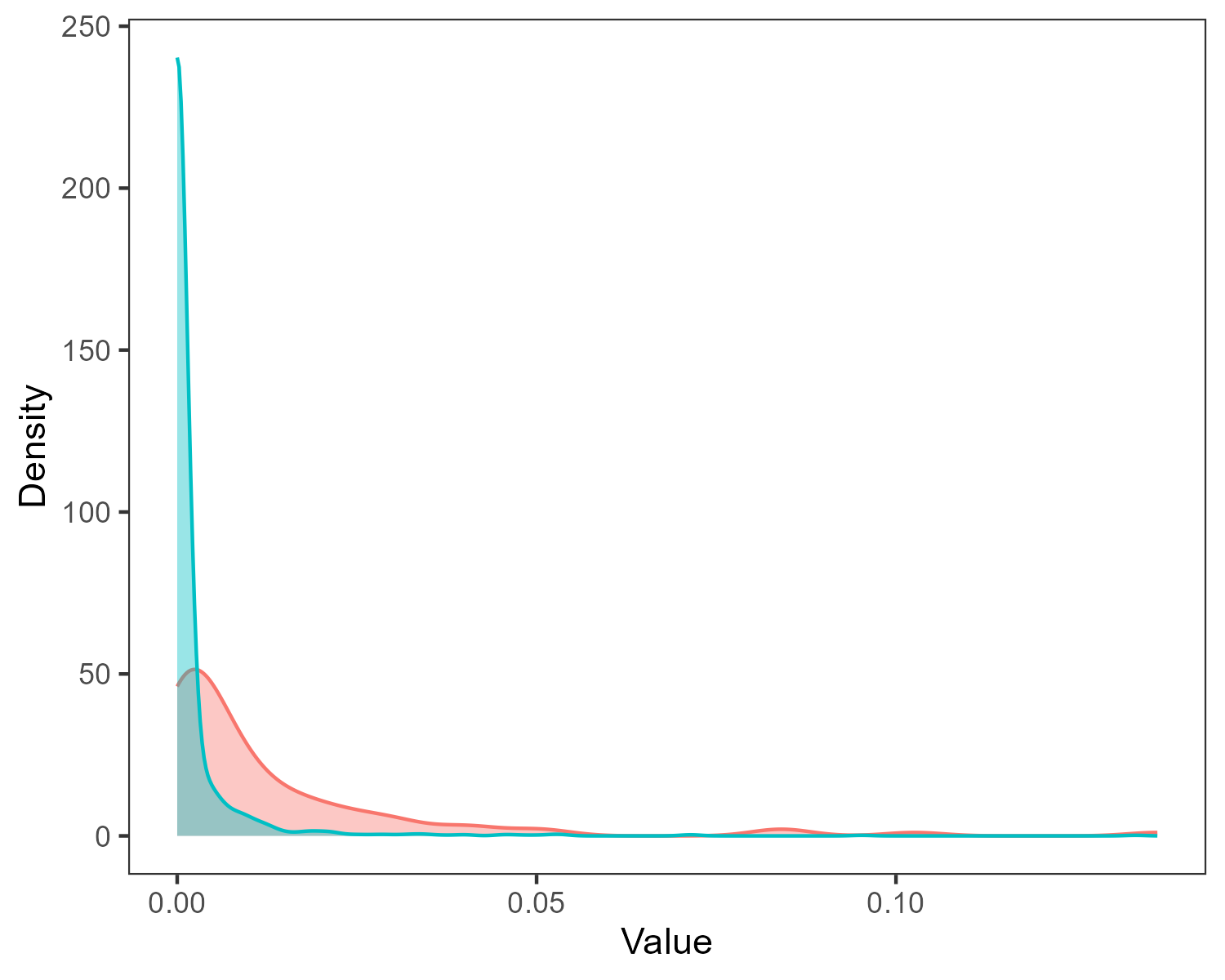}
        \caption{Raw index for longitudinal acceleration}
        \label{fig:plot3}
    \end{subfigure}
    \begin{subfigure}[b]{0.45\textwidth}
        \centering
        \includegraphics[width=\textwidth]{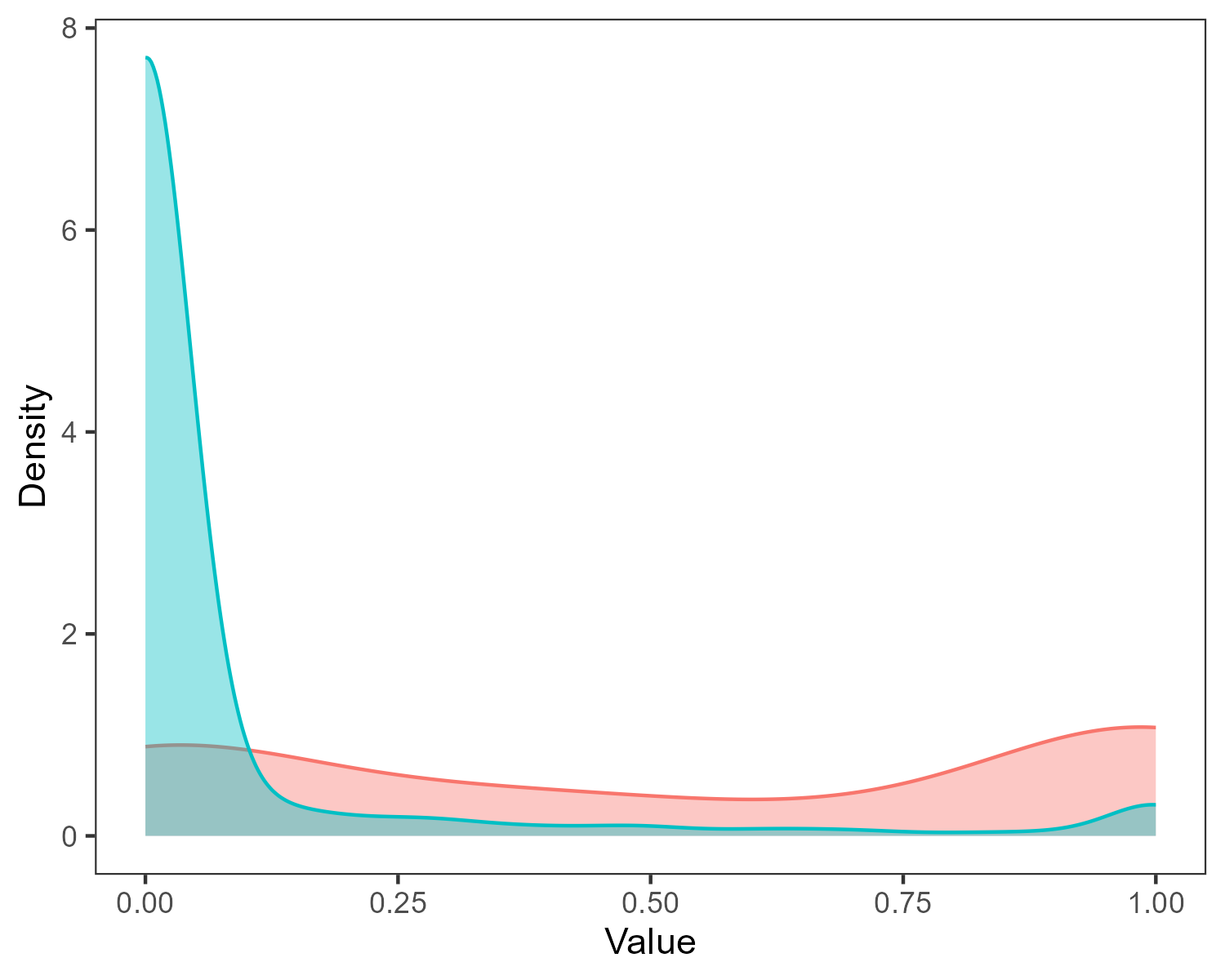}
        \caption{Normalized index for longitudinal acceleration}
        \label{fig:plot4}
    \end{subfigure}
    \begin{subfigure}[b]{0.45\textwidth}
        \centering
        \includegraphics[width=\textwidth]{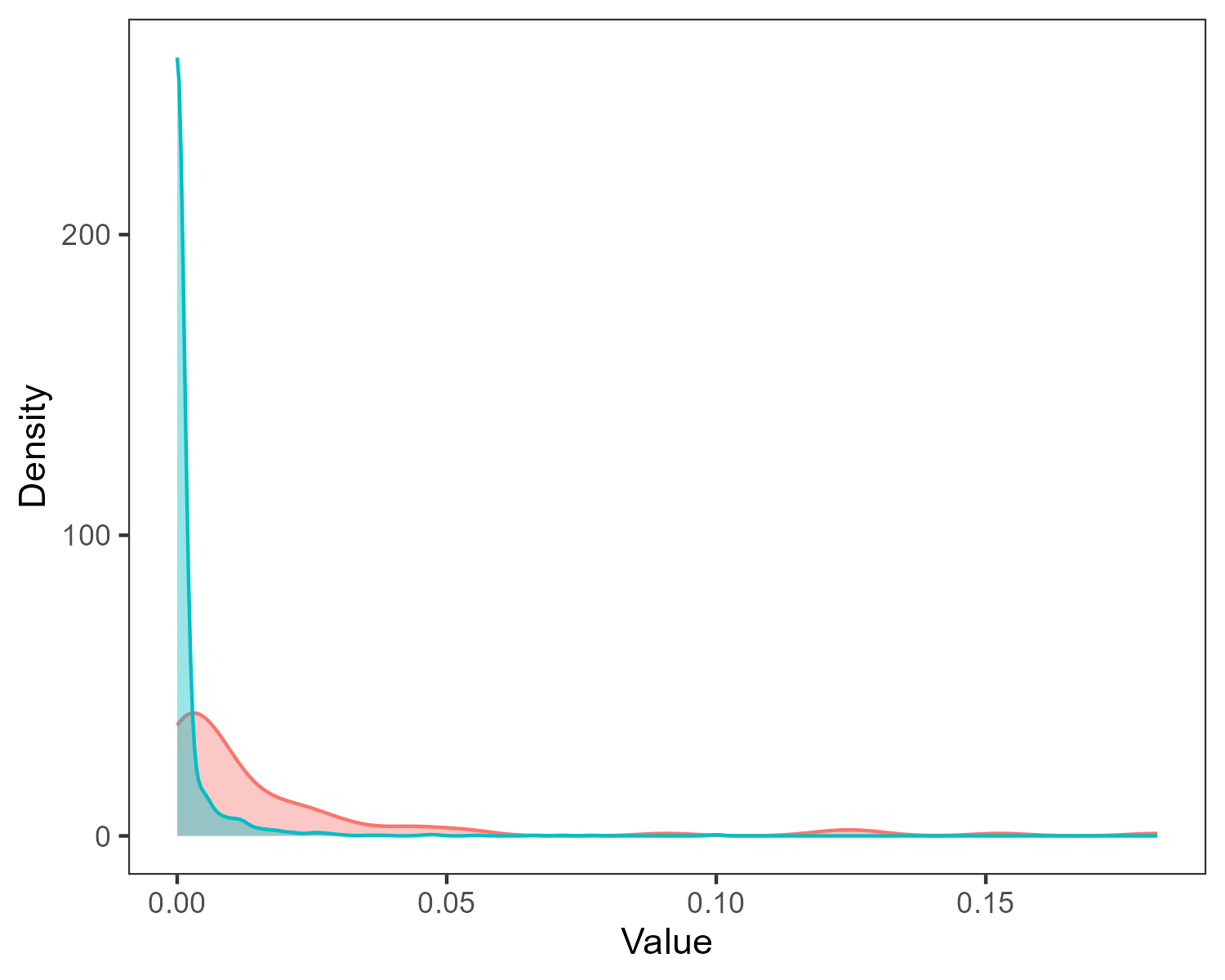}
        \caption{Raw index for lateral acceleration}
        \label{fig:plot5}
    \end{subfigure}
    \begin{subfigure}[b]{0.45\textwidth}
        \centering
        \includegraphics[width=\textwidth]{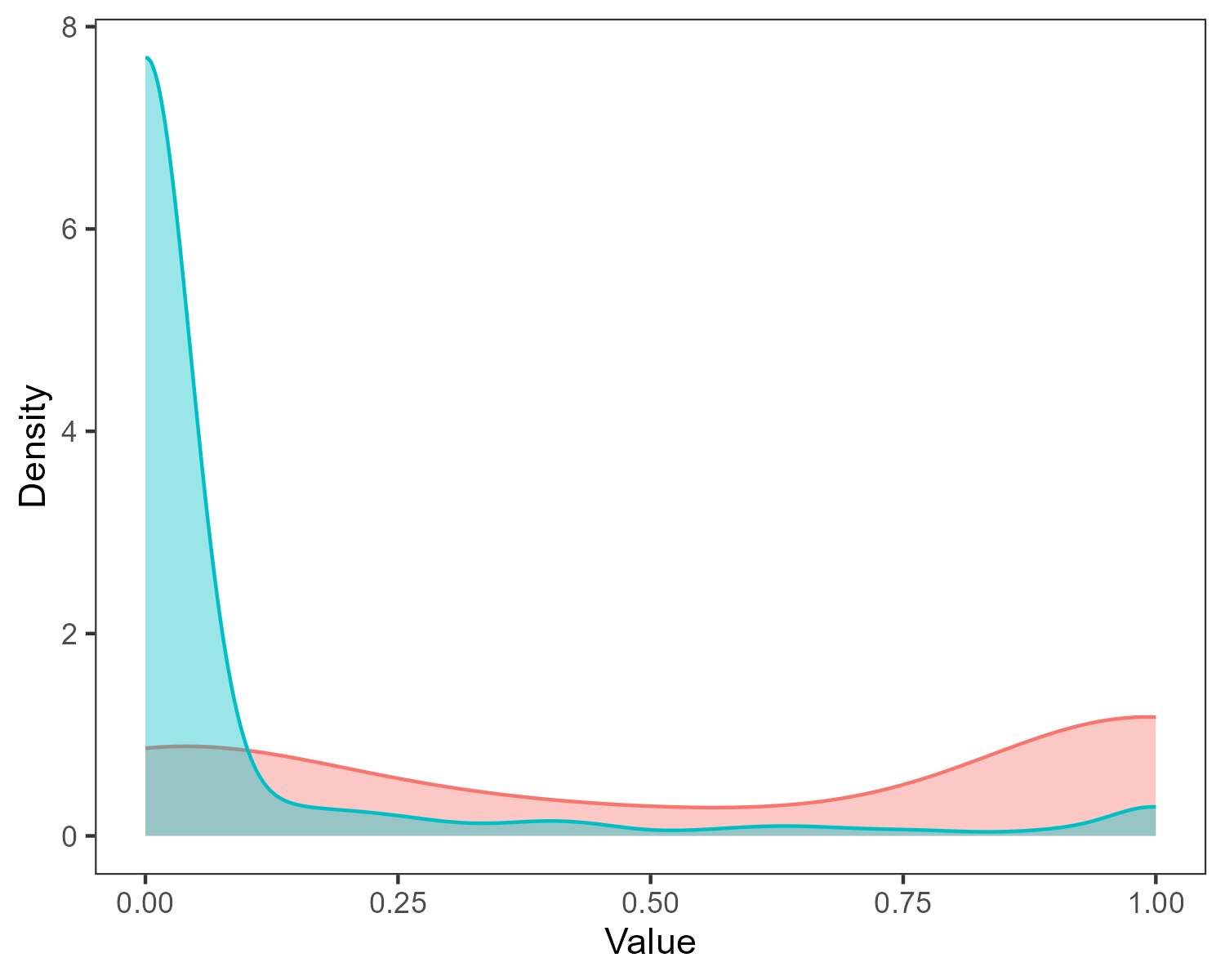}
        \caption{Normalized index for lateral acceleration}
        \label{fig:plot6}
    \end{subfigure}
    \caption{Empirical densities of target trips and all other trips for their anomaly indices.  The left column shows the raw indices, while the right column shows the normalized indices.  Red: target trips, and blue: all other trips.}
    \label{fig: individualized-emp-densities}
\end{figure}

After model fitting, we compute the anomaly indices of each trip under its corresponding model.  In Table \ref{tab: Sum-Stats-Prop-Residuals-20}, we present summary statistics of these indices for the 94 target trips and all other trips.  We first observe that the target trips exhibit a slightly lower average anomaly index in speed compared to the other trips; however, this difference appears minimal as indicated by the other quartiles.  As we learnt from the controlled study, speed alone does not fully capture driving behaviour and the analysis should focus on the accelerations.  The summary statistics support this, revealing that the target trips have significantly higher anomaly indices in both acceleration components, with averages nearly 10 times higher.  To further illustrate these differences, the left column of Figure \ref{fig: individualized-emp-densities} presents the empirical density plots of the anomaly indices for the two groups, with target trips shown in red and all other trips in blue.  These plots help visualize the different distributional characteristics between the two groups, particularly in the acceleration components.  Additionally, as we would like to rank the trips in each 3-day window by their level of anomalousness and ultimately identify the target trip, we can further consider the normalized anomaly indices given in Equation \ref{eqn: normalized-anomaly-index}, which is the ratio of each trip's indices to the maximum indices within that window.

We present the summary statistics of these normalized indices in Table \ref{tab: Sum-Stats-Ratio-Residuals-20}.  The target trips have higher indices across all three dimensions compared to the other trips, with the differences much more pronounced for both acceleration components.  This suggests that the target trips contain more outliers among the trips completed by the same driver, which will facilitate our identification of the target trips.  It is important to note that we did not specify the target trips during model training nor pseudo-residuals computation, yet we observe distinct differences in anomaly indices and their normalized versions between the target and all other trips.  The right column of Figure \ref{fig: individualized-emp-densities} displays the empirical density plots of the normalized anomaly indices for the two groups, with target trips represented in red and all other trips in blue.

\begin{table}[h]
    \caption{Summary statistics of anomaly indices (\%), computed from CTHMMs with 20 states}
    \centering
    \begin{tabular}{c|cc|cc|cc}
         &  \multicolumn{2}{c|}{\textbf{Speed}}&  \multicolumn{2}{c|}{\textbf{Longitudinal Acceleration}}&  \multicolumn{2}{c}{\textbf{Lateral Acceleration}}\\
         &  Target&  Other&  Target&  Other&  Target&  Other\\\hline
         Min&  0&  0&  0&  0&  0&  0\\
         1st Qu.&  0&  0&  0&  0&  0&  0\\
         Median&  0&  0&  0.5333&  0&  0.6944&  0\\
         Mean&  0.1286&  0.1749&  1.3403&  0.1621&  1.8249&  0.1596\\
         3rd Qu.&  0&  0&  1.4983&  0&  1.9423&  0\\
         Max&  2.0833&  28.5714&  13.6364&  13.3333&  18.1818&  10.0000\\\hline
    \end{tabular}
    \label{tab: Sum-Stats-Prop-Residuals-20}
\end{table}

\begin{table}[h]
    \caption{Summary statistics of normalized anomaly indices, computed from CTHMMs with 20 states}
    \centering
    \begin{tabular}{c|cc|cc|cc}
         &  \multicolumn{2}{c|}{\textbf{Speed}}&  \multicolumn{2}{c|}{\textbf{Longitudinal Acceleration}}&  \multicolumn{2}{c}{\textbf{Lateral Acceleration}}\\
         &  Target&  Other&  Target&  Other&  Target&  Other\\\hline
         Min&  0&  0&  0&  0&  0&  0\\
         1st Qu.&  0&  0&  0&  0&  0&  0\\
         Median&  0&  0&  0.5100&  0&  0.6087&  0\\
         Mean&  0.0622&  0.0424&  0.5310&  0.0708&  0.5574&  0.0720\\
         3rd Qu.&  0&  0&  1&  0&  1&  0\\
         Max&  1&  1&  1&  1&  1&  1\\\hline
    \end{tabular}
    \label{tab: Sum-Stats-Ratio-Residuals-20}
\end{table}

With the three raw or normalized anomaly indices as covariates, we consider three models to predict the probability that a trip is actually a target trip.  The candidates are logistic regressions with and without cubic splines (i.e., Generalized Linear Model (GLM) and Generalized Additive Model (GAM), respectively) and XGBoost (eXtreme Gradient Boosting).  We implement these models with standard R packages \texttt{stats}, \texttt{mgcv} and \texttt{xgboost}, respectively.  For GAM, we specify a cubic regression spline (\texttt{bs = `cr'}), while for XGBoost, we use the package's default parameter settings, adjusting only the number of iteration rounds to 20.  This choice ensures reproducibility and consistency, as training log loss stabilizes around 0.001 at this point, indicating sufficient convergence without overfitting.  While further optimization of the XGBoost model is possible, our primary focus is on demonstrating the usefulness of the proposed anomaly index in identifying accident-related trips.

We then evaluate model performance based on ROC-AUC (the area under the receiver operating characteristic curve).  ROC-AUC is a performance metric frequently used to evaluate the effectiveness for a binary classification model, which quantifies the model's ability to distinguish between the two classes.  A perfect model has an AUC of 1, while a model that randomly guesses has an AUC of 0.5.  To robustify the performance evaluation, we perform 5-fold cross-validations (CV) for each set of covariates and model.  In a $k$-fold cross validation, the dataset is split into $k$ subsets.  In each fold, the model is trained on $k-1$ subsets, and the trained model is tested on the remaining subset.  The $k$-fold CV then reports the average of the $k$ results.  Our dataset is randomly partitioned into 5 subsets such that: first, all trips from a window are in the same subset; and second, the proportion of target trips are close (0.05234, 0.05263, 0.05882, 0.05740 and 0.04651).

\begin{table}[h]
    \caption{5-fold cross-validation ROC-AUC of logistic models with raw or normalized anomaly indices as covariates.}
    \centering
    \begin{tabular}{c|cc}
         &  \textbf{Raw}&  \textbf{Normalized}\\\hline
         GLM&  0.849&  0.863\\
         GAM&  0.857&  0.853\\
         XGBoost&  0.825&  0.826\\\hline
    \end{tabular}
    \label{tab: AUC-20}
\end{table}

\begin{table}[h]
    \caption{5-fold cross-validation window-based ROC-AUC of logistic models with raw or normalized anomaly indices as covariates.}
    \centering
    \begin{tabular}{c|ccc|ccc}
     & \multicolumn{3}{c|}{\textbf{Raw}} & \multicolumn{3}{c}{\textbf{Normalized}} \\
     & Mean & Median & Accuracy & Mean & Median & Accuracy \\ \hline
    GLM & 0.849 & 0.957 & 0.436 & 0.856 & 0.966 & 0.499 \\
    GAM & 0.855 & 0.978 & 0.488 & 0.857 & 0.973 & 0.532 \\
    XGBoost & 0.844 & 0.952 & 0.445 & 0.854 & 0.962 & 0.531\\\hline
    \end{tabular}
    \label{tab: AUC-window-20}
\end{table}

In Table \ref{tab: AUC-20}, we report the cross-validation performance of the candidate models with raw or normalized anomaly indices as covariates.  We observe that the average ROC-AUCs are all above 0.82, with the ones using normalized indices as covariates performing slightly better for GLM and XGBoost.

Recall that in practice, the goal is to rank the trips in each 3-day window based on their level of anomalousness, ideally identifying the target trip.  To assess the effectiveness of this ranking, we evaluate the ROC-AUC for each 3-day window.  In Table \ref{tab: AUC-window-20}, we present the mean and median window-based ROC-AUC from cross-validation.  Furthermore, we report the average accuracy, which we define as the proportion of windows where the ROC-AUC is 1 --- that is, when the target trip has the highest predicted probability among trips within the same window and is considered the most anomalous.  The results show that models using normalized indices as covariates again perform slightly better on average.  Although the average accuracy ranges only between 0.43 and 0.53, achieving a classifier with an AUC of 1 is inherently challenging.  This difficulty is amplified in our context, where identifying the target trip requires selecting it from a group that can range from 2 to 63 trips.  Nevertheless, it is noteworthy that more than half of the cases achieve perfect classification.  Moreover, the relatively high (average) population ROC-AUCs (above 0.82), along with the mean and median window-based ROC-AUCs (above 0.84 and 0.95, respectively), all indicate that the models effectively rank trips by their anomalousness, both overall and within each window, with the target trips often ranked near or at the top.

\begin{table}[h]
\caption{Model summary of logistic GAMs, with raw or normalized anomaly indices as covariates.  Note: x1, x2 and x3 denote indices for speed, longitudinal and lateral accelerations respectively.}
\centering
\begin{threeparttable}
\begin{subtable}{1\textwidth}
    \centering
    \begin{tabular}{c|rrrrl}
         &  \textbf{Estimate}&  \textbf{Std. Error}&  \textbf{z value}&  \textbf{p value}&  \\\hline
         (Intercept)&  -4.054&  0.746&  -5.437&  5.41e-08&   ***\\\\
         &  \textbf{edf}&  \textbf{Ref.df}&  \textbf{Chi.sq}&  \textbf{p value}&  \\\hline
         s(x1)&  3.124&  3.722&  3.098&  0.586&  \\
         s(x2)&  4.858&  5.406&  71.446&  <2e-16&  ***\\
         s(x3)& 6.944& 7.612& 86.267& <2e-16&***\\\hline\hline
    \end{tabular}
\end{subtable}

\bigskip

\begin{subtable}{1\textwidth}
    \centering
    \begin{tabular}{c|rrrrl}
         &  \textbf{Estimate}&  \textbf{Std. Error}&  \textbf{z value}&  \textbf{p value}&  \\\hline
         (Intercept)&  -3.829&  0.184&  -20.820&  <2e-16&   ***\\\\
         &  \textbf{edf}&  \textbf{Ref.df}&  \textbf{Chi.sq}&  \textbf{p value}&  \\\hline
         s(norm x1)&  1.002&  1.004&  0.363& 0.549& \\
         s(norm x2)&  2.721&  3.252&  82.759& <2e-16& ***\\
         s(norm x3)&  2.960&  3.530&  94.035& <2e-16& ***\\\hline\hline
    \end{tabular}
\begin{tablenotes}
\small
\item Significance codes: 0 `***', 0.001 `**', 0.01 `*', 0.05 `.', 0.1 ` ', 1
\end{tablenotes}
\end{subtable}
\end{threeparttable}
\label{tab: GAM-coefficients}
\end{table}

Finally, we fit the models to the whole dataset, and report the model summary of GAM in Table \ref{tab: GAM-coefficients}, while model summary of GLM and feature importance of XGBoost are provided in Appendix C.  The findings show that outliers in both longitudinal and lateral accelerations are more significant predictors of target trips than outliers in speed, which is in line with observations from the controlled study.

Our proposed CTHMM modelling and anomaly detection framework hold great potential for practical applications, particularly for companies aiming to enhance safety and operational efficiency.  For instance, it can be employed to detect anomalous driving behaviors, assess risk factors, and provide actionable feedback to improve driver safety.

In our rental-car dataset, recall that our collaborator's collision detection method produces accident triggers for only around 10\% of claims.  By applying the proposed method, we can effectively pinpoint target trips for investigation in the remaining 90\% of cases.  Moreover, leveraging the analysis of pseudo-residual points, as detailed in Section \ref{Controlled-Study-Individual-Specific-Model}, allows us to locate each outlier and delve deeper into its nature.  This facilitates the identification of underlying causes, such as harsh braking or cornering events relative to each driver's driving patterns, and the provision of individualized feedback to mitigate such actions and improve driving practices.

\subsection{Pooled Model}
\label{Real-Data-Analysis-Pooled-Model}

Lastly, we apply the pooled model to compare trips across individuals, and investigate differences in driving behaviour between drivers with and without (at-fault) claims.  While we can fit a pooled model using all 35 trips in the controlled study, the sheer number of trips in real-world settings makes it infeasible to train the model on each and every trip.  Hence, sampling of trips is necessary for practical model training.

To ensure the robustness of our results, we experiment with multiple training sets.  As the claim rate in the dataset is approximately 10\%, we decide to create training sets with random stratified samples reflecting this proportion.  Each training set consists of 100 rentals, including 10 rentals with at-fault claims and 90 rentals without claims.  For the claim rentals, we consider trips within the 3-day window around the date of loss, while for the no-claim rentals, we consider the 3 days immediately before the rental ends.  Additionally, to test the extreme cases, we construct two alternative training sets: one with 0\% claim rate (100 no-claim rentals) and one with 100\% claim rate (94 rentals with at-fault claims, discussed in Section \ref{Real-Data-Analysis-Individual-Specific-Model}), using the same respective 3-day windows.

For our primary analysis, we focus on the 10\% claim rate sample, as it aligns with the portfolio average.  We train a single 20-state CTHMM with state-dependent distributions Gamma for speed and Normal for both acceleration components.  Training pooled models, such as for 100 rentals, is more computationally intensive than training 100 individual-specific models.  This is because each iteration of the EM algorithm must process all trips from the 100 rentals.  Based on our experience, fitting such a pooled model takes about a day to complete 5,000 iterations of the EM algorithm.  However, once the model is fitted, it can evaluate both existing and newly observed trips.

Using the fitted pooled model, we determine the anomaly indices of \textbf{all} trips from \textbf{all} rentals over their \textbf{entire} rental periods, rather than just a 3-day window.  This allows us to examine any differences in driving behaviour, focusing particularly on differences in driving anomalies, between rentals with and without (at-fault) claims.  The analysis is repeated with four different random training sets, each with a 10\% claim rate, and results are reported in Table \ref{tab: Sum-Stats-pooled-hmm-10}.

\setcounter{table}{12}
\begin{table}[h!]
\caption{Comparison of anomaly indices (\%) in each telematics response dimension for claimed and no-claim rentals, computed from pooled CTHMMs trained on training sets with a 10\% claim rate.}
\centering
\centerline{
\begin{threeparttable}
    \begin{tabular}{l|rr|rr|rr|rr}
    \multicolumn{9}{c}{\textbf{Trip-basis}} \\\hline
     & \multicolumn{2}{c|}{\textbf{Training Set 1}} & \multicolumn{2}{c|}{\textbf{Training Set 2}} & \multicolumn{2}{c|}{\textbf{Training Set 3}} & \multicolumn{2}{c}{\textbf{Training Set 4}} \\
     & claimed & no claim & claimed & no claim & claimed & no claim & claimed & no claim \\\hline
    speed & 0.0466 & 0.0613 & 0.0722 & 0.0984 & 0.0642 & 0.0940 & 0.4480 & 0.4582 \\
    longitudinal & 0.4391 & 0.5303 & 0.4949 & 0.6037 & 0.5071 & 0.6189 & 0.8317 & 0.9802 \\
    lateral & 0.2714 & 0.3033 & 0.3241 & 0.3712 & 0.3319 & 0.3832 & 0.3180 & 0.3635 \\\hline
    \end{tabular}
    
    \bigskip
    
    \begin{tabular}{l|rr|rr|rr|rr}
    \multicolumn{9}{c}{\textbf{Rental-Basis}} \\\hline
     & \multicolumn{2}{c|}{\textbf{Training Set 1}} & \multicolumn{2}{c|}{\textbf{Training Set 2}} & \multicolumn{2}{c|}{\textbf{Training Set 3}} & \multicolumn{2}{c}{\textbf{Training Set 4}} \\
     & claimed & no claim & claimed & no claim & claimed & no claim & claimed & no claim \\\hline
    \textbf{Average} &  &  &  &  &  &  &  &  \\
    speed & 0.0500 & 0.0721 & 0.0770 & 0.1048 & 0.0634 & 0.0988 & 0.6035 & 0.9938 \\
    longitudinal & 0.5136 & 0.7373 & 0.5803 & 0.8254 & 0.5841 & 0.8485 & 0.9253 & 1.3156 \\
    lateral & 0.2807 & 0.3459 & 0.3429 & 0.4091 & 0.3455 & 0.4179 & 0.3293 & 0.3847 \\\hline
    \textbf{Maximum} &  &  &  &  &  &  &  &  \\
    speed & 25.5600 & 13.3159 & 31.4940 & 17.0440 & 30.3340 & 16.8740 & 46.6200 & 26.2900 \\
    longitudinal & 30.0000 & 17.2080 & 34.8800 & 20.3670 & 34.2000 & 20.3170 & 36.0900 & 21.5540 \\
    lateral & 27.8200 & 15.2737 & 33.0760 & 18.6550 & 32.1820 & 18.5560 & 32.9830 & 18.2270 \\\hline
    \end{tabular}
\end{threeparttable}
}
\label{tab: Sum-Stats-pooled-hmm-10}
\end{table}

The first question we address is whether there are notable differences in anomaly indices for individual trips performed by the claimed and no-claim groups.  In the sub-table titled `trip-basis', we calculate the average indices across all trips completed by each group.  We find that the differences are minimal.  In fact, the claimed group shows slightly lower average anomaly indices than the no-claim group, which may seem counterintuitive.  A plausible explanation is that the model was trained on only 300 days of trip data, which may not capture all major driving behaviour.  In contrast, the evaluation spans entire rental periods of over 17,000 rentals, providing a broader context that may dilute the trip-specific differences.

\begin{figure}[hp]
    \centering
    \begin{subfigure}[b]{0.45\textwidth}
        \centering
        \includegraphics[width=\textwidth]{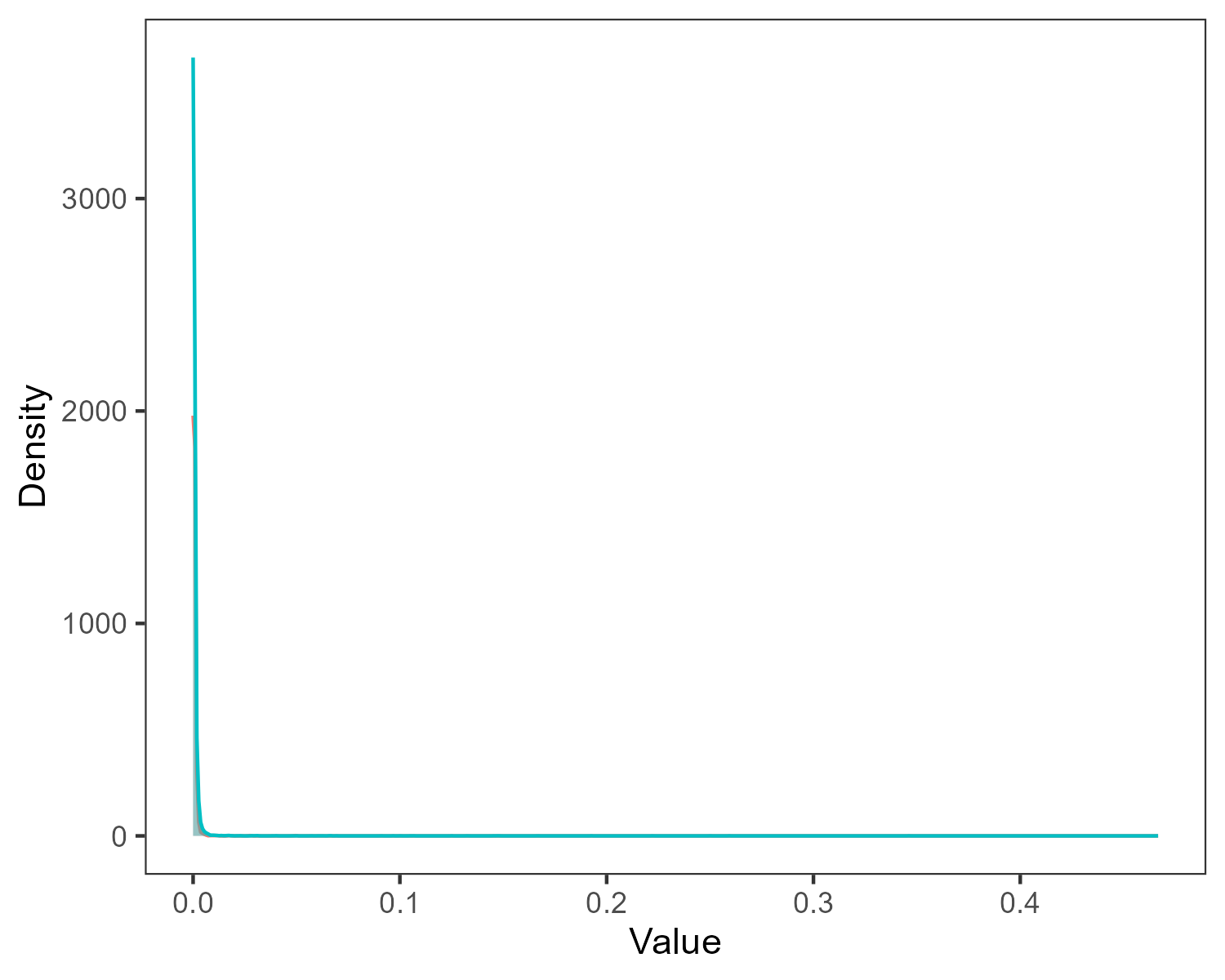}
        \caption{Mean index for speed}
        \label{fig:plot1}
    \end{subfigure}
    \begin{subfigure}[b]{0.45\textwidth}
        \centering
        \includegraphics[width=\textwidth]{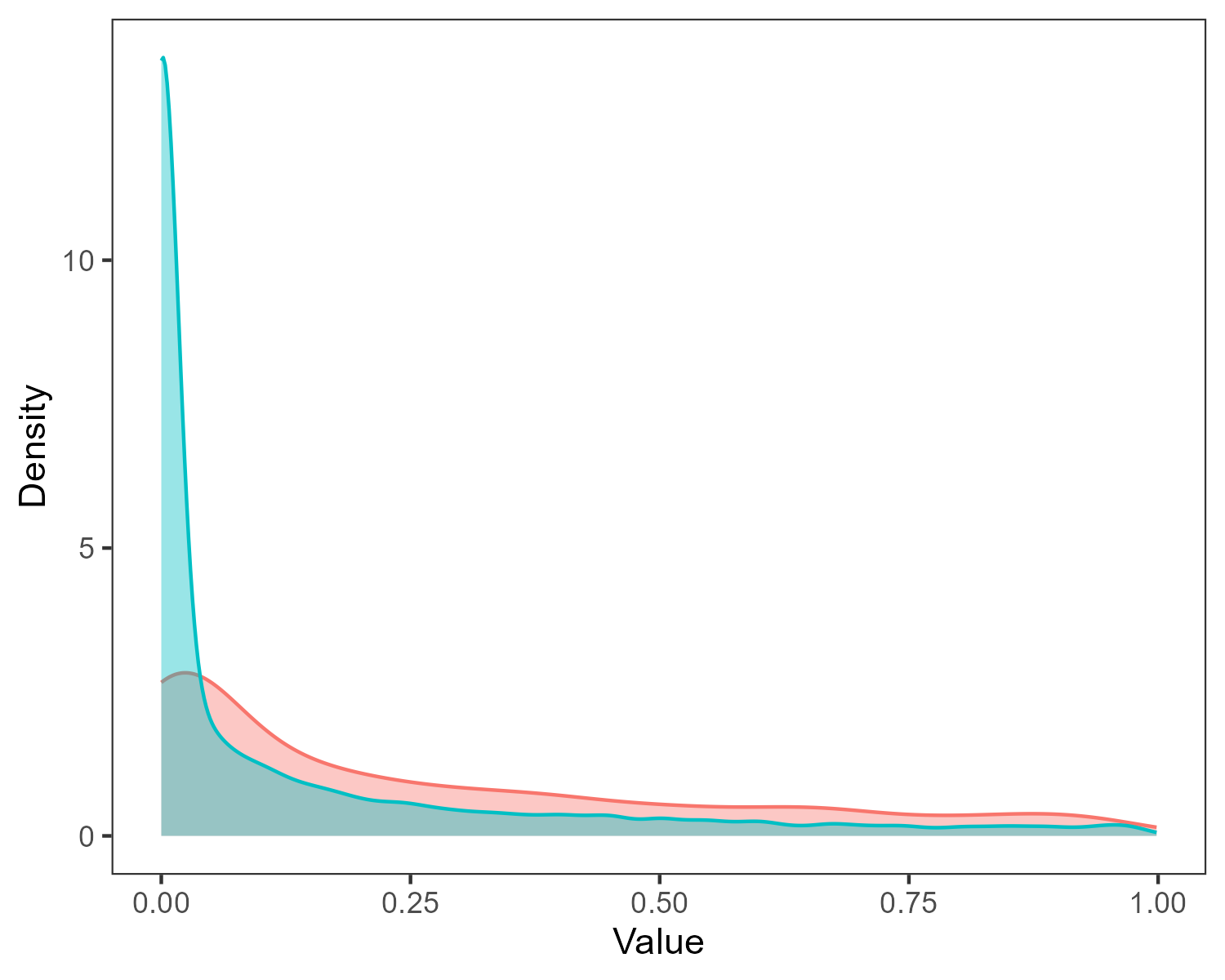}
        \caption{Max index for speed}
        \label{fig:plot2}
    \end{subfigure}
    \begin{subfigure}[b]{0.45\textwidth}
        \centering
        \includegraphics[width=\textwidth]{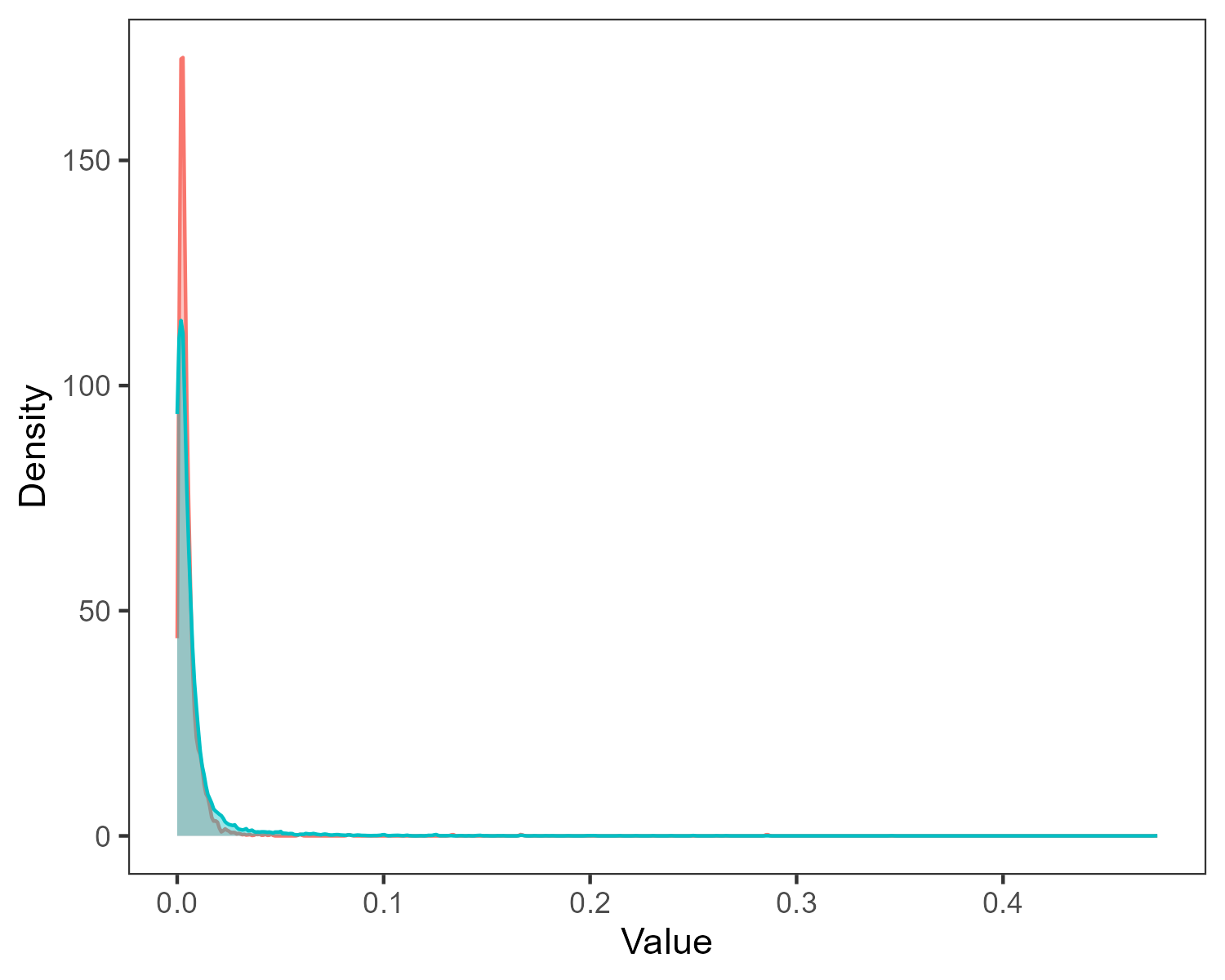}
        \caption{Mean index for longitudinal acceleration}
        \label{fig:plot3}
    \end{subfigure}
    \begin{subfigure}[b]{0.45\textwidth}
        \centering
        \includegraphics[width=\textwidth]{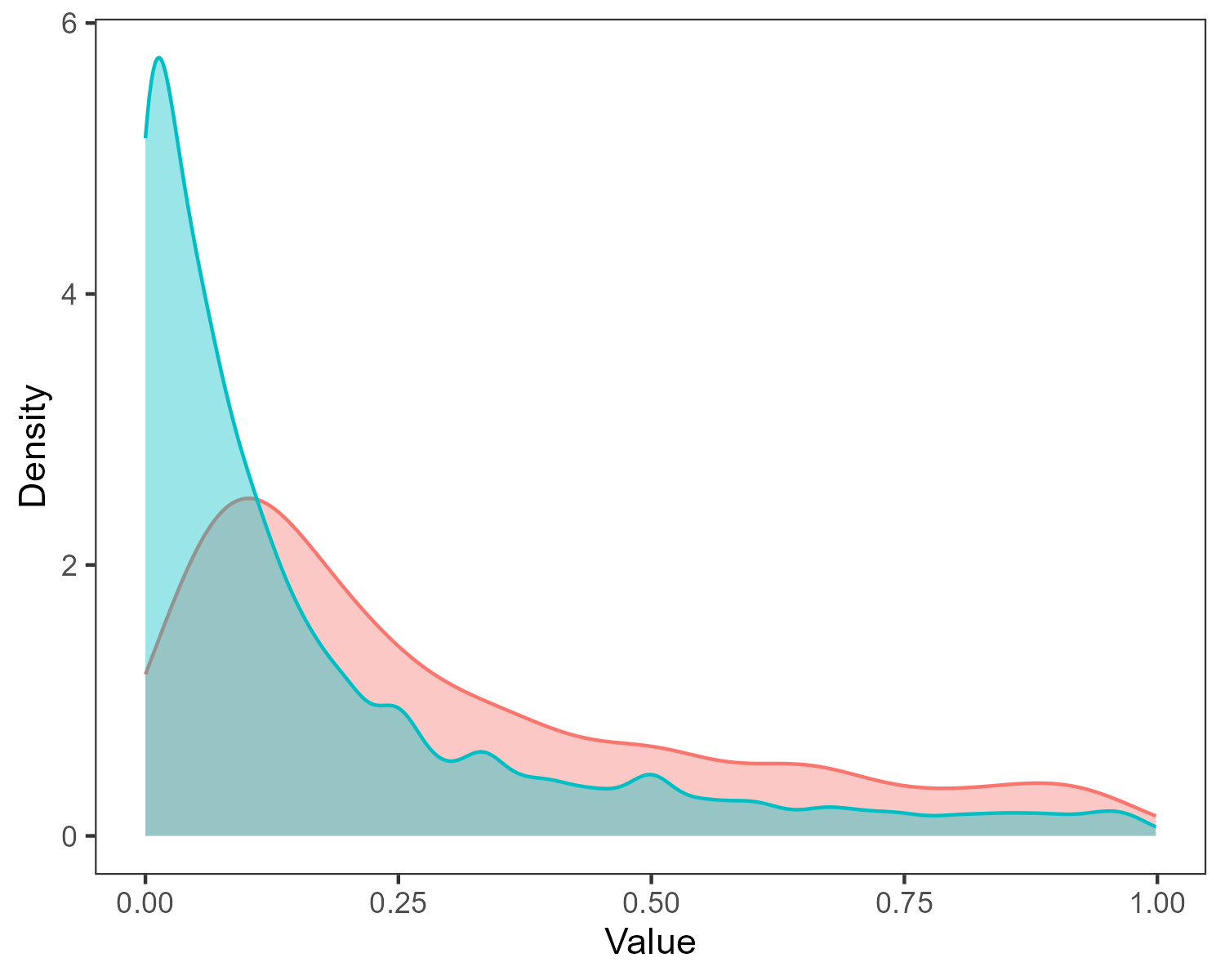}
        \caption{Max index for longitudinal acceleration}
        \label{fig:plot4}
    \end{subfigure}
    \begin{subfigure}[b]{0.45\textwidth}
        \centering
        \includegraphics[width=\textwidth]{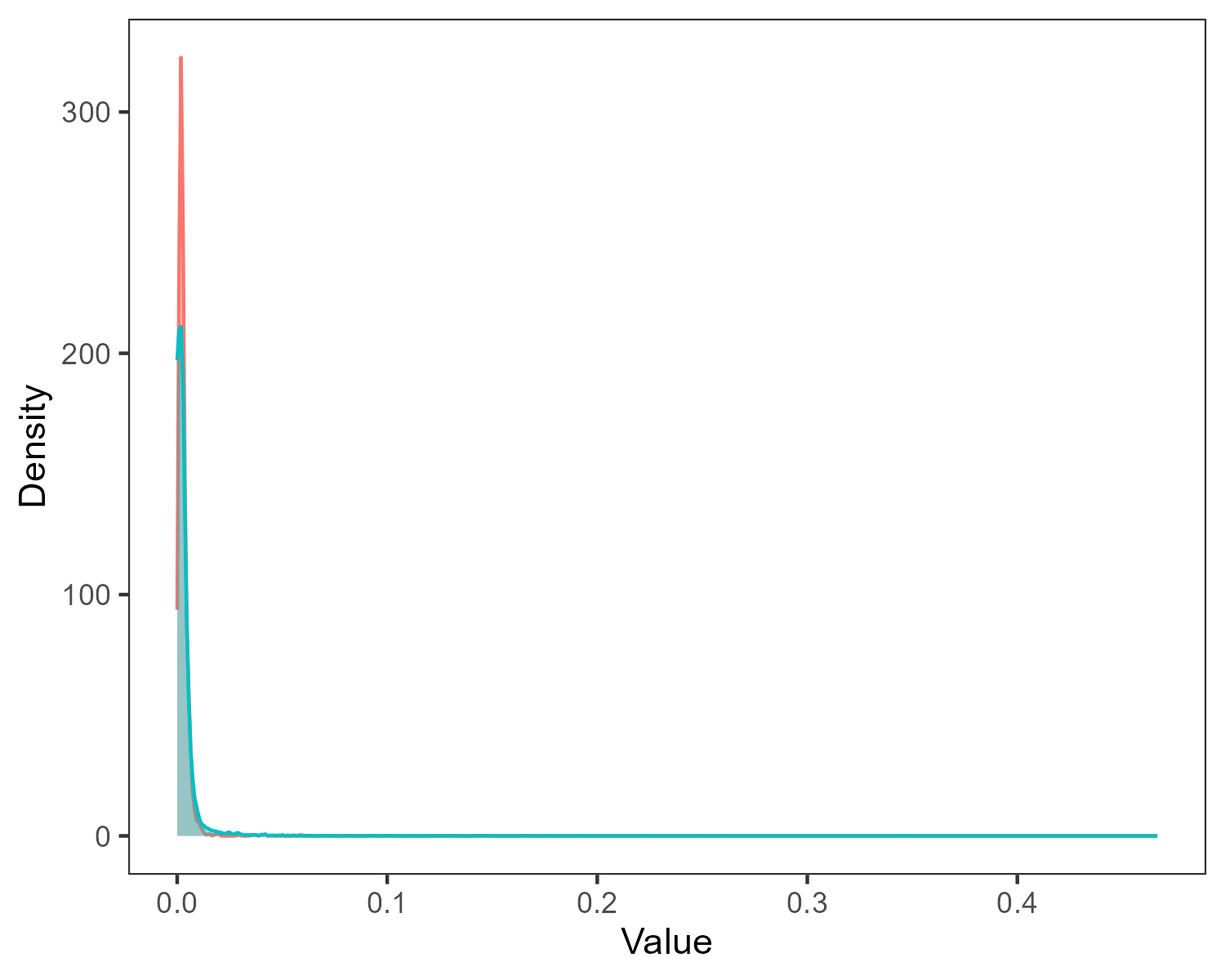}
        \caption{Mean index for lateral acceleration}
        \label{fig:plot5}
    \end{subfigure}
    \begin{subfigure}[b]{0.45\textwidth}
        \centering
        \includegraphics[width=\textwidth]{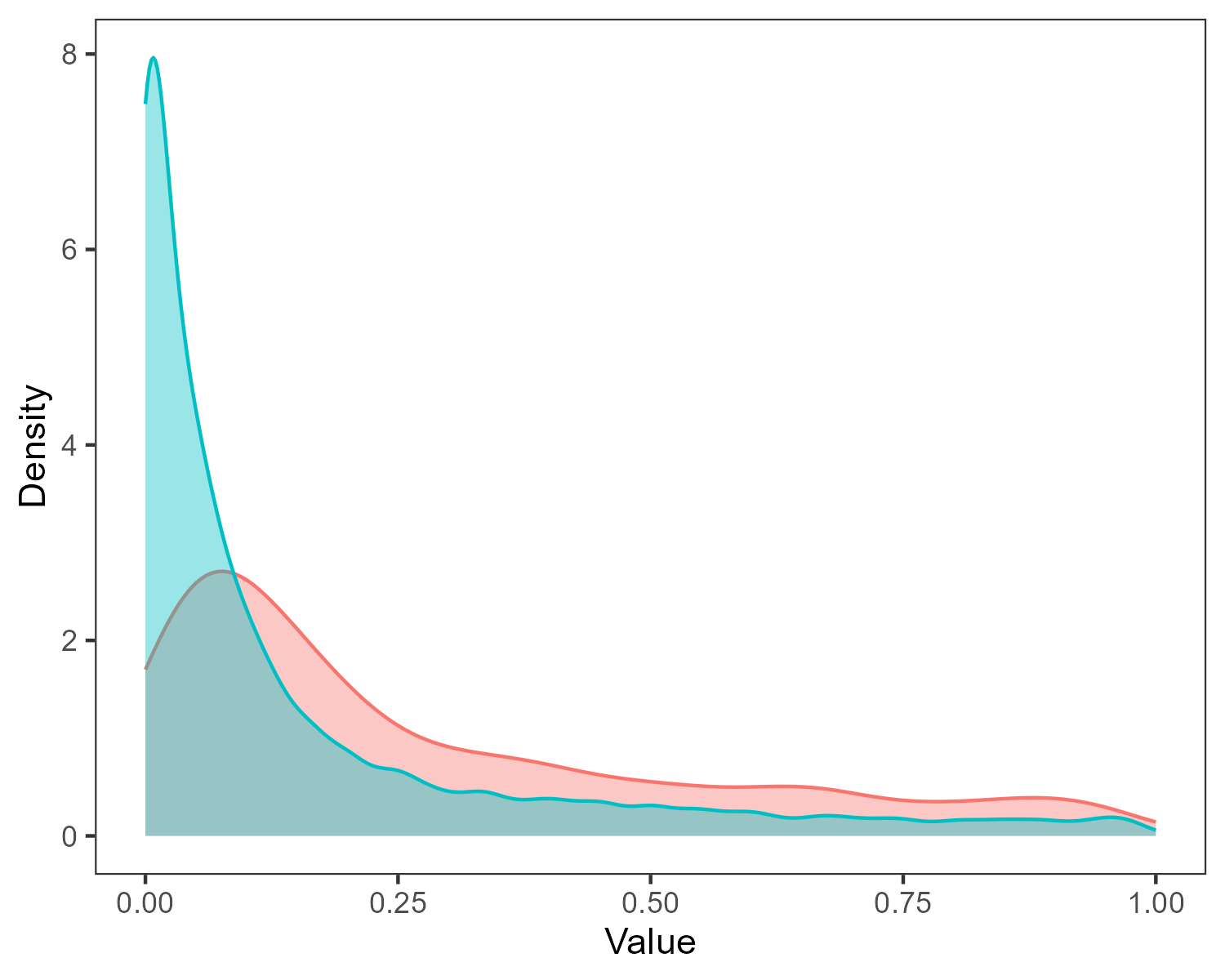}
        \caption{Max index for lateral acceleration}
        \label{fig:plot6}
    \end{subfigure}
    \caption{Empirical densities of claimed and no-claim groups for their anomaly indices on a rental basis.  The left column shows the mean indices over rentals, while the right column shows the maximum over rentals.  Red: with claims, and blue: without claims.}
    \label{fig: pooled-emp-densities}
\end{figure}

Next, we investigate whether anomaly indices differ when aggregated over entire rental periods for the claimed and no-claim groups.  This analysis examines the empirical distribution of anomaly indices for each rental.  In the sub-table titled `rental-basis', we present the averages of two metrics for each group: the rental-wise average indices and the rental-wise maximum indices.  Specifically, we calculate the average of the rental averages and the average of the rental maximums for each group.  While the average indices between the groups remain similar, the claimed group exhibits significantly higher maximums.  On average, the maximum indices for the claimed group are roughly double those of the no-claim group across all response dimensions.  To help visualize these differences, Figure \ref{fig: pooled-emp-densities} presents the empirical density plots of the anomaly indices for the claimed and no-claim groups on a rental basis.  The left column displays the rental averages, while the right column illustrates the rental maximums, with the claimed group represented in red and the no-claim group in blue.

While it may seem counterintuitive that the claimed group has slightly lower average anomaly indices per trip, it is important to recognize that the no-claim group is ten times larger than the claimed group, leading to greater variability in their driving behaviour.  This highlights the importance of trip-based risk evaluation, as not every trip by the claimed group is dangerous, and similarly, not all trips by the no-claim group are safe.  Over the entire rental period, however, the claimed group consistently exhibits more outliers, with this difference remains significant across all quantiles.  This is similar to the finding that higher maximum speed attained is associated with increased driving risk, while average speed is not (\cite{chan_data_2024}).  It only takes one very aggressive or anomalous trip to result in an accident and incur a claim.

As mentioned, to further validate the robustness of these findings, we test two alternative training sets with 0\% and 100\% claim rates, representing two limiting cases.  Despite the change in claim rate, the observed trends remain consistent across all scenarios.  The claim group consistently shows slightly lower average trip-based anomaly indices but higher maximum indices across rentals.  Additionally, the anomaly indices for longitudinal and lateral acceleration in the claim group are 1.5 to 2 times greater than those in the no-claim group.  These persistent observations suggest that the findings are robust and indicative of true underlying behaviour trends.  Detailed results from analyses performed with the two alternative training sets (0\% and 100\% claim rates) are provided in Appendix D.

The next goal of our numerical investigation is to assess the risk associated with each rental.  To achieve this, we aim to differentiate between claimed and no-claim rentals and predict the probability of a rental resulting in an at-fault claim.  Our analysis relies solely on telematics data, with no additional covariates such as demographic or vehicle-specific information included.  Specifically, we consider the following three sets of telematics covariates for predicting the claim probability of each rental:
\begin{enumerate}
    \item Set 1 includes maximum anomaly indices across all telematics response dimensions;
    \item Set 2 includes the right tail of the anomaly index empirical distribution ($\alpha$th-percentiles $P_{\alpha,d}$ for $\alpha=90,\, 95,\, 97,\, 97.5,\, 98,\, 98.5,\, 99,\, 99.5$ and the maximum); and finally
    \item Set 3 combines the right tail from Set 2 and the number of trips as an offset (i.e., with regression coefficient 1).
\end{enumerate}

For each covariate set, we fit a logistic regression and evaluate the model performance based on ROC-AUC.  To robustify the evaluation, we perform 5-fold CV for each set of covariates computed from different pooled CTHMMs (fitted to 4 different training sets).  Our dataset is randomly partitioned into 5 subsets with close claim rates (0.1007, 0.0996, 0.0968, 0.0989 and 0.0986).

\begin{table}[h!]
\caption{5-fold cross-validation ROC-AUC of logistic GLMs with different sets of covariates.  Set 1: maximum anomaly indices, Set 2: right tail of the anomaly index empirical distribution, Set 3: right tail and the number of trips as an offset.}
\centering
\begin{tabular}{c|rrr}
 & \multicolumn{3}{c}{\textbf{Covariate Sets}}\\
 & \textbf{Set 1}& \textbf{Set 2}& \textbf{Set 3}\\\hline
Training Set 1& 0.6901 & 0.7447 & 0.7806 \\
Training Set 2& 0.6878 & 0.7439 & 0.7834 \\
Training Set 3& 0.6824& 0.7467& 0.7869\\
Training Set 4& 0.6967& 0.7174& 0.7707\\\hline
\end{tabular}
\label{tab: AUC-pooled-HMM-10}
\end{table}

In Table \ref{tab: AUC-pooled-HMM-10}, we report the CV performance.  We observe that the average ROC-AUCs are generally stable across anomaly indices computed from pooled CTHMMs trained on different training sets.  (Covariate) Set 1 achieves ROC-AUCs around 0.69, while Set 2 improves to 0.74.  Since the tail and its smoothness depend on the number of observed data points, we incorporate the exposure, which is the number of trips, as an offset.  The resulting ROC-AUCs further improve to around 0.78.  We would like to highlight that first, our ROC-AUC values consistently outperform our collaborator's current benchmark of 0.65, demonstrating the effectiveness of our proposed method; and second, we perform our analysis using solely telematics data, and inclusion of traditional covariates such as driver and vehicle information can definitely further improve the results.

\begin{table}[h!]
\caption{Model summary of logistic GLM with the right tail of the anomaly index empirical distribution and the number of trips as an offset.}
\centering
\begin{threeparttable}
\begin{tabular}{ll|rrrrl}
  && \textbf{Estimate} & \textbf{Std. Error} & \textbf{z value} & \textbf{p value} &  \\\hline\hline
 &(Intercept) & -7.1969 & 0.0726 & -99.1720 & < 2e-16 & *** \\\hline
 Speed&p\_x1\_90 & -83.1564 & 14.6050 & -5.6940 & 0.0000 & *** \\
 &p\_x1\_95 & 20.7476 & 26.3087 & 0.7890 & 0.4303 &  \\
 &p\_x1\_97 & 64.4520 & 81.0186 & 0.7960 & 0.4263 &  \\
 &p\_x1\_97.5 & -63.8849 & 79.6475 & -0.8020 & 0.4225 &  \\
 &p\_x1\_98 & 22.1807 & 42.4384 & 0.5230 & 0.6012 &  \\
 &p\_x1\_98.5 & 16.8307 & 25.0453 & 0.6720 & 0.5016 &  \\
 &p\_x1\_99 & 30.1601 & 10.3361 & 2.9180 & 0.0035 & ** \\
 &p\_x1\_99.5 & -24.7184 & 2.8232 & -8.7560 & < 2e-16 & *** \\
 &max\_x1 & -2.3644 & 0.7786 & -3.0370 & 0.0024 & ** \\\hline
 Longitudinal&p\_x2\_90 & 36.1721 & 8.1520 & 4.4370 & 0.0000 & *** \\
 acceleration&p\_x2\_95 & -35.9972 & 12.6930 & -2.8360 & 0.0046 & ** \\
 &p\_x2\_97 & 9.1265 & 17.2016 & 0.5310 & 0.5957 &  \\
 &p\_x2\_97.5 & -17.8940 & 18.1615 & -0.9850 & 0.3245 &  \\
 &p\_x2\_98 & 29.6963 & 14.4371 & 2.0570 & 0.0397 & * \\
 &p\_x2\_98.5 & 2.2637 & 10.7571 & 0.2100 & 0.8333 &  \\
 &p\_x2\_99 & -22.4719 & 5.9504 & -3.7770 & 0.0002 & *** \\
 &p\_x2\_99.5 & 8.2769 & 1.9463 & 4.2530 & 0.0000 & *** \\
 &max\_x2 & -0.2077 & 0.5316 & -0.3910 & 0.6960 &  \\\hline
 Lateral&p\_x3\_90 & 51.7715 & 12.4382 & 4.1620 & 0.0000 & *** \\
 acceleration&p\_x3\_95 & -24.9194 & 17.2206 & -1.4470 & 0.1479 &  \\
 &p\_x3\_97 & 70.9683 & 26.8603 & 2.6420 & 0.0082 & ** \\
 &p\_x3\_97.5 & -65.0287 & 28.9889 & -2.2430 & 0.0249 & * \\
 &p\_x3\_98 & 2.1981 & 21.3955 & 0.1030 & 0.9182 &  \\
 &p\_x3\_98.5 & -25.7424 & 14.2144 & -1.8110 & 0.0701 & . \\
 &p\_x3\_99 & -4.5599 & 7.5234 & -0.6060 & 0.5445 &  \\
 &p\_x3\_99.5 & 6.1297 & 2.8084 & 2.1830 & 0.0291 & * \\
 &max\_x3 & 1.3960 & 0.8951 & 1.5600 & 0.1189 & \\\hline\hline
\end{tabular}
\begin{tablenotes}
\small
\item Significance codes: 0 `***', 0.001 `**', 0.01 `*', 0.05 `.', 0.1 ` ', 1
\end{tablenotes}
\end{threeparttable}
\label{tab: pooled-HMM-GLM-coefficients}
\end{table}

Finally, we fit the logistic GLM to (covariate) Set 3 of the entire portfolio.  The model summary is presented in Table \ref{tab: pooled-HMM-GLM-coefficients}.  In general, larger percentile values of the anomaly indices increase the claim probability.  Moreover, the right tails of anomaly index empirical distribution in both longitudinal and lateral accelerations appear to have more significant predictive power for claim probability than that in speed, aligning with our previous findings.

\section{Conclusion}
\label{Conclusion}

In this paper, we proposed a CTHMM framework for modelling trip-level vehicle telematics data, enabling analysis and driving risk assessment at both trip and driver levels.  Through a detailed discussion of the telematics data structure and challenges such as irregular time intervals and varying trip lengths, we have highlighted the complexities involved in processing and modelling telematics data.  We stated and explained the key adjustments to the CTHMM and necessary preparation of the telematics data to ensure effective modelling and analysis.

Moreover, the proposed anomaly detection method demonstrated the CTHMM's ability to learn the major, normal driving behaviour and identify outliers in behaviour and trips, even in the absence of direct collision data.  We introduced the anomaly index to quantify the level of anomalousness for each trip, providing a measure of deviation from normal driving patterns.  This method effectively detects anomalous behaviour associated with higher driving risks.  Across both controlled and real-world datasets, the proposed model successfully identifies trips that produce claims for an individual or classifies dangerous rentals within the portfolio.  As our findings are based solely on telematics data, without the use of any covariates, they highlight the value of telematics in trip risk evaluation, collision detection, and driver risk classification.

While the current work enhances our understanding of telematics-based risk assessment, it also suggests directions for future research.  One direction is to incorporate covariates into the transition rate matrix and/or state-dependent distributions.  This will allow the model to better capture the different driving behaviour and enhance model performance.  With a larger pool of drivers or vehicles, we can include covariates which control for driver and vehicle characteristics and unify the individual-specific models into a single regression framework.  Moreover, as telematics data becomes more detailed, including contextual features such as road and weather conditions can further refine the model to detect trip-specific driving risks more effectively.

Another direction is to explore automatic methods for selecting the optimal number of states.  In this work, the number of states in the CTHMM was chosen based on expert judgment.  Developing methods for automated state selection will reduce dependence on manual tuning and improve model adaptability.

A further direction is to extend the current static/offline training approach by incorporating online learning for HMMs.  In this work, models were trained on trip data over relatively short periods of time, such as a 3-day window for individual-specific models in the real data analysis.  Future research can allow the model to dynamically adapt to evolving individual and population driving behaviour.  This will enable continuous risk assessment over longer periods and provide more comprehensive insights into long-term driving risks.

Finally, the proposed anomaly index presents a direction for integration into future ratemaking frameworks to enhance insurance premium pricing.  By leveraging telematics-based risk assessment alongside traditional covariates, insurers can achieve more precise risk classification and fairer pricing, improving underwriting profitability and customer satisfaction. Future research can explore the practical implementation of these methods in ratemaking and evaluate their impact on premium determination.

\section*{Acknowledgement}

The authors acknowledge the financial support provided by the Natural Sciences and Engineering Research Council of Canada [RGPIN 284246, RGPIN-2023-04326] and Canadian Institute of Actuaries [CS000271].

\section*{Conflicts of interest or Competing interests }

The authors declare no conflicts of interest or competing interests in this paper, with no financial or personal affiliations that could compromise the objectivity or integrity of the presented work.

\section*{Data Availability}

The UAH-DriveSet used in the controlled study is publicly available, while the rental car dataset studied in the real data analysis is proprietary data from a telematics company that we are unable to share.


\bibliographystyle{apalike}
\bibliography{references}  

\clearpage
\appendix
\renewcommand{\thetable}{\Alph{section}\arabic{table}}

\newpage
\setcounter{table}{0}
\section{Maximum Likelihood Estimators of Selected State-Dependent Distributions}
\label{appendix-MLE}

Without loss of generality, we provide the MLE for the distribution parameters of the $u$-th latent state in the $m$-th iteration.  To simplify the notation, we let
$\hat{z}^{(m)}_{ijl,u} = \mathbb{E}(\mathbbm{1}_{\{Z_{ijl} = u\}} | \mathbf{Y}^*, \mathbf{T}^*, \boldsymbol{\Phi}^{(m-1)})$.
\begin{itemize}
    \item Gamma($k$, $\theta$) (shape-scale parametrization)
    \begin{align*}
        k^{(m)} &= k \quad\text{such that}\quad \sum_{i,j,l} \hat{z}^{(m)}_{ijl,u} \left\{ \log\Gamma(k) + k \log(\theta^{(m-1)}) - (k-1)\log(y_{ijl}) + \frac{1}{\theta^{(m-1)}} y_{ijl} \right\} = 0 \\
        \theta^{(m)} &= \frac{\sum_{i,j,l} \hat{z}^{(m)}_{ijl,u} y_{ijl}}{{k}^{(m)}\sum_{i,j,l} \hat{z}^{(m)}_{ijl,u}}
    \end{align*}
    \item Laplace($\mu$, $\theta$)
    \begin{align*}
        \mu^{(m)} &= \text{weighted median of $y_{ijl}$ with weights $\hat{z}^{(m)}_{ijl,u}$}\\
        \theta^{(m)} &= \frac{\sum_{i,j,l} \hat{z}^{(m)}_{ijl,u} |y_{ijl} - \mu^{(m)}|}{\sum_{i,j,l} \hat{z}^{(m)}_{ijl,u}}
    \end{align*}
    \item Log-Normal($\mu$, $\sigma^2$)
    \begin{align*}
        \mu^{(m)} &= \frac{\sum_{i,j,l} \hat{z}^{(m)}_{ijl,u} \log(y_{ijl})}{\sum_{i,j,l} \hat{z}^{(m)}_{ijl,u}}\\
        (\sigma^2)^{(m)} &= \frac{\sum_{i,j,l} \hat{z}^{(m)}_{ijl,u} (\log(y_{ijl}) - \mu^{(m)})^2}{\sum_{i,j,l} \hat{z}^{(m)}_{ijl,u}}
    \end{align*}
    \item Normal($\mu$, $\sigma^2$)
    \begin{align*}
        \mu^{(m)} &= \frac{\sum_{i,j,l} \hat{z}^{(m)}_{ijl,u} y_{ijl}}{\sum_{i,j,l} \hat{z}^{(m)}_{ijl,u}}\\
        (\sigma^2)^{(m)} &= \frac{\sum_{i,j,l} \hat{z}^{(m)}_{ijl,u} (y_{ijl} - \mu^{(m)})^2}{\sum_{i,j,l} \hat{z}^{(m)}_{ijl,u}}
    \end{align*}
    \item von Mises($\mu$, $\kappa$)
    \begin{align*}
        \mu^{(m)} &= \arctan\left(\frac{\sum_{j,k}\hat{z}^{(m)}_{ijl,u} \sin(y_{ijl})}{\sum_{j,k}\hat{z}^{(m)}_{ijl,u} \cos(y_{ijl})}\right)\\
        \kappa^{(m)} &= \kappa \quad\text{such that}\quad A(\kappa) = \frac{I_1(\kappa)}{I_0(\kappa)} = \frac{\sum_{i,j,l} \hat{z}^{(m)}_{ijl,u} \cos(y_{ijl} - \mu^{(m)})}{\sum_{i,j,l} \hat{z}^{(m)}_{ijl,u}}
    \end{align*}
    where $I_0(\kappa)$ and $I_1(\kappa)$ are the modified Bessel function of the first kind of orders 0 and 1 respectively.
\end{itemize}

\newpage
\setcounter{table}{0}
\section{Controlled Study: Anomaly Indices from Individual-Specific Models with Fewer States}
\label{appendix-Controlled-Study-Individual-Specific-Model}

\begin{table}[htp]
\caption{UAH-DriveSet: Anomaly index in each dimension of telematics observations, computed from individual-specific CTHMMs with 5 states for each driver and averaged over 100 random initializations.  Note: S stands for Secondary Road and M stands for Motorway; Z and Y are longitudinal and lateral accelerations, respectively.}
\centering
    \begin{tabular}{c|l|rrr}
    \textbf{Driver} & \textbf{Trip} & \textbf{Y\_KF} & \textbf{Z\_KF} & \textbf{Speed}\\\hline
    D1 & S-Normal 1 & 0.0025 & 0.0014 & 0.0000 \\
     & S-Normal 2 & 0.0027 & 0.0032 & 0.0000 \\
     & S-Aggressive & 0.0112 & \textbf{0.0104} & 0.0000 \\
     & S-Drowsy & 0.0043 & \textbf{0.0083} & 0.0000 \\
     & M-Normal & 0.0135 & 0.0020 & 0.0002 \\
     & M-Aggressive & 0.0032 & \textit{0.0079} & 0.0010 \\
     & M-Drowsy & 0.0013 & 0.0034 & 0.0000 \\\hline
    D2 & S-Normal 1 & 0.0031 & 0.0015 & 0.0000 \\
     & S-Normal 2 & 0.0047 & 0.0000 & 0.0000 \\
     & S-Aggressive & 0.0016 & \textbf{0.0317} & 0.0000 \\
     & S-Drowsy & 0.0258 & 0.0078 & 0.0000 \\
     & M-Normal & 0.0063 & 0.0019 & 0.0019 \\
     & M-Aggressive & 0.0129 & \textbf{0.0158} & 0.0108 \\
     & M-Drowsy & 0.0180 & 0.0043 & 0.0001 \\\hline
    D3 & S-Normal 1 & 0.0065 & 0.0029 & 0.0000 \\
     & S-Normal 2 & 0.0121 & 0.0105 & 0.0000 \\
     & S-Aggressive & 0.0114 & \textbf{0.0436} & 0.0000 \\
     & S-Drowsy & 0.0114 & 0.0062 & 0.0000 \\
     & M-Normal & 0.0067 & 0.0038 & 0.0119 \\
     & M-Aggressive & 0.0103 & \textbf{0.0305} & 0.0404 \\
     & M-Drowsy & 0.0150 & 0.0010 & 0.0000 \\\hline
    D4 & S-Normal 1 & 0.0056 & 0.0021 & 0.0000 \\
     & S-Normal 2 & 0.0103 & 0.0022 & 0.0000 \\
     & S-Aggressive & 0.0112 & \textbf{0.0168} & 0.0000 \\
     & S-Drowsy & 0.0235 & 0.0059 & 0.0000 \\
     & M-Normal & 0.0035 & 0.0030 & 0.0070 \\
     & M-Aggressive & 0.0242 & \textbf{0.0275} & 0.0000 \\
     & M-Drowsy & 0.0107 & 0.0019 & 0.0000 \\\hline
    D5 & S-Normal 1 & 0.0000 & 0.0000 & 0.0000 \\
     & S-Normal 2 & 0.0000 & 0.0000 & 0.0044 \\
     & S-Aggressive & 0.0067 & \textbf{0.0116} & 0.0000 \\
     & S-Drowsy & 0.0040 & 0.0013 & 0.0000 \\
     & M-Normal & 0.0033 & 0.0030 & 0.0011 \\
     & M-Aggressive & 0.0088 & \textbf{0.0138} & 0.0000 \\
     & M-Drowsy & 0.0046 & 0.0009 & 0.0000 \\\hline
    \end{tabular}
\label{tab: UAH-individual-5}
\end{table}

\begin{table}[htp]
\caption{UAH-DriveSet: Anomaly index in each dimension of telematics observations, computed from individual-specific CTHMMs with 10 states for each driver and averaged over 100 random initializations.  Note: S stands for Secondary Road and M stands for Motorway; Z and Y are longitudinal and lateral accelerations, respectively.}
\centering
    \begin{tabular}{c|l|rrr}
    \textbf{Driver} & \textbf{Trip} & \textbf{Y\_KF} & \textbf{Z\_KF} & \textbf{Speed}\\\hline
    D1 & S-Normal 1 & 0.0076 & 0.0032 & 0.0000 \\
     & S-Normal 2 & 0.0033 & 0.0000 & 0.0000 \\
     & S-Aggressive & 0.0064 & \textit{0.0056} & 0.0000 \\
     & S-Drowsy & 0.0053 & \textbf{0.0057} & 0.0000 \\
     & M-Normal & 0.0023 & 0.0004 & 0.0010 \\
     & M-Aggressive & 0.0042 & \textbf{0.0064} & 0.0039 \\
     & M-Drowsy & 0.0018 & 0.0032 & 0.0000 \\\hline
    D2 & S-Normal 1 & 0.0026 & 0.0003 & 0.0000 \\
     & S-Normal 2 & 0.0053 & 0.0003 & 0.0000 \\
     & S-Aggressive & 0.0006 & \textbf{0.0174} & 0.0000 \\
     & S-Drowsy & 0.0245 & 0.0051 & 0.0000 \\
     & M-Normal & 0.0051 & 0.0022 & 0.0019 \\
     & M-Aggressive & 0.0104 & \textbf{0.0136} & 0.0060 \\
     & M-Drowsy & 0.0151 & 0.0038 & 0.0000 \\\hline
    D3 & S-Normal 1 & 0.0049 & 0.0028 & 0.0000 \\
     & S-Normal 2 & 0.0073 & 0.0082 & 0.0000 \\
     & S-Aggressive & 0.0092 & \textbf{0.0301} & 0.0000 \\
     & S-Drowsy & 0.0073 & 0.0043 & 0.0000 \\
     & M-Normal & 0.0062 & 0.0017 & 0.0016 \\
     & M-Aggressive & 0.0078 & \textbf{0.0283} & 0.0138 \\
     & M-Drowsy & 0.0083 & 0.0008 & 0.0000 \\\hline
    D4 & S-Normal 1 & 0.0046 & 0.0028 & 0.0000 \\
     & S-Normal 2 & 0.0068 & 0.0023 & 0.0000 \\
     & S-Aggressive & 0.0082 & \textbf{0.0145} & 0.0000 \\
     & S-Drowsy & 0.0184 & 0.0054 & 0.0000 \\
     & M-Normal & 0.0023 & 0.0015 & 0.0028 \\
     & M-Aggressive & 0.0112 & \textbf{0.0239} & 0.0002 \\
     & M-Drowsy & 0.0062 & 0.0005 & 0.0000 \\\hline
    D5 & S-Normal 1 & 0.0001 & 0.0003 & 0.0000 \\
     & S-Normal 2 & 0.0030 & 0.0027 & 0.0046 \\
     & S-Aggressive & 0.0056 & \textbf{0.0147} & 0.0000 \\
     & S-Drowsy & 0.0042 & 0.0019 & 0.0000 \\
     & M-Normal & 0.0009 & 0.0025 & 0.0010 \\
     & M-Aggressive & 0.0105 & \textbf{0.0112} & 0.0000 \\
     & M-Drowsy & 0.0056 & 0.0011 & 0.0000 \\\hline
    \end{tabular}
\label{tab: UAH-individual-10}
\end{table}

\begin{table}[htp]
\caption{UAH-DriveSet: Anomaly index in each dimension of telematics observations, computed from individual-specific CTHMMs with 15 states for each driver and averaged over 100 random initializations.  Note: S stands for Secondary Road and M stands for Motorway; Z and Y are longitudinal and lateral accelerations, respectively.}
\centering
    \begin{tabular}{c|l|rrr}
    \textbf{Driver} & \textbf{Trip} & \textbf{Y\_KF} & \textbf{Z\_KF} & \textbf{Speed}\\\hline
    D1 & S-Normal 1 & 0.0105 & 0.0043 & 0.0000 \\
     & S-Normal 2 & 0.0055 & 0.0015 & 0.0000 \\
     & S-Aggressive & 0.0085 & \textit{0.0060} & 0.0000 \\
     & S-Drowsy & 0.0090 & \textbf{0.0062} & 0.0000 \\
     & M-Normal & 0.0033 & 0.0012 & 0.0001 \\
     & M-Aggressive & 0.0049 & \textbf{0.0092} & 0.0027 \\
     & M-Drowsy & 0.0021 & 0.0038 & 0.0000 \\\hline
    D2 & S-Normal 1 & 0.0022 & 0.0006 & 0.0000 \\
     & S-Normal 2 & 0.0043 & 0.0002 & 0.0000 \\
     & S-Aggressive & 0.0017 & \textbf{0.0197} & 0.0003 \\
     & S-Drowsy & 0.0226 & 0.0048 & 0.0000 \\
     & M-Normal & 0.0032 & 0.0019 & 0.0011 \\
     & M-Aggressive & 0.0082 & \textbf{0.0163} & 0.0019 \\
     & M-Drowsy & 0.0146 & 0.0041 & 0.0000 \\\hline
    D3 & S-Normal 1 & 0.0047 & 0.0021 & 0.0000 \\
     & S-Normal 2 & 0.0072 & 0.0056 & 0.0000 \\
     & S-Aggressive & 0.0099 & \textbf{0.0216} & 0.0000 \\
     & S-Drowsy & 0.0075 & 0.0040 & 0.0000 \\
     & M-Normal & 0.0057 & 0.0020 & 0.0049 \\
     & M-Aggressive & 0.0057 & \textbf{0.0225} & 0.0004 \\
     & M-Drowsy & 0.0080 & 0.0006 & 0.0000 \\\hline
    D4 & S-Normal 1 & 0.0046 & 0.0031 & 0.0000 \\
     & S-Normal 2 & 0.0083 & 0.0015 & 0.0000 \\
     & S-Aggressive & 0.0098 & \textbf{0.0143} & 0.0000 \\
     & S-Drowsy & 0.0169 & 0.0032 & 0.0000 \\
     & M-Normal & 0.0022 & 0.0020 & 0.0021 \\
     & M-Aggressive & 0.0146 & \textbf{0.0187} & 0.0002 \\
     & M-Drowsy & 0.0059 & 0.0011 & 0.0000 \\\hline
    D5 & S-Normal 1 & 0.0002 & 0.0009 & 0.0000 \\
     & S-Normal 2 & 0.0022 & 0.0022 & 0.0043 \\
     & S-Aggressive & 0.0065 & \textbf{0.0128} & 0.0000 \\
     & S-Drowsy & 0.0056 & 0.0027 & 0.0000 \\
     & M-Normal & 0.0011 & 0.0031 & 0.0005 \\
     & M-Aggressive & 0.0102 & \textbf{0.0072} & 0.0000 \\
     & M-Drowsy & 0.0059 & 0.0011 & 0.0000 \\\hline
    \end{tabular}
\label{tab: UAH-individual-15}
\end{table}

\clearpage
\newpage
\mbox{~}
\setcounter{table}{0}
\section{Real-Data Analysis: Summaries of Logistic Models from Individual-Specific Models}
\label{appendix-Real-Data-Analysis-Individual-Specific-Model}

\begin{table}[hp]
\caption{Model summary of logistic GLMs, with raw or normalized anomaly indices as covariates.  Note: x1, x2 and x3 denote indices for speed, longitudinal and lateral accelerations respectively.}
\label{tab: GLM-coefficients}
\centering
\begin{threeparttable}
\begin{subtable}{1\textwidth}
    \centering
    \begin{tabular}{c|rrrrl}
         &  \textbf{Estimate}&  \textbf{Std. Error}&  \textbf{z value}&  \textbf{p value}&  \\\hline
         (Intercept)&  -3.395&  0.137&  -24.825&  < 2e-16&  ***\\
         x1&  -23.146&  20.594&  -1.124&  0.261&  \\
         x2&  51.219&  8.630&  5.935&  2.93e-09&  ***\\
         x3&  63.827&  9.416&  6.779&  1.21e-11&  ***\\\hline\hline
    \end{tabular}
\end{subtable}

\bigskip
\begin{subtable}{1\textwidth}
    \centering
    \begin{tabular}{c|rrrrl}
         &  \textbf{Estimate}&  \textbf{Std. Error}&  \textbf{z value}&  \textbf{p value}&  \\\hline
         (Intercept)&  -4.316&  0.206&  -20.961&  <2e-16&  ***\\
         norm x1&  0.309&  0.557&  0.555&  0.579&  \\
         norm x2&  2.770&  0.298&  9.305&  <2e-16&  ***\\
         norm x3&  2.903&  0.292&  9.957&  <2e-16&  ***\\\hline\hline
    \end{tabular}
\begin{tablenotes}
\small
\item Significance codes: 0 `***', 0.001 `**', 0.01 `*', 0.05 `.', 0.1 ` ', 1
\end{tablenotes}
\end{subtable}
\end{threeparttable}
\end{table}

\setcounter{table}{1}
\begin{table}[hp]
\caption{Feature importance of logistic XGBoost, with raw or normalized anomaly as covariates.  Note: x1, x2 and x3 denote indices for speed, longitudinal and lateral accelerations respectively.}
\label{tab: XGB-feature-importance}
\begin{subtable}{1\textwidth}
    \centering
    \begin{tabular}{c|rrr}
         &  \textbf{Gain}&  \textbf{Cover}&  \textbf{Frequency}\\\hline
         x3&  0.557&  0.492&  0.447\\
         x2&  0.382&  0.401&  0.392\\
         x1&  0.061&  0.106&  0.161\\\hline\hline
    \end{tabular}
\end{subtable}

\bigskip
\centering
\begin{subtable}{1\textwidth}
    \centering
    \begin{tabular}{c|rrr}
         &  \textbf{Gain}&  \textbf{Cover}&  \textbf{Frequency}\\\hline
         norm x3&  0.537&  0.530&  0.389\\
         norm x2&  0.430&  0.414&  0.465\\
         norm x1&  0.033&  0.056&  0.146\\\hline\hline
    \end{tabular}
\end{subtable}
\end{table}

\clearpage
\newpage
\mbox{~}
\setcounter{table}{0}
\section{Real-Data Analysis: Pooled Model Analysis with Alternative Training Sets}
\label{appendix-Real-Data-Analysis-Pooled-Model}

\begin{table}[hp]
\caption{Comparison of anomaly indices (\%) in each telematics response dimension for claimed and no-claim rentals, computed from pooled CTHMMs trained on samples with 0\% and 100\% claim rates.}
\centering
\begin{threeparttable}

    \begin{tabular}{l|rr|rr}
    \multicolumn{5}{c}{\textbf{Trip-Basis}} \\\hline
     & \multicolumn{2}{c|}{\textbf{0\%}} & \multicolumn{2}{c}{\textbf{100\%}} \\
     & claimed & no claim & claimed & no claim \\\hline
    speed & 0.0716 & 0.0995 & 0.0462 & 0.0523 \\
    longitudinal & 0.5405 & 0.6549 & 0.2473 & 0.3008 \\
    lateral & 0.3048 & 0.3517 & 0.2178 & 0.2384 \\\hline
    \end{tabular}

    \vspace{0.5cm} 

    \begin{tabular}{l|rr|rr}
    \multicolumn{5}{c}{\textbf{Rental-Basis}} \\\hline
     & \multicolumn{2}{c|}{\textbf{0\%}} & \multicolumn{2}{c}{\textbf{100\%}} \\
     & claimed & no claim & claimed & no claim \\\hline
    \textbf{Average} & & & & \\
    speed & 0.0713 & 0.0968 & 0.0433 & 0.0559 \\
    longitudinal & 0.6246 & 0.8882 & 0.2680 & 0.3936 \\
    lateral & 0.3190 & 0.3801 & 0.2219 & 0.2711 \\\hline
    \textbf{Maximum} & & & & \\
    speed & 32.3730 & 17.2720 & 25.2180 & 12.0900 \\
    longitudinal & 35.8500 & 20.8000 & 28.2700 & 14.8470 \\
    lateral & 33.9200 & 18.8010 & 27.2260 & 13.9027 \\\hline
    \end{tabular}

\end{threeparttable}
\label{tab: Sum-Stats-pooled-hmm-0-100}
\end{table}

\begin{table}[hp]
\caption{5-fold cross-validation ROC-AUC of logistic GLMs with different sets of covariates.  Set 1: maximum anomaly indices, Set 2: right tail of the anomaly index empirical distribution, Set 3: right tail and the number of trips as an offset.  Note: first column indicates the claim rate of the training set the pooled CTHMM is fitted on.}
\centering
\begin{tabular}{c|rrr}
 & \multicolumn{3}{c}{\textbf{Covariate Sets}}\\
 & \textbf{Set 1} & \textbf{Set 2} & \textbf{Set 3} \\\hline
0\% Training Set& 0.6905 & 0.7509 & 0.7861\\
100\% Training Set& 0.6962 & 0.7286 & 0.7669 \\\hline
\end{tabular}
\label{tab: AUC-pooled-HMM-0-100}
\end{table}

\end{document}